\documentclass[letterpaper]{article}

\usepackage[T1]{fontenc}

\usepackage{geometry}
\usepackage{setspace}

\newcommand{\eps}{\varepsilon}
\newcommand{\veps}{\varepsilon_{\mathbf{0}}}
\newcommand{\bfg}{\mathbf{g}}
\newcommand{\bfk}{\mathbf{k}}
\newcommand{\bfr}{\mathbf{r}}
\newcommand{\bfe}{\mathbf{E}}

\newcommand{\iu}{\mathrm{i}\mkern1mu}
\newcommand{\eu}{\mathrm{e}\mkern1mu}
\newcommand{\opp}[1]{\hat{\mathbf{#1}}} % 
\newcommand{\intz}{\int_{-h}^{h} \mathrm{d}z\,}
\newcommand{\intzp}{\int_{-h}^{h} \mathrm{d}z'\,}

\usepackage{braket}
\usepackage{verbatim}

\usepackage{achemso}

\usepackage{graphicx}
\usepackage{float}
\newfloat{scheme}{htbp}{los}
\floatname{scheme}{Scheme}
\floatname{chart}{Chart}
\newfloat{graph}{htbp}{loh}

\usepackage{chemformula} % Formulas using \ch{}
\usepackage[version = 4]{mhchem} % Formulas using \ce{}

\usepackage{authblk}
\author[a]{Polina Pantiukhina}
\author[a,b,*]{Daria Smirnova}
\author[a,*]{Kirill Koshelev}
\affil[a]{Department of Electronic Materials Engineering, Research School of Physics, Australian National University, Canberra, ACT 2601, Australia}
\affil[b]{Department of Fundamental and Theoretical Physics, Research School of Physics, Australian National University, Canberra, ACT 2601, Australia}

\title{Band Engineering of Exciton Polaritons in Resonant Polaritonic Metasurfaces}
\date{}
\begin{document}

\maketitle

\begin{abstract}
Polaritonic metasurfaces provide a versatile platform for engineering hybrid light-matter states through the interplay of optical resonances and excitonic excitations. Yet, predictive models often remain phenomenological and rely on coupled-mode equations. Here, we develop an effective Hamiltonian framework for exciton polaritons in resonant polaritonic metasurfaces, derived from a semiclassical single-pole description of excitonic polarization and a Green’s-function description of guided-mode resonances.
The resulting non-Hermitian Hamiltonian rigorously incorporates resonant photonic harmonics, multiple excitonic degrees of freedom, and radiative losses. 
The model establishes selection rules for photonic-excitonic coupling governed by orbital multipole index and polarization and identifies one bright excitonic degree of freedom per retained guided harmonic.
We apply the framework to a bulk van der Waals WS$_2$ metasurface patterned into a hexagonal lattice of triangular holes and uncover 
a new 
{\it geometry-controlled} topological transition 
driven by dipole-quadrupole band inversion, distinct from the conventional breathing-honeycomb-lattice transition. 
Full-wave simulations confirm the predicted topological phase diagram and the emergence of photonic and polaritonic edge states at a topological interface.
Our results establish a theoretical multimode framework for geometry-controlled bandstructure engineering in polaritonic metasurfaces, with applications in topological, chiral, and quantum integrated photonics.
\end{abstract}

\section{Introduction}

Polaritonic metasurfaces created by the coupling of excitons with optical modes of structured materials recently opened new avenues for light manipulation at the nanoscale through engineered light-matter interactions~\cite{basov2016polaritons,huang2026advances}. In this context, optically resonant metasurfaces integrated with monolayers of semiconductor materials~\cite{wang2018colloquium,kang2023exciton} offered a route to achieving strong light-matter coupling~\cite{zhang2018photonic,chen2020metasurface}, leading to ultrahigh nonlinearities~\cite{kravtsov2020nonlinear} and boson condensation~\cite{ardizzone2022polariton,wu2024exciton}. More recently, nanofabrication advances enabled metasurfaces and individual metastructures made directly of polaritonic materials~\cite{biechteler2025fabrication}, such as bulk van der Waals (vdW) metasurfaces~\cite{verre2019transition,weber2023intrinsic,bouteyre2025simultaneous,guddala2025topological}. Such vdW polaritonic metasurfaces % unify 
combine excitonic and photonic degrees of freedom within a single structure, % expanding light-matter-coupling functionalities for the enhancement of
facilitating enhanced light–matter coupling for optical harmonic generation~\cite{zograf2024combining,ling2025nonlinear,zograf2025ultrathin}, spontaneous parametric downconversion~\cite{fan2025enhanced}, and chiral response~\cite{tonkaev2024nonlinear,heimig2026chiral}.

Optical properties of polaritonic metasurfaces are commonly analyzed in the context of semiclassical excitations of the electromagnetic field in a polarizable medium~\cite{kwek2013strong}. In this approach, the excitonic response is introduced as a single-pole nonlocal polarization function defined by the exciton frequency, non-radiative lifetime, and oscillator strength~\cite{ivchenko2005optical,deych2007exciton}. In the case of non-resonant photonic metastructures, this model predicts exciton fine structure and spin decoherence due to long-range interactions~\cite{glazov2014exciton,glazov2015spin,prazdnichnykh2021control}. In the case of resonant metasurfaces, the model allows to compute the photonic response via modal expansion methods, such as plane-wave expansion~\cite{takeda2012exciton}, guided-mode expansion~\cite{andreani2006photonic}, or through resonant Green's function~\cite{pantyukhina2025excitonic,wurdack2026intrinsically}. On the other hand, a more precise approach based on Hopfield quantum theory allows one to account for excitonic fine structure arising from periodic-potential quantization and highlight the importance of multi-excitonic Hamiltonians~\cite{gerace2007quantum,zanotti2022theory}. These approaches often rely either on phenomenological parameters and simplified descriptions or on complex numerical methods resulting in large Hamiltonian matrices.

A simple framework for constructing effective Hamiltonians for polaritonic metasurfaces that accurately captures the essential physics and establishes the minimal number of interacting modes has not yet been investigated.  The need for such a framework is further underscored by the active development of topological photonics~\cite{smirnova2024topological} and, more recently, topological polaritonics~\cite{karzig2015topological,kim2025van}. In topological systems, the effective Hamiltonian parameters, commonly obtained using the Bernevig–Hughes–Zhang-like model~\cite{bernevig2006quantum} or tight-binding models~\cite{gorlach2018far,li2021experimental}, allow one to % establish 
define the topological phase and (spin) Chern number, which can be used to %characterize 
describe propagating edge states and localized defect modes~\cite{smirnova2024polaritonic}. Topological polaritonic metasurfaces provide a platform for realizing helical exciton-polariton edge waves~\cite{liu2020generation,guddala2025topological} and polaritonic flatbands~\cite{heimig2025topological}.

In this paper, we develop a systematic effective Hamiltonian framework for exciton-polaritons in optically resonant polaritonic metasurfaces. Using a bulk vdW metasurface with $C_{6v}$ point group symmetry as an example, we transform Maxwell’s equations into the form of a non-Hermitian effective Hamiltonian that describes the interaction of metasurface quasi-guided modes and excitonic degrees of freedom via near-field and radiative coupling. For this, we utilize the expansion of Green’s function into guided modes and apply the spatial coupled-mode theory to establish the non-Hermitian radiative terms. For the $C_{Nv}$ model considered here, the $N$ retained harmonics of a single guided mode couple to $N$ bright excitonic combinations. With degenerate excitons, the orthogonal excitonic combinations remain dark. At the $\Gamma$ point, the Hamiltonian separates into $N$ photon-exciton blocks:
\begin{equation}
\hat{H}=\left(\begin{array}{ccccc}
\Omega_1 & V^* & \cdots & 0 & 0 \\
V & {E}_{\rm X}  & \cdots & 0 & 0 \\
\vdots & \vdots & \ddots & 0  & 0 \\
0 & 0 & 0 & \Omega_N & V^* \\
0 & 0 &  0 & V & {E}_{\rm X}
\end{array}\right),  
\label{eq:0}
\end{equation}
where $\Omega_1,\ldots,\Omega_N$ and ${E}_{\rm X}$ are photonic and excitonic mode frequencies, respectively, and $V$ is the light-matter coupling strength. The model, therefore, complements the recent findings for polaritonic states in planar non-periodic multimode cavities~\cite{balasubrahmaniyam2021coupling,mandal2023microscopic} with the symmetry- and polarization-imposed selection rules specific for polaritonic metasurfaces. We validate the model with full-wave numerical simulations and apply it to predict a new form of topological transition for a dipole-quadrupole pair of bands that depends on the meta-element and unit-cell geometry rather than the lattice arrangement. We confirm numerically that the predicted {\it geometry-controlled} topological transition enables the formation of photonic and polaritonic edge states at a topological interface.

\section{Model description} 

\begin{figure}[t]
\centering
\includegraphics[width=1\linewidth]{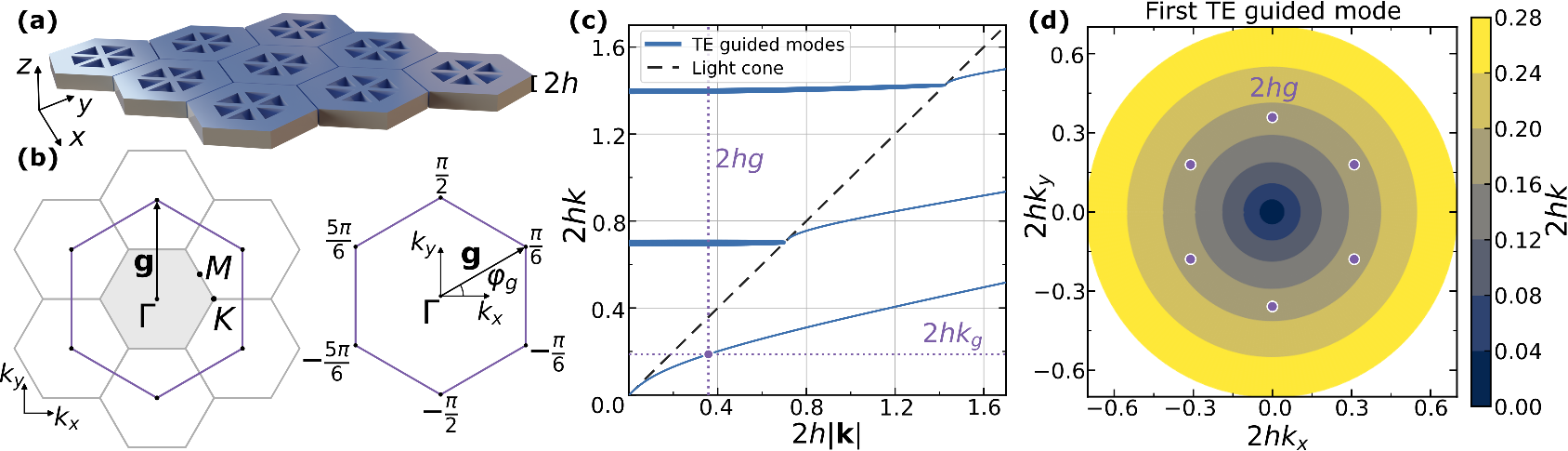}
\caption{{\bf Metasurface design and effective slab model.} (a) Schematic of polaritonic WS$_2$ metasurface of $2h$ thickness with $C_{6v}$ symmetry of the lattice and unit cell; (b) Reciprocal-space geometry. The purple hexagon denotes $|\bfg|=g$, and the first Brillouin zone is shaded in gray. (c) $\rm s$-polarized guided modes of the effective homogeneous slab with $\veps=19$ and $2h=20$ nm, the area average for the metasurface of Fig.~\ref{fig:3} with the slab material permittivity $\eps=20.25$. The dashed line shows the light cone, and the purple dotted lines indicate the intersection of $|\mathbf{k}|= g$ and the fundamental guided mode dispersion. The linewidth is proportional to mode radiative losses, while the guided branches below the light cone are lossless; (d) Equal frequency contours of the fundamental guided mode. Purple markers show $|\bfg|=g$. 
%\textcolor{blue}{comments to Fig.1: change $k_{\parallel}$  to $|\mathbf{k}|$; remove $k_g^{(2)}$ and second purple dotted line; change $k_g^{(1)}$ to $k_g$; is x-y in (a) aligned with $k_x$,$k_y$ in (b)?; remove $0$ from $\Gamma_0$; increase all fonts; $2h$ in (a) and $M,K,\bfg$ in (b) overlap with other parts of the panel -- fix.}
}
\label{fig:1}
\end{figure}

In this section, we develop a semiclassical effective Hamiltonian describing the light-matter interaction between excitons and optical modes of polaritonic metasurfaces based on the framework that combines the theory of exciton-polaritons for periodic photonic crystals with quantum wells~\cite{ardizzone2022polariton}, spatial coupled-mode theory~\cite{liang2011three} and non-Hermitian approaches inspired by the resonant state expansion~\cite{neale2020resonant}. The structure schematically shown in Fig.~\ref{fig:1}(a) is periodic in the $xy$-plane and confined within a slab of finite thickness $2h$ along the $z$-axis. We focus on structures with $C_{6v}$ symmetry, which offer additional opportunities for realizing topological polaritonic phenomena, as discussed below, while the results are generally applicable to metasurfaces with other symmetries.

\subsection{Photonic response} The permittivity profile $\eps(\bfr)$ is defined as $\eps(x,y)$ inside the slab and $1$ outside of it. Assuming harmonic time-dependence of fields $\eu^{-\iu\omega t}$, the vector Helmholtz equation for the electric field $\bfe(\textbf{r})$ is~\cite{ivchenko2005optical}
\begin{equation}
\left[ k^2\eps(\textbf{r})- \nabla\times\nabla\times\right]\bfe(\textbf{r}) = -4\pi k^2 \mathbf{P}(\bfr),\label{eq:1}
\end{equation}
where $k=\omega/c$ and $\mathbf{P}(\bfr)$ is the nonlocal excitonic contribution to the dielectric polarization that will be discussed later.

To exploit the in-plane periodicity, we apply Bloch's theorem. The electric field, polarization, permittivity, and del operator are expanded into Fourier series over the reciprocal lattice vectors $\bfg=(g_x, g_y, 0)$ with the in-plane Bloch wavevector $\bfk=(k_x, k_y, 0)$:
\begin{equation}
\begin{split}
\bfe(\textbf{r})=\frac{1}{\sqrt{S_{\rm u}}}&\sum_{\bfg}\eu^{\iu (\bfg+\bfk)\cdot\textbf{r}}\bfe_{\bfg}(z), \quad \mathbf{P}(\bfr)=\frac{1}{\sqrt{S_{\rm u}}}\sum_{\bfg}\eu^{\iu (\bfg+\bfk)\cdot\textbf{r}}\mathbf{P}_{\bfg}(z),\\
\eps(\textbf{r})=&\sum_\bfg\eu^{\iu \bfg\cdot\textbf{r}}\eps_{\bfg}(z),\quad \nabla \times \bfe(\bfr)= \sum_\bfg\eu^{\iu (\bfg+\bfk)\cdot\textbf{r}}\nabla_{z,\bfg+\bfk}\times \bfe_{\bfg+\bfk}(z),
\end{split}
\label{eq:2}
\end{equation}
where $S_{\rm u}$ represents the unit cell area, the modified gradient operator is $\nabla_{z,\bfg}=\left(\iu g_x, \iu g_y, \partial_z\right)$. We focus on the physical effects governed by the symmetry of the resonant modes at the $\Gamma$-point ($\bfk=\mathbf 0$), while finite-$\bfk$ corrections can be included perturbatively as described in Sec.~S1A of the Supplemental Material (SM). Following the coupled-mode approaches~\cite{andreani2006photonic,liang2011three,neale2020resonant}, we separate the permittivity into one of a homogeneous effective slab $\veps(z)$ and a periodic perturbation $\Delta\eps_{\bfg}(z)=\eps_{\bfg}(z) - \veps(z) \delta_{\bfg,\bf0}$.
Substituting Eq.~\eqref{eq:2} into Eq.~\eqref{eq:1} and expressing $\bfe_{\bfg}(z)$ in the form of the Lippmann--Schwinger equation yields
\begin{equation}
\bfe_{\bfg}(z) = -k^2\sum_{\bfg'}\intzp\opp{G}_{\bfg}(z,z')\Delta\eps_{\bfg-\bfg'}(z')\bfe_{\bfg'}(z')-4\pi k^2 \intzp\opp{G}_{\bfg}(z,z')\mathbf{P}_\bfg(z'),
\label{eq:3}
\end{equation}
where the dyadic Green's function $\opp{G}_{\bfg}(z,z')$ satisfies $[k^2\veps(z) - \nabla_{z,\bfg}\times\nabla_{z,\bfg}\times]\opp{G}_{\bfg}(z,z')=\hat{\mathbf{1}}\delta(z-z')$.
For the selected $C_{6v}$ symmetry, Eq.~\eqref{eq:3} provides the description of the metasurface resonant response in terms of six guided modes of the effective slab with $\veps(z)$ at $|\bfg|=g$ and multiple radiative leaky modes of the slab at $\bfg=\mathbf{0}$ coupled via the periodic perturbation $\Delta\eps_{\bfg-\bfg'}$, see Figs.~\ref{fig:1}(b,c).

In the following, we assume that the GF  of the effective slab at $|\bfg|=g$ is dominated by a single $\rm s$-polarized guided mode with the frequency  $k_{g}=\omega_g/c$,
\begin{equation}
\opp{G}_{\bfg}(z,z')
\simeq
\frac{
\boldsymbol{\theta}_{\bfg}(z)
\otimes
\boldsymbol{\theta}_{\bfg}(z')
}{
k(k-k_{g})
}.
\label{eq:4}
\end{equation}
An example of such $\rm s$-polarized mode dispersion is shown in Fig.~\ref{fig:1}(c,d) for the effective slab of the metasurface in Fig.~\ref{fig:3}, where the slab material has $\eps = 20.25$ that corresponds to WS$_2$ in the optical range, as discussed  further in Eq.~\eqref{eq:18}, and its area average over the unit cell gives $\eps_{\bf0}\simeq19$. The mode profile can be therefore written as
$\boldsymbol{\theta}_{\bfg}(z) = \theta(z)\hat{\mathbf e}_{\rm s,\bfg}$ with the polarization vector  $\hat{\mathbf e}_{\rm s,\bfg} = -\sin(\varphi_{\textbf{g}})\hat{\mathbf e}_x  +\cos(\varphi_{\textbf{g}})\hat{\mathbf e}_y$ determined by the angle
$\varphi_{\bfg}$, illustrated in Fig.~\ref{fig:1}(b). In this approximation, the full electric field at $|\bfg|=g$ can be projected onto the guided-mode profile as $\bfe_{\bfg}(z) =\eta\hbar c  \boldsymbol{\theta}_{\bfg}(z)a_{{\rm s},\bfg}$,
where $\eta=\sqrt{4\pi k_{g}/(\hbar c)}$, and $a_{{\rm s},\bfg}$ is the semiclassical modal amplitude associated with a bosonic annihilation operator in the quantum limit. 

% All prefactors in the effective Hamiltonian are evaluated at the first-shell TE guided-mode wavenumber $k_g$, while $k$ remains the spectral parameter.

Substituting Eq.~\eqref{eq:4} and field expansion into Eq.~\eqref{eq:3}, we obtain for $|\bfg|=g$
\begin{equation}
(k-k_{g})a_{{\rm s},\bfg} = -{k_{g} N_h}\sum_{\bfg'\neq 0}(\hat{\mathbf e}_{\rm s,\bfg}\cdot\hat{\mathbf e}_{\rm s,\bfg'})\Delta\eps_{\bfg-\bfg^{\prime}}a_{{\rm s},\bfg'}-\intz \boldsymbol{\theta}_{\bfg}(z)\left[\dfrac{k_{g}\Delta\eps_{\bfg}}{(\eta\hbar c)}\bfe_{\bf 0}(z)+{\eta} \mathbf{P}_\bfg(z)\right],
\label{eq:5}
\end{equation}
where $N_h=\intz \theta^2(z)$. Below, we show that the non-Hermitian contribution of $\bfe_{\bf 0}(z)$ can be eliminated from equations due to its non-resonant nature similar to the approach used in spatial coupled mode theory~\cite{liang2011three}.

\subsection{Excitonic response} 

We treat excitons within the envelope function approximation, yielding the two-body wavefunction of electron and hole. It can be separated into $\Phi(x,y)$ and $\varphi(r,z)=\varphi(r)F(z)$ describing the center-of-mass in-plane electron-hole motion along $x,y$, relative radial electron-hole motion along $r$, and out-of-plane exciton motion as a whole along $z$~\cite{ivchenko2005optical}. The standard expression for exciton-induced polarization yields~\cite{glazov2014exciton}
\begin{equation}
\mathbf{P}(x,y,z)=\frac{\sum_\alpha\textbf{d}_\alpha^*\otimes\textbf{d}_\alpha }{\hbar(\omega_{\rm X}-\omega-\iu\Gamma)}F(z)\intzp F^*(z')\bfe(x,y,z'),
\label{eq:6}
\end{equation}
where $\omega_{\rm X}$ and $\Gamma$  are bare exciton frequency and nonradiative loss rate, respectively, $\alpha$ is the exciton polarization index. The dipole moment $\mathbf{d}_\alpha=d\hat{\mathbf{e}}_\alpha$, $d=-\iu  \varphi^*(0) ep_{\rm cv}/(m_0\omega_{\rm X})$ is defined via the polarization vector $\hat{\mathbf{e}}_\alpha$ and microscopic parameters $m_0$, $p_{\rm cv}$. For a bulk polaritonic metasurface,  $F(z)=\cos{[\pi z/(2h)]}/\sqrt{h}$~\cite{ivchenko2005optical}. The model can be also applied to a two-dimensional excitonic layer integrated with a photonic metasurface by replacing $F(z)$ with  $\delta(z)$~\cite{pantyukhina2025excitonic} (see more details in Sec.~S1B of the SM). 

In our model, we extend the approach of Eq.~\eqref{eq:6} by considering the effect of the periodic potential $U(x,y)$ induced by the permittivity profile on the center-of-mass in-plane envelope
wavefunction $\Phi_\alpha(x,y)$ following the established framework for polaritonic photonic crystal slabs with embedded quantum wells~\cite{gerace2007quantum,zanotti2022theory}. Since excitons exist only inside the material, $U(x,y)=0$ inside the material and $U(x,y)=U_{\rm max}$ inside the holes, where $U_{\rm max}$ is the maximal quantum barrier height. We can conveniently write this form as $U(x,y)=U_{\rm max}[1-\xi(x,y)]$, where $\xi(x,y)$ is an auxiliary step function equal to $1$ inside the material and $0$ inside the holes. By expressing the polarization as
$\mathbf{P}(\bfr)=F(z)\sum_\alpha\textbf{d}_\alpha^*\Phi_\alpha(x,y)$, we can write a Schr\"odinger-like equation for $\Phi_\alpha(x,y)$ from Eq.~\eqref{eq:6} as (see Sec.~S1B of the SM for more details):
\begin{equation*}
\left[\hbar\omega-\hbar\omega_{\rm X}+\iu\hbar\Gamma+\frac{\hbar^2 \nabla^2}{2 M}   -U(x,y)\right]  \Phi_\alpha(x,y) = -\intz\, F^*(z)\textbf{d}_\alpha\cdot\bfe(\bfr),
\label{eq:65}
\end{equation*}
where $M$ is the exciton translational mass. For the sake of simplicity, we consider the hard wall limit $U_{\rm max}\to\infty$, in which the potential reduces to the boundary condition $\Phi_\alpha=0$ in the holes. Then, Eq.~\eqref{eq:65} can be written as
\begin{equation}
\left[\hbar\omega-\hbar\omega_{\rm X}+\iu\hbar\Gamma+\frac{\hbar^2 \nabla^2}{2 M}  \right]  \Phi_\alpha(x,y) = -\xi(x,y)\intz\, F^*(z)\textbf{d}_\alpha\cdot\bfe(\bfr),
\label{eq:7}
\end{equation}
i.e., excitons are driven only by the field inside the material. Comparing with Eq.~\eqref{eq:6}, one can see that $\mathbf{P}(x,y,z)$ is now proportional to $\xi(x,y)$, thus, the exciton-induced polarization is nonzero only within the material.

% the Bohr radius $a_{\rm B}\ll a$ yielding $\varphi(0)=1/{\sqrt{\pi a^3_{\rm B}}}$ for a hydrogen-like 1s-exciton and 

We further apply the Fourier expansion to Eq.~\eqref{eq:7} with $\Phi_\alpha(x,y)\to\Phi_{\alpha,\bfg}$, $\xi(x,y)\to \xi_\bfg$. Since the background permittivity of the slab is $\eps(x,y)=1+(\eps-1)\xi(x,y)$, the step function has the same Fourier coefficients as the permittivity: $\xi_{\mathbf 0}=(\veps-1)/(\eps-1)$ is the area fraction of the material, and $\xi_{\bfg}=\Delta\eps_{\bfg}/(\eps-1)$ for $\bfg\neq\mathbf 0$. The confinement therefore introduces no additional parameters. Instead of using the polarization basis fixed to the coordinate frame, we use the basis of $\alpha=\{\rm s,p\}$ polarized excitons with dipole moments $\textbf{d}_{\alpha, \bfg} =d \hat{\mathbf e}_{\alpha,\bfg}$. The term $\xi_{\bfg}$ with $\bfg'=\bfg$ restricts the excitons to material on average over the unit cell. The terms with $\bfg'\neq\bfg$ are of relative order $\xi_{\bfg-\bfg'}/\xi_{\mathbf 0}=\Delta\eps_{\bfg-\bfg'}/(\veps-1)$, i.e., of the order of the dielectric modulation. For the photons, the same $\Delta\eps_{\bfg-\bfg'}$ couples six degenerate guided modes in Eq.~\eqref{eq:5} and therefore determines their splitting already in the first order. For the excitons at $|\bfg|=g$, these terms do not couple the excitons to each other. They only correct the light-matter coupling by a small relative amount of order $\Delta\eps_{\bfg-\bfg'}/(\veps-1)$. We therefore neglect these terms and introduce the semiclassical excitonic (bosonic) amplitudes 
$b_{\alpha,\bfg}=\Phi_{\alpha,\bfg}/\sqrt{\xi_{\bf 0}}$. Equation~\eqref{eq:7}, therefore, transforms to 
\begin{equation}
(k-E_{\rm X})b_{\alpha,\bfg} =-\dfrac{\sqrt{\xi_{\bf 0}}}{\hbar c}\intz F(z)\textbf{d}_{\alpha,\bfg}\cdot\bfe_\bfg(z),
\label{eq:8}
\end{equation}
where $E_{\rm X} = (
\omega_{\rm X} -\iu\Gamma )/c$ and $\hat{\mathbf e}_{\rm p,\bfg} = \cos(\varphi_{\textbf{g}})\hat{\mathbf e}_x  +\sin(\varphi_{\textbf{g}})\hat{\mathbf e}_y$. The center-of-mass kinetic energy of the exciton
${\hbar^2(\bfk+\bfg)^2}/(2M)$ can be neglected at the scale of photonic dispersion.

% \begin{equation}
% \left[ \hbar\omega-\hbar\omega_{\rm X}+\iu\hbar\Gamma-\frac{\hbar^2 (\textbf{k}+\bfg)^2}{2 M}      - U_{\mathbf{0}} \right] {\Phi}_{\alpha,\bfg} -\sum_{\bfg'\neq\bfg} U_{\bfg-\bfg'}{\Phi}_{\alpha,\bfg'} = -\intz\, F^*(z)\textbf{d}_{\alpha, \bfg}\cdot\bfe_\bfg(z),
% \label{sch2}
% \end{equation}

\subsection{Elimination of radiative photonic and excitonic modes}

% \begin{equation}
% \bfe_{\bf 0}(z) = -\xi k^2\sum_{\bfg}a_{\bfg}\intzp\opp{G}_{0}(z,z')\Delta\eps_{-\bfg}(z')
% \boldsymbol{\theta}_{\bfg}(z')-4\pi k^2 \intzp\opp{G}_{0}(z,z')\mathbf{P}_{\bf 0}(z'),
% \label{eq:E0_eq}
% \end{equation}

% Equations~\eqref{eq:5},~\eqref{eq:E0_eq}, and~\eqref{eq:8} show that the zeroth photonic harmonic and the zeroth excitonic harmonics form a nonresonant sector which can be eliminated from the equations for the resonant first-shell modes. The zeroth photonic harmonic represents the radiative channel and is therefore not part of the guided-mode resonance forming the quasiguided states. 

% At the same time, the zeroth excitonic harmonics couple directly to this radiative photonic component. 
The resonant contributions of $\opp{G}_{\mathbf 0}$ in Eq.~\eqref{eq:3} at $\bfg=\mathbf{0}$ describe the non-Hermitian radiative response of the homogeneous slab and are generally detuned from $k_g$, as can be seen from Fig.~\ref{fig:1}(c). We therefore eliminate $\bfe_{\mathbf 0}$ from Eqs.~(\ref{eq:5},\,\ref{eq:8}) following an approach similar to Ref.~\cite{liang2011three}, see Sec.~S1C of the SM for more details. It results in the formation of the radiative losses $\gamma_{\rm ph}
=\iu k_g^3 \eps_1^2
\iint {\rm d}z{\rm d}z'\,
\theta(z){G}_{\mathbf 0}(z,z')\theta(z')$ of six guided photonic modes at $|\mathbf{g}|=g$, where $\eps_1=\Delta\eps_{g}$ and ${G}_{\mathbf 0}$ is the scalar GF of the in-plane part of $\opp{G}_{\mathbf 0}$ at $\bfk=0$. Two degenerate excitonic modes at $\bfg=\mathbf{0}$ acquire radiative losses
$\gamma_X=\iu \xi_{\bf 0}{\eta^2 |d|^2  k_g } 
\iint  {\rm d}z{\rm d}z'\,F(z){G}_{\mathbf 0}(z,z')F(z')$ that detune them from twelve excitons at $|\bfg|=g$. Therefore, $b_{\alpha,\mathbf{0}}$ can be also eliminated from the coupled light-matter equations, see Sec.~S1C of the SM for more details. Following this procedure, we achieve coupled equations for six photonic and twelve excitonic resonant amplitudes at $|\bfg|=g$ for our example with $C_{6v}$ symmetry:
\begin{align}
({k-k_g})a_{{\rm s},\bfg}
&={ V^*}b_{{\rm s},\bfg}+
\sum_{|\bfg'|=g}
\left[
{-\iu\gamma_{\rm ph}}-{k_{g}N_h}\Delta\eps_{\bfg-\bfg'}+\frac{|d|^2 J^2}{{D}}\right]
(
\hat{\mathbf e}_{{\rm s},\bfg}
\cdot
\hat{\mathbf e}_{{\rm s},\bfg'}
)
a_{{\rm s},\bfg'},\label{eq:9}\\
(k-E_{\rm X}) b_{\alpha,\bfg}
&=
V\,\delta_{\alpha,\rm s}\,a_{{\rm s},\bfg}.
\label{eq:10}
\end{align}
% We keep the shell-dependent parameters $\Delta\eps_{\bfg-\bfg'}$ and $u_{\bfg-\bfg'}$ explicitly, because the difference $\bfg-\bfg'$ can belong to the zeroth, first, second, or third reciprocal-lattice shell.
%\begin{split}
%(k-E_{\rm X}) b_{{\rm s},\bfg}=\sum_{\substack{\bfg'\neq\bfg\\ \bfg'\neq\mathbf 0}}\left(u_{\bfg-\bfg'}+\frac{u_{\bfg}u_{-\bfg'}}{D}\right)&\left[b_{{\rm s},\bfg'}\cos\Delta\varphi_{\bfg\bfg'} -b_{p,\bfg'}\sin\Delta\varphi_{\bfg\bfg'}\right]\\ &-{V}a_{\bfg}+\frac{u_{\bfg}}{D} d J \sum_{\bfg'\neq\mathbf 0}\Delta\eps_{-\bfg'} a_{\bfg'} \cos\Delta\varphi_{\bfg\bfg'},
%\end{split}
%\label{bsg_without0}
%\end{equation}
%and
%\begin{equation}
%\begin{split}
%(k-E_{\rm X}) b_{p,\bfg}=\sum_{\substack{\bfg'\neq\bfg\\ \bfg'\neq\mathbf 0}}\left(u_{\bfg-\bfg'}+\frac{u_{\bfg}u_{-\bfg'}}{D}\right) &\left[b_{{\rm s},\bfg'}\sin\Delta\varphi_{\bfg\bfg'} +b_{p,\bfg'}\cos\Delta\varphi_{\bfg\bfg'}\right]\\&+\frac{u_{\bfg}}{D}d J\sum_{\bfg'\neq\mathbf 0}\Delta\eps_{-\bfg'} a_{\bfg'}\sin\Delta\varphi_{\bfg\bfg'} .
%\end{split}
%\label{bpg_without0}
%\end{equation}

Light-matter interaction terms in Eqs.~(\ref{eq:9},\ref{eq:10}) include the near-field coupling via $V=-\sqrt{\xi_{\bf 0}}d \eta\int{\rm d}z\, \theta(z)F(z)$ and radiative coupling via $J=\sqrt{\xi_{\bf0}}\eta k_g^2  \eps_1 \iint {\rm d}z{\rm d}z'\,F(z){G}_{\mathbf 0}(z,z')\theta(z')$. The frequency detuning of eliminated excitons at $\bfg=\mathbf{0}$ enters the equations as $D = k_g - (E_{\rm X}  - \iu\gamma_X) $. Comparison with the system retaining the zeroth-harmonic excitons supports the approximation $D(k)\simeq D(k_g)$ within the accuracy of the reduced model (Sec.~S4A of the SM). This allows us to retain frequency-independent coefficients in the effective Hamiltonian.

% The eliminated zero harmonics enter the effective first-shell equations through several distinct self-energy contributions. The purely radiative photonic correction is proportional to $\iu\gamma_0$ and describes scattering of a first-shell guided harmonic into the zeroth radiative channel and back. The mixed overlap $J$ appears when the zeroth radiative harmonic couples to the eliminated zeroth excitonic harmonics. This produces corrections proportional to $J/D$ and $J^2/D$, corresponding respectively to effective photon-exciton coupling and photon-photon self-energy mediated by the zeroth exciton. The terms proportional to $u_{\bfg}u_{-\bfg'}/D$ are excitonic self-energy corrections arising from virtual transitions through the eliminated zeroth excitonic sector.

% In contrast to previous studies, where radiative losses and light-matter coupling are introduced at the level of the effective modal description~\cite{gorlach2018far,li2021experimental}, here we obtain non-Hermitian radiative contributions explicitly.

\subsection{Effective Hamiltonian in the vicinity of $\Gamma$ point}

The developed model in Eqs.~(\ref{eq:9},\ref{eq:10}) represents the full effective Hamiltonian at the $\Gamma$-point ($\mathbf{k} = {\bf{0}}$) that consists of six $\rm s$-polarized guided modes, six $\rm s$- and six $\rm p$-polarized excitonic modes. According to Eq.~\eqref{eq:10}, the excitons are not coupled to each other, and each guided photon drives only the $\rm s$-polarized exciton of the same harmonic, whereas the $\rm p$-polarized excitons are not driven and remain dark. We therefore consider the reduced model of six photonic and six $\rm s$-polarized excitonic modes.

The photonic and excitonic subblocks of the reduced system are symmetric circulant matrices~\cite{gray2006toeplitz}. Applying the  discrete Fourier transform to the full Hamiltonian, we diagonalize the photonic and excitonic blocks that transforms the basis from the Fourier diffraction orders $|\mathbf{g}|=g$ to the modes with a defined orbital multipole index $q=0,\pm 1,\pm 2, 3$. The photonic eigenvalues $\Omega_q$ then can be written as
\begin{equation}
\Omega_q
=
k_g+3\delta_{q,\pm 1}\left[-\iu \gamma_{\rm ph}+\frac{ |d|^2J^2}{D}\right]+k_gN_h\left[
-\eps_1 \cos\left(\frac{\pi q}{3}\right)
+
\eps_2 \cos\left(\frac{2\pi q}{3}\right)
+
\eps_3  (-1)^q\right],
\label{eq:11}
\end{equation}
where the permittivity parameters $\eps_2=\Delta\eps_{\sqrt{3}g}$ and $\eps_3=\Delta\eps_{2g}$ describe mixing between second- and third-order neighbors, respectively. The excitonic eigenvalues $E_q$ are all equal, $E_q=E_{\rm X}$. The light-matter coupling induces mixing between the photonic and excitonic singlets (monopole $q=0$ and octupole $q=3$) and doublets (dipoles $q=\pm 1$ and quadrupoles $q=\pm 2$). The corresponding sub-blocks $\hat H^{(q)}$ of the full Hamiltonian are thus determined by the selection rules with respect to the multipole index $q$, similar to Eq.~\eqref{eq:0},
\begin{equation}
\hat H^{(q)}
=
\begin{pmatrix}
\Omega_q & V^*\\
V& E_q
\end{pmatrix}.
\label{eq:12}
\end{equation}
The condition $V_q=V$ results from retaining only $\xi_\mathbf{0}$ in the model. Restoring other coefficients $\xi_{\bfg\ne\mathbf{0}}$ results in multipole-dependent light-matter coupling $V_q$ and modifies the dipolar self-energy. We note that even in this case the full effective Hamiltonian contains only six excitonic degrees of freedom that can be constructed from the bright ${\rm s}$ combinations and their orthogonal dark ${\rm p}$ complements  (Sec.~S2B of the SM).

In the vicinity of the $\Gamma$ point ($\mathbf{k} \to {\bf{0}}$), the structure of the Hamiltonian becomes more complex due to the mixing of different multipole indices. Using the perturbation approach, we recover the first- and second-order corrections to the Hamiltonian
depending on $k_\pm=(k_x\pm\iu k_y)/\sqrt{2}$ and $\mathbf{k}^2=2k_+k_-$, see details in Sec.~S1A,~S1C, and~S2A of the SM. First, the guided mode pole of each harmonic follows the dispersion of the fundamental $\rm s$-polarized mode of the effective slab at $|\bfg+\bfk|$, which is characterized by its group velocity $v_{\rm g}$ and curvature $\mu_\parallel$ at $|\bfg+\bfk|=g$; the displacement of $\bfk$ perpendicular to $\bfg$ changes $|\bfg+\bfk|$ only in second order, which gives the transverse coefficient $\mu_\perp=v_{\rm g}/(2g)$. Second, the polarization vector $\hat{\mathbf e}_{{\rm s},\bfg}$ rotates with $\bfg+\bfk$, which modifies all coupling coefficients proportional to $(\hat{\mathbf e}_{{\rm s},\bfg}\cdot\hat{\mathbf e}_{{\rm s},\bfg'})$. Third, the eliminated zero harmonic splits into two radiation channels $\alpha=\rm s,p$, which modifies the non-Hermitian coupling parameters as $\gamma_{\rm ph}\rightarrow \gamma_{\rm ph}+ \iu\Delta\gamma_{{\rm ph},\alpha} \mathbf{k}^2$, $J\rightarrow J+\Delta J_\alpha \mathbf{k}^2$, and $D\rightarrow D + \Delta D_\alpha \mathbf{k}^2$, and generates the exciton-mediated correction $\Lambda_{{\rm X},\alpha}=|d|^2(2J\Delta J_\alpha/D-J^2\Delta D_\alpha/D^2)$. Consequently, the rescaled photonic eigenvalues are $\tilde{\Omega}_q(\mathbf{k})= \Omega_q+\Delta \Omega_q \mathbf{k}^2$. As a result, the full exciton-photon Hamiltonian takes form:
\begin{equation}
\left(
\begin{array}{cccccccccccc}
\tilde{\Omega}_{0}&\nu_{01} k_-&-\mu_{02} k_-^2&0&-\mu_{02} k_+^2&-\nu_{01} k_+&V^*&0&0&0&0&0\\
-\nu_{01} k_+&\tilde{\Omega}_{1}&\nu_{12} k_-&-\mu_{13} k_-^2&0 &-\mu_{11} k_+^2&0&V^*&0&0&0&0\\
-\mu_{02} k_+^2&-\nu_{12} k_+&\tilde{\Omega}_{2}&\nu_{23} k_-&-\mu_{22} k_-^2&0&0&0&V^*&0&0&0\\
0&-\mu_{13} k_+^2&-\nu_{23} k_+&\tilde{\Omega}_{3}&\nu_{23} k_-&-\mu_{13} k_-^2&0&0&0&V^*&0&0\\
-\mu_{02} k_-^2&0&-\mu_{22} k_+^2&-\nu_{23} k_+&\tilde{\Omega}_{2}&\nu_{12} k_-&0&0&0&0&V^*&0\\
\nu_{01} k_-&-\mu_{11} k_-^2&0&-\mu_{13} k_+^2&-\nu_{12} k_+&\tilde{\Omega}_{1}&0&0&0&0&0&V^*\\
V&0&0&0&0&0&E_{\rm X}&0&0&0&0&0\\
0&V&0&0&0&0&0&E_{\rm X}&0&0&0&0\\
0&0&V&0&0&0&0&0&E_{\rm X}&0&0&0\\
0&0&0&V&0&0&0&0&0&E_{\rm X}&0&0\\
0&0&0&0&V&0&0&0&0&0&E_{\rm X}&0\\
0&0&0&0&0&V&0&0&0&0&0&E_{\rm X}\\
\end{array}
\right),
\label{eq:13}
\end{equation}
where $\tilde{\Omega}_{-q}(\mathbf{k})=\tilde{\Omega}_q(\mathbf{k})$, and the coefficients $\Delta\Omega_q$, $\nu_{qq'}$, and $\mu_{qq'}$ are given in Eqs.~(S26)--(S28) of the SM.
In the limit of vanishing modulation, $\eps_m\to0$ ($m=1,2,3$), all linear coupling coefficients reduce to $\nu=\iu v_{\rm g}/\sqrt2$, all quadratic coupling coefficients to $(\mu_\parallel-\mu_\perp)/2$, and all $\Delta\Omega_q$ to $(\mu_\parallel+\mu_\perp)/2$, so that the photonic eigenmodes in Eq.~\eqref{eq:13} reproduce the folded dispersion of the reference guided mode.
% \begin{equation}
% \hat H_{\rm ph}=  
% \left(
% \begin{array}{cccccc}
% \Omega_{0}(\mathbf{k})
% &
% \nu k_-
% &
% -\mu k_-^2
% &
% 0
% &
% -\mu k_+^2
% &
% -\nu k_+
% \\
% -\nu k_+
% &
% \Omega_{1}(\mathbf{k})
% &
% \nu k_-
% &
% -\mu k_-^2
% &
% 0
% &
% -\mu k_+^2
% \\
% -\mu k_+^2
% &
% -\nu k_+
% &
% \Omega_{2}(\mathbf{k})
% &
% \nu k_-
% &
% -\mu k_-^2
% &
% 0
% \\
% 0
% &
% -\mu k_+^2
% &
% -\nu k_+
% &
% \Omega_{3}(\mathbf{k})
% &
% \nu k_-
% &
% -\mu k_-^2
% \\
% -\mu k_-^2
% &
% 0
% &
% -\mu k_+^2
% &
% -\nu k_+
% &
% \Omega_{2}(\mathbf{k})
% &
% \nu k_-
% \\
% \nu k_-
% &
% -\mu k_-^2
% &
% 0
% &
% -\mu k_+^2
% &
% -\nu k_+
% &
% \Omega_{1}(\mathbf{k})
% \end{array}
% \right),
% \label{eq:13}
% \end{equation}

\begin{figure}
\centering
\includegraphics[width=0.6\linewidth]{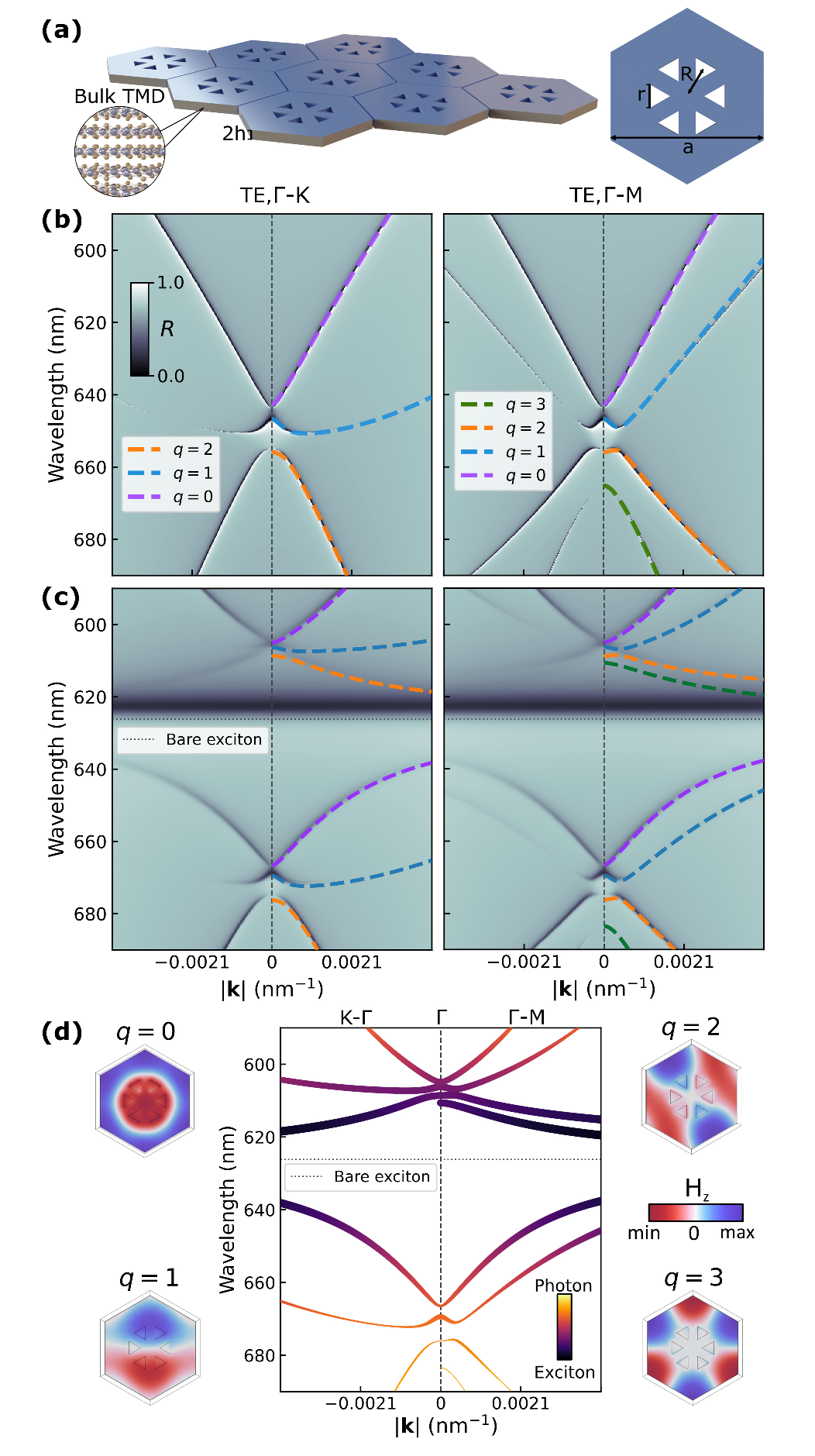}
\caption{{\bf Numerical validation of the effective Hamiltonian (TE polarization).} (a) Schematic of the polaritonic TMD metasurface and the unit cell. The thickness $2h=20$ nm, lattice period $a=405$ nm, triangular holes with side length $r=60$ nm, and radial displacement $R=77$ nm from the unit cell center. (b) RCWA reflectance maps of the photonic structure with $\eps(\lambda)=\eps_\infty=20.25$ and $f=0$ (no excitonic response) for TE-polarized incidence along the $\Gamma-K$ and $\Gamma-M$ directions. Dashed lines show the analytical branches obtained by fitting to the effective photonic Hamiltonian given in Sec. S2A of the SM. (c) Reflectance maps of the polaritonic structure with $\eps(\lambda)$ given by Eq.~\eqref{eq:18} for $\eps_\infty=20.25$ and $f = 0.52~\mu$m$^{-2}$ (realistic WS$_2$ dispersion). The bare exciton wavelength is indicated by the dotted horizontal line. Dashed lines show the corresponding exciton-polariton branches fitted to the respective polaritonic Hamiltonian, see Sec.~S2A of the SM. (d) Dispersion of polariton branches extracted from the fitting procedure. The color indicates the photon/exciton composition of each branch, while the line thickness shows the total decay rate. The insets show simulated $H_z$ field profiles at the $\Gamma$ point for the modes with the orbital index $q=0,1,2,3$. }
\label{fig:3}
\end{figure}

\begin{figure}
\centering
\includegraphics[width=0.6\linewidth]{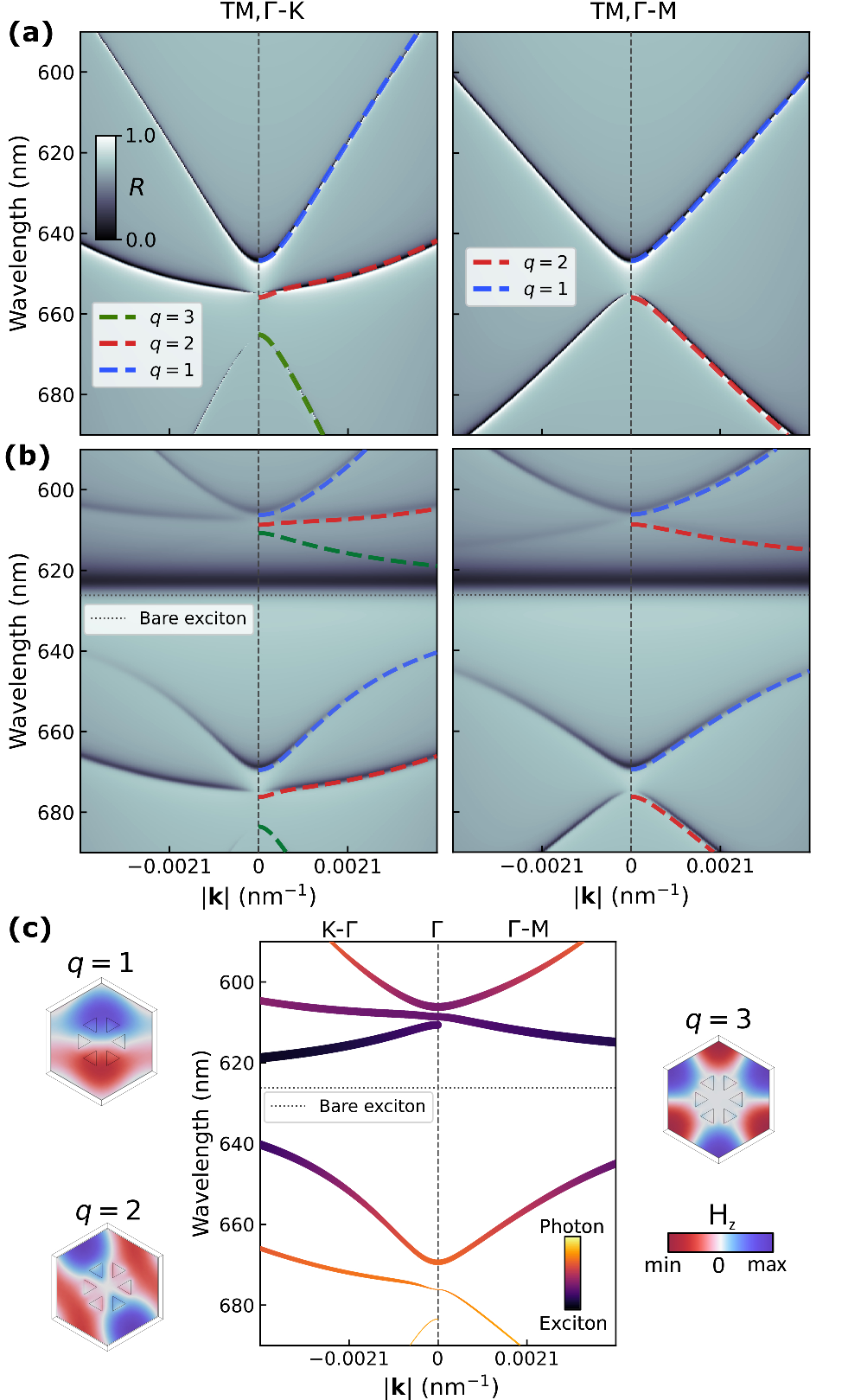}
\caption{{\bf Numerical validation of the effective Hamiltonian (TM polarization).} (a-c) Same data as Fig.~\ref{fig:3}(b-d) calculated for the TM polarization of incident light. (a) The analytical branches are obtained by fitting to the effective photonic Hamiltonian given by Eqs.~(\ref{eq:15},\, \ref{eq:17}). (b) The analytical branches are obtained by fitting to the respective polaritonic Hamiltonian, see Sec.~S2A of the SM. }
\label{fig:4}
\end{figure}

For the sake of simplicity of further numerical analysis, the photonic sub-block of the Hamiltonian in Eq.~\eqref{eq:13} can be split into independent blocks with a defined mirror parity along the high-symmetry directions $\Gamma-K$ and  $\Gamma-M$, see Sec.~S2A of the SM for more details. This basis corresponds to modes that are coupled to TE- and TM-polarized radiation channels at oblique incidence~\cite{overvig2020selection}. Along the $\Gamma -K$ direction, $\hat H_{\rm ph}$ sub-block in Eq.~\eqref{eq:13} splits into a three-dimensional TE and a three-dimensional TM block, and along the $\Gamma -M$ direction it splits into a four-dimensional TE and a two-dimensional TM blocks. The TM blocks are given by
\begin{equation}
\hat H_{{\rm ph},K}^{({\rm TM})}
=
\left(
\begin{array}{ccc}
\tilde{\Omega}_3(\mathbf{k})
&
-\dfrac{\mu_{13}}{\sqrt{2}} \mathbf{k}^2
&
-\nu_{23} |\mathbf{k}|
\\[2mm]
-\dfrac{\mu_{13}}{\sqrt{2}} \mathbf{k}^2
&
\tilde{\Omega}_1(\mathbf{k})-\dfrac{\mu_{11}}{2} \mathbf{k}^2
&
\dfrac{\nu_{12}}{\sqrt{2}} |\mathbf{k}|
\\[2mm]
\nu_{23} |\mathbf{k}|
&
-\dfrac{\nu_{12}}{\sqrt{2}}|\mathbf{k}|
&
\tilde{\Omega}_2(\mathbf{k})+\dfrac{\mu_{22}}{2}\mathbf{k}^2
\end{array}
\right),
\label{eq:15}
\end{equation}
and
\begin{equation}
\hat H_{{\rm ph},M}^{({\rm TM})}
=
\left(
\begin{array}{cc}
\tilde{\Omega}_1(\mathbf{k})-\dfrac{\mu_{11}}{2} \mathbf{k}^2
&
\dfrac{\nu_{12}}{\sqrt{2}}|\mathbf{k}|
\\[2mm]
-\dfrac{\nu_{12}}{\sqrt{2}}|\mathbf{k}|
&
\tilde{\Omega}_2(\mathbf{k})-\dfrac{\mu_{22}}{2} \mathbf{k}^2
\end{array}
\right),
\label{eq:17}
\end{equation}
while the respective TE blocks are given in Sec. S2A of the SM. The excitonic and light-matter coupling block modify accordingly, preserving the selection rules with respect to the parity and orbital index similar to Eq.~\eqref{eq:13}. 

% As an example, the full Hamiltonian along the $\Gamma -M$ direction that describes modes active in TM polarization can be obtained from Eqs.~(\ref{eq:12},\,\ref{eq:17}) as
% \begin{equation}
% \hat H_{M}^{({\rm TM})}
% =
% \left(
% \begin{array}{cccc}
% \tilde{\Omega}_1(\mathbf{k})-\dfrac{\mu}{2} \mathbf{k}^2
% &
% \dfrac{\nu}{\sqrt{2}}|\mathbf{k}| & V^* & 0
% \\
% -\dfrac{\nu}{\sqrt{2}}|\mathbf{k}|
% &
% \tilde{\Omega}_2(\mathbf{k})-\dfrac{\mu}{2} \mathbf{k}^2& 0 & V^*\\
% V & 0 & E_{\rm X} & 0 \\
% 0 & V & 0 & E_{\rm X} 
% \end{array}
% \right) .
% \label{eq:18}
% \end{equation}

\section{Numerical validation of the effective Hamiltonian}
We next numerically validate the effective model given by Eqs.~(\ref{eq:13}--\ref{eq:17}) and the respective TE matrices in Sec. S2A of the SM. We use the rigorous coupled-wave analysis (RCWA) based on numerical implementation in FMMAX~\cite{schubert2023fourier}. The polaritonic metasurface geometry is shown in Fig.~\ref{fig:3}(a). The accuracy of the developed effective model increases for small perturbation parameters $\eps_1,\eps_2,\eps_3$ that are defined by the ratio of hole side length $r$ to the lattice period $a$. We therefore choose $r=60$ nm that is much smaller than $a=405$ nm. The metasurface thickness is $2h=20$ nm, and radial displacement of individual holes from the unit cell center is $R=77$ nm. 

The permittivity function of excitonic material is generally anisotropic and nonlocal, as can be seen from Eq.~\eqref{eq:6}. In our reduced Hamiltonian model, however, the excitonic polarization enters via the projection onto the s-polarized guided mode of the effective slab. We, therefore, can use an equivalent local isotropic model for $\eps$ by projecting $\mathbf{P}(\bfr)=F(z)\sum_\alpha\textbf{d}_\alpha^*\Phi_\alpha(x,y)$ onto $\boldsymbol{\theta}_{\bfg}(z)$ and using $\Phi_\alpha(x,y)$ from Eq.~\eqref{eq:7} (Sec. S4D of the SM). The resulting $\eps(\lambda)$ coincides with the tabulated the permittivity of bulk WS$_2$~\cite{bouteyre2025simultaneous},
\begin{equation}
\eps(\lambda)=\eps_\infty + \frac{f}{\lambda_{\rm X}^{-2}-\lambda^{-2}-\iu{\Gamma}\lambda^{-1}/{( \pi c )}},   
\label{eq:18}
\end{equation}
where $\lambda=2 \pi/k$, $\eps_\infty=20.25$, $\lambda_{\rm X}=626.2$ nm, $\Gamma= 7.6$ ps$^{-1}$  (full width at half maximum $2\hbar\Gamma=10$~meV), and $f = 0.52~\mu$m$^{-2}$. Here, the oscillator strength is directly connected to the light-matter coupling amplitude from Eq.~\eqref{eq:13} as $f\simeq \veps|V|^2/(\pi^2\xi_{\mathbf 0})$ (Sec. S4D of the SM). We note that the comparison of RCWA calculations for a model based on Eq.~\eqref{eq:18} with a uniaxial permittivity model shows negligible differences in results of Figs.~\ref{fig:3} and~\ref{fig:4} (Sec.~S4A of the SM).

Figure~\ref{fig:3}(b) shows the calculated reflectance map along $\Gamma-K$ and $\Gamma-M$ directions for the photonic metasurface with $\eps(\lambda)=\eps_\infty$ and no excitonic response ($f=0$) for the TE polarization of incidence. The dashed lines show fitting of the dispersion curves of modes with defined $q$ to the respective matrices in Sec. S2A of the SM, with the model parameters listed in Sec.~S4 of the SM. The analytical dispersions reproduce the ordering of the modes at the $\Gamma$ point and follow the spectral features along the high-symmetry directions. We attribute the deviations at large incident angles to higher-order finite-$\bfk$ corrections that are not included in the model. The model shows that all quasiguided modes originate from an $\rm s$-polarized guided mode with the dominant out-of-plane magnetic field $H_z$ field component. We use eigenmode solver of full-wave simulation software COMSOL Multiphysics to calculate $H_z$ for modes with different $q$, as shown in the inset of Fig.~\ref{fig:3}(d). 

Figure~\ref{fig:3}(c) shows the respective reflectance map for the realistic polaritonic WS$_2$ metasurface with the permittivity given by Eq.~\eqref{eq:18}. The spectrum shows the formation of exciton-polariton avoided resonance crossings near the bare exciton resonance wavelength. The dispersion curves fit to the corresponding full Hamiltonian matches the spectral features in the vicinity of the $\Gamma$ point. The photonic parameters are kept fixed, and $V$ is the only additional fitted parameter (Sec.~S4A of the SM). The oscillator-strength estimate and the multipole-dependent $V_q$ fits yield coupling strengths close to the common fitted value, supporting the common-$V$ approximation for these spectra (Sec.~S4D of the SM). We note that the number of polaritons is double the number of photonic modes that supports our multi-excitonic effective model. The extracted dispersion curves for the polaritonic metasurface with the relative exciton and photon components are shown in Fig.~\ref{fig:3}(d). It shows that upper branches possess excitonic-like behavior and lower branches show photonic-like behavior in the vicinity of $\Gamma$ point. The photon and exciton fractions are obtained by summing the squared magnitudes of the corresponding components of each normalized right eigenvector of the polaritonic Hamiltonian. The two fractions sum to unity.

Figure~\ref{fig:4} shows the corresponding reflectance and fitting data for TM incident polarization with all other parameters as in Fig.~\ref{fig:3}. The photonic dispersion curves are fitted to Eqs.~(\ref{eq:15},\,\ref{eq:17}). The root mean square (RMS) deviations from the reflectance minima are $0.59$~nm for the photonic maps and $0.80$~nm for the polaritonic maps (Sec.~S4A of the SM).
For comparison, we also calculate the respective TM reflectance maps and fitting data for a WS$_2$ monolayer on a dielectric metasurface, see Sec.~S4B of the SM.

\section{Topological polaritonic effects}

We further apply the developed model in the context of topological photonics, demonstrating the generation of polaritonic edge states at interfaces between differently patterned metasurface domains. Specifically, we describe and exploit the topological transition between trivial and nontrivial phases arising from band inversion between the dipolar ($|q|=1$) and quadrupolar ($|q|=2$) mode pairs near the $\Gamma$ point.

\subsection{Topological transition diagram}

\begin{figure}[t]
\centering
\includegraphics[width=1\linewidth]{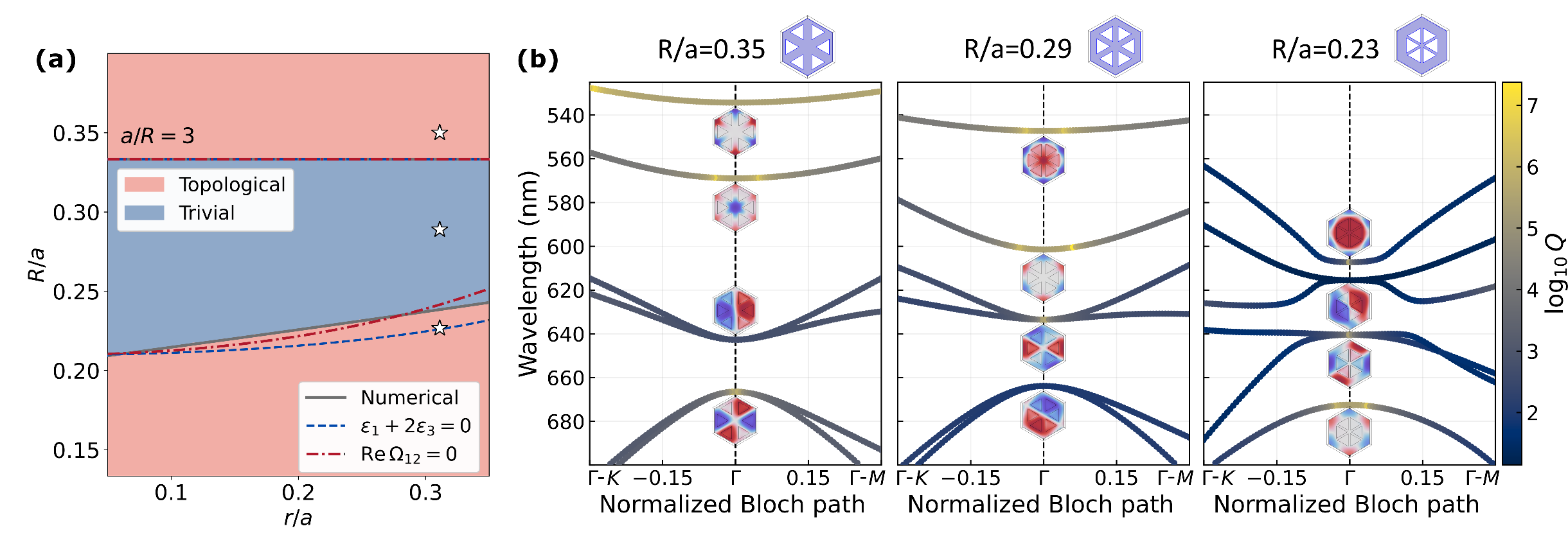}
\caption{{\bf Topological phase map and verification of photonic band inversion.}
(a) Topological phase map in the $(r/a,R/a)$ parameter plane. The color shading is determined by the numerical transition boundaries for a metasurface with lattice period $a=450$~nm and thickness $2h=23$~nm. The solid gray lines show the numerical boundaries, while the blue dashed lines show the analytical predictions obtained from the approximate condition in Eq.~\eqref{eq:20}, and the red dash-dotted lines show the full condition ${\rm Re}\,\Omega_{12}=0$. The starred points indicate the geometries selected for full-wave eigenmode calculations shown in panel (b).
(b) Calculated band structures for three representative geometries at fixed
$a=450$~nm, $r=140$~nm, and $2h=23$~nm describing topological and trivial regimes. The line color shows $ \log_{10}Q$, where $Q$ is the mode quality factor. The insets show the eigenmode $H_z$ field profiles at the $\Gamma$ point.
}
\label{fig:5}
\end{figure}
We first consider the photonic limit ($d=0$) and determine the
dipole-quadrupole band-inversion boundaries. At the $\Gamma$ point, Eq.~\eqref{eq:11} gives
\begin{equation}
\Omega_{12}\equiv\frac{\Omega_1-\Omega_2}{2}
=-\frac{k_gN_h}{2}(\eps_1+2\eps_3)
-\frac{3\iu}{2}\gamma_{\rm ph}.
\label{eq:19}
\end{equation}
The real-frequency ordering of the two doublets reverses at
${\rm Re}\,\Omega_{12}=0$. Neglecting the radiative frequency shift $\gamma_{\rm ph}\propto\eps_1^2$ in the small-hole limit $\eps_i\ll \veps$ yields the band inversion criterion
\begin{equation}
\eps_1+2\eps_3\simeq0.
\label{eq:20}
\end{equation}
This inversion criterion provides a starting point for the topological
analysis of the full finite-$\bfk$ Hamiltonian in
Eq.~\eqref{eq:13}.

Equation~\eqref{eq:20} admits two solutions as a function of the ratio between the lattice period $a$ and the radial displacement $R$. These solutions follow from the analytical expressions for $\eps_1$ and $\eps_3$ evaluated for the geometry shown in Fig.~\ref{fig:3}(a); further details are provided in Sec.~S3 of the SM. The first solution is independent of the individual hole shape and corresponds to the conventional breathing-honeycomb-lattice transition at $R=a/3$~\cite{wu2015scheme,guddala2025topological}. At this point, the relevant structure factors vanish simultaneously, yielding $\eps_1=\eps_3=0$. The second solution depends on the individual hole shape and, in the limit $r\ll a$, generally exists for arbitrary hole geometries, including those considered in Figs.~\ref{fig:3} and~\ref{fig:4}. For the photonic block of Eq.~\eqref{eq:13}, the quadratic
diagonal half-splitting is characterized by $\Delta\Omega_{12}=(\Delta\Omega_1-\Delta\Omega_2)/2$.
To leading order in the dielectric modulation
$\Delta\Omega_{12}\simeq k_gN_h(\eps_1+8\eps_3)/({8g^2}).$
At the additional inversion, $\eps_1\simeq-2\eps_3$ with
$\eps_3\ne0$, so this coefficient remains nonzero. The full expression, including radiative corrections,
is given in Sec.~S2A of the SM.

Because $\gamma_{\rm ph}\propto\eps_1^2$, the radiative shift vanishes at the conventional transition $R=a/3$, where $\eps_1=\eps_3=0$. For the geometry-controlled inversion, this shift generally remains finite and depends on the slab thickness, so the full condition ${\rm Re}\,\Omega_{12}=0$ must be used beyond the small-hole limit (Sec.~S3 of the SM). 

Figure~\ref{fig:5}(a) compares the boundaries given by Eq.~\eqref{eq:20} and the full condition ${\rm Re}\,\Omega_{12}=0$ with those extracted from full-wave eigenfrequency calculations. The full condition is evaluated without fitted parameters, using $k_g$, $N_h$, and $\gamma_{\rm ph}$ of the fundamental $\rm s$-polarized mode of the area-averaged effective slab (Sec.~S3B of the SM). We performed full-wave eigenfrequency calculations at the $\Gamma$ point for a range of hole radii $r$ and radial displacements $R$, using $a=450$~nm and $2h=23$~nm. For each geometry, the dipolar and quadrupolar doublets were identified from their field symmetries and quality factors. The transition points were then extracted from the calculated spectra and fitted with smooth quadratic curves. For the geometry-dependent transition, the numerical and analytical boundaries coincide in the small-hole limit but progressively deviate as $r/a$ increases. For the conventional transition at $R=a/3$, the numerical and analytical boundaries coincide throughout the considered parameter range.

We further verify the phase diagram using full-wave band structures calculated for fixed $a=450$~nm and $r=140$~nm while varying the radial displacement $R$, as shown in Fig.~\ref{fig:5}(b). The geometries with $R/a=0.35$ and $R/a=0.23$ belong to the topological phase, whereas the geometry with $R/a=0.29$ lies within the intermediate trivial region. The ordering of the mode wavelengths and the corresponding $H_z$ field profiles at the $\Gamma$ point confirm the inversion of the dipolar and quadrupolar states across both phase boundaries, in agreement with the phase diagram in Fig.~\ref{fig:5}(a). Additional full-wave validation is provided in Figs.~S1 and~S2 of the SM for small holes with $r/a\simeq0.15$, where the analytical model is expected to be most accurate.

For the polaritonic structure, the resulting effective topological Hamiltionian $\hat H_{\rm eff}$ of the reduced dipole-quadrupole model retains both photonic doublets and four bright excitonic combinations, while singlet contributions are eliminated as separate degrees of freedom via coefficient renormalization:
\begin{equation}
\hat H_{\rm eff}=
\begin{pmatrix}
\Omega_1+\Delta\tilde\Omega_1\bfk^2&\nu_{12}k_-&0&-\tilde\mu_{11}k_+^2& V^*&0&0&0\\
-\nu_{12}k_+&\Omega_2+\Delta\tilde\Omega_2\bfk^2&-\tilde\mu_{22}k_-^2&0&0 &V^*&0&0\\
0&-\tilde\mu_{22}k_+^2&\Omega_2+\Delta\tilde\Omega_2\bfk^2&\nu_{12}k_-&0&0&V^*&0\\
-\tilde\mu_{11}k_-^2&0&-\nu_{12}k_+&\Omega_1+\Delta\tilde\Omega_1\bfk^2&0&0&0&V^*\\
V&0&0&0&E_X&0&0&0\\
0&V&0&0&0&E_X&0&0\\
0&0&V&0&0&0&E_X&0\\
0&0&0&V&0&0&0&E_X
\end{pmatrix}.
\label{eq:last}
\end{equation}
Here, the parameters with a tilde are renormalized coefficients (Sec.~S4C of the SM).

\begin{figure}[H]
\centering
\includegraphics[width=0.65\linewidth]{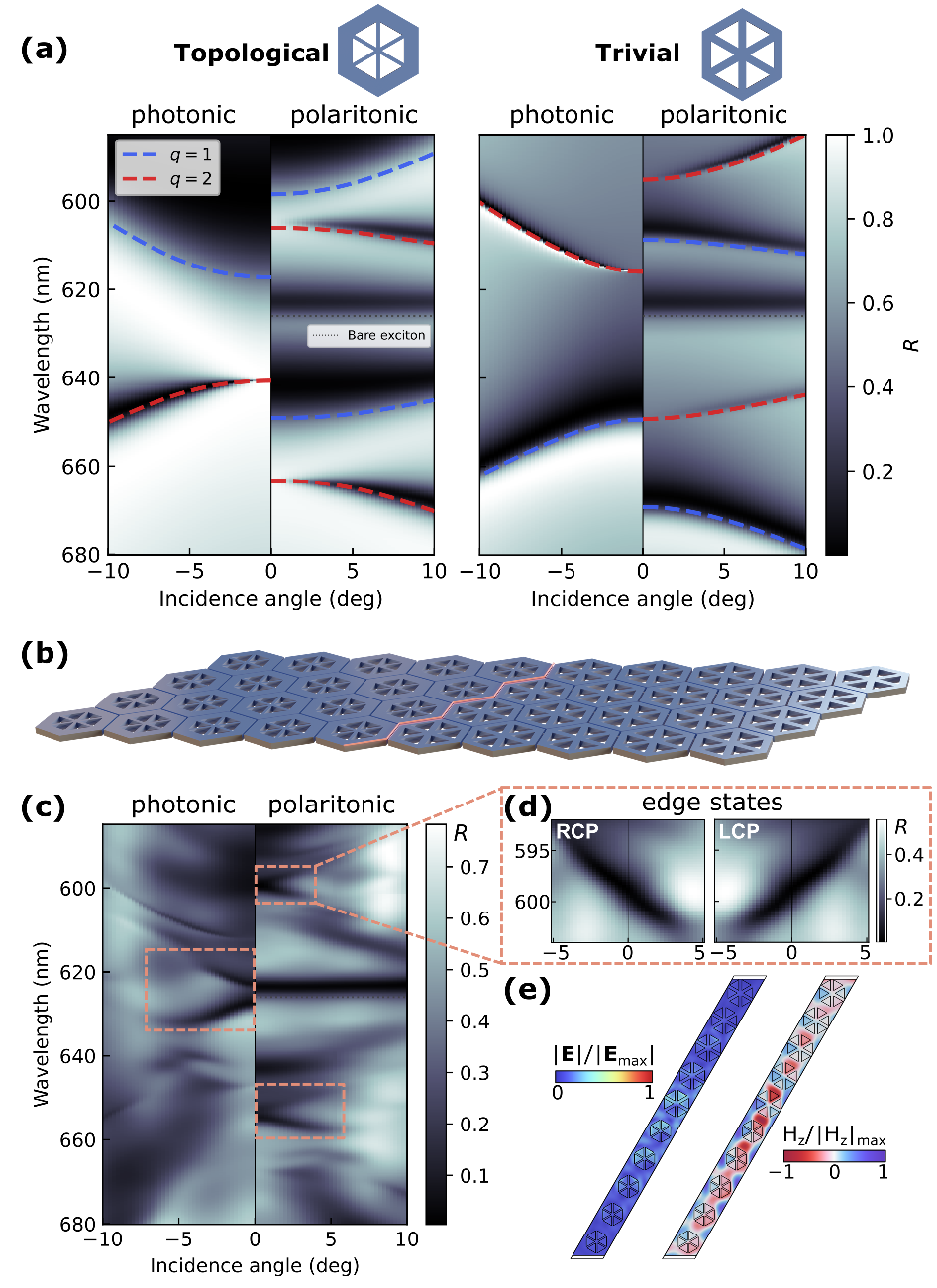}
\caption{{\bf Photonic and polaritonic edge states.}
(a) TM-polarized reflectance maps along the $\Gamma$-$M$ direction for isolated topological and trivial domains, shown for purely photonic ($\varepsilon_{\infty}=20.25$, $f=0$) and polaritonic ($f\ne 0$) structures. Dashed curves mark the analytical dipole ($q=1$) and quadrupole ($q=2$) branches, the dotted line indicates the bare exciton wavelength.
(b) Schematic of the finite-size metasurface at the interface of topological and trivial domains. The domain wall is highlighted with the pink line.
(c) TM-polarized reflectance maps for the structure shown in (b), calculated along the $\Gamma$-$M$ direction. The left and right panels correspond to the purely photonic and polaritonic cases, respectively. Orange dashed boxes indicate edge state regions.
(d) Reflectance maps of the polaritonic edge state region for right- and left-circularly polarized (RCP/LCP) incident light.
(e) Field profiles of the polaritonic edge state excited at normal incidence under TM-polarized illumination at $599~\mathrm{nm}$: electric-field amplitude $|\mathbf E|$ (left) and out-of-plane magnetic-field component $H_z$ (right).
}
\label{fig:6}
\end{figure}

\subsection{Edge states at the hole-shape-controlled topological transition}
We next demonstrate the emergence of photonic and polaritonic edge states associated with the hole-shape-controlled topological transition. We choose a pair of geometries across the lower transition boundary in Fig.~\ref{fig:5}(a), where the topological structure is more shrunken than the trivial one. The two domains have the same lattice period $a=450$ nm and slab thickness $2h=23$ nm, and differ only by the in-plane geometry of the triangular holes, defined in Fig.~\ref{fig:3}(a). For the trivial domain we use side length $r=150$ nm and radial displacement $R=130$ nm, while for the topological domain we use $r=140$ nm and $R=102$ nm.

For these representative geometries, the topological and trivial domains were first analyzed separately through full-wave simulations of their unit cells using COMSOL Multiphysics. Their TM-polarized reflectance maps along the $\Gamma$-$M$ direction are shown in Fig.~\ref{fig:6}(a) for both the purely photonic structure with $\varepsilon_{\infty}=20.25$, $f=0$, and the polaritonic structure including the excitonic response, Eq.~\eqref{eq:18}. For these large-hole domains, we use the effective dipole-quadrupole model of Eq.~\eqref{eq:last} (Sec.~S4C of the SM). Its complex photonic coefficients are fitted to COMSOL eigenfrequencies along both $\Gamma$-$K$ and $\Gamma$-$M$ for the TE and TM channels. Within $|\mathbf{k}|\leq0.12g$, the RMS deviations of the photonic eigenmode wavelengths are $0.44$~nm and $1.85$~nm for the trivial and topological domains, respectively. With the photonic coefficients fixed, a common $V$ is fitted to the upper and lower polariton branches in the TM $\Gamma$-$M$ reflectance maps Fig.~\ref{fig:6}(a). The fitted coefficients and coupling are listed in Sec.~S4C of the SM.
The fitted dispersion curves reproduce the ordering and dispersion of the dipole and quadrupole modes in both domains. Finite-domain Berry flux calculations for these Hamiltonians in the form of Eq.~\eqref{eq:last} give spin Chern numbers close to $-1$ for the topological domain and close to $0$ for the trivial domain (Sec.~S4C of the SM). Consequently, the interface between the two domains is expected to support edge states.

% To verify the existence of localized modes at the interface between the two domains
To verify this prediction, we construct a supercell geometry, as shown in Fig.~\ref{fig:6}(b). The corresponding reflectance maps are presented in Fig.~\ref{fig:6}(c). In the purely photonic case, edge-state resonances appear inside the photonic gap opened by the dipole-quadrupole inversion. 
%Upon inclusion of the excitonic response, these modes hybridize with the bulk exciton resonance, giving rise to polaritonic edge states. 
Upon inclusion of the excitonic response, hybrid polaritonic edge states are formed.
The relevant spectral regions are highlighted by orange dashed boxes.

The polaritonic edge-state region is shown enlarged in Fig.~\ref{fig:6}(d), where it is probed with right- and left-circularly polarized incident light. The two circular polarizations excite the edge state branches with different efficiencies, consistent with the pseudospin character expected for helical edge states. Finally, Fig.~\ref{fig:6}(e) shows the field distribution of polaritonic edge state excited at normal incidence under TM-polarized illumination at $599~\mathrm{nm}$. Both the electric-field amplitude $|\mathbf E|$ and the out-of-plane magnetic-field component $H_z$ are localized along the domain wall. Full-wave simulations in Sec.~S4C of the SM further show that the photonic and polaritonic edge states propagate along the domain wall.

% \section{Other potential effects}
% \subsection{Exciton-induced topological transition}
% \subsection{Change in radiative losses }
\section{Discussion and Conclusion}

Effective Hamiltonians of polaritonic periodic metastructures obtained within the semiclassical approach from a single-pole polarization function, i.e., Eqs.~(\ref{eq:6},\,\ref{eq:18}), commonly utilize a single excitonic degree of freedom\cite{zhang2018photonic,verre2019transition,chen2020metasurface,guddala2025topological,wurdack2026intrinsically}. More recent studies show that even for a single-pole excitonic polarization, multiple excitonic degrees of freedom are required in periodic structures for a correct description of the observed polariton branches and the light-matter interaction coefficient $V$\cite{lu2020engineering,balasubrahmaniyam2021coupling,mandal2023microscopic,sigurdhsson2024dirac,bouteyre2025simultaneous}, as also confirmed by the Hopfield quantum-theory derivation\cite{gerace2007quantum,zanotti2022theory}. For the symmetry-related guided harmonics and degenerate excitons considered here, Eqs.~(\ref{eq:9},~\ref{eq:10}) and the selection rules in Eq.~\eqref{eq:12} identify $N$ bright birght excitonic combinations coupled to the $N$ retained photonic modes. These combinations are retained in the effective dipole-quadrupole model used for Fig.~\ref{fig:6}(a) (Sec.~S4C of the SM). 

Extension of the model in Eqs.~(\ref{eq:9},~\ref{eq:10}) that includes $\xi_{\bfg\ne\mathbf{0}}$ (Sec.~S2B of the SM) shows that resonantly interacting excitonic degrees of freedom are constructed from the bright ${\rm s}$ combinations and their orthogonal dark ${\rm p}$ complements. This approach generalizes the description of the long-range electron-hole exchange interaction and the associated longitudinal-transverse splitting for a non-resonant photonic background\cite{glazov2014exciton,prazdnichnykh2021control}.

\begin{comment}
The model developed in Eq.~\eqref{eq:20} predicts the existence of a new shape-controlled topological transition. Previous effective topological Hamiltonian models constructed photonic modes using plane-wave expansion and were unable to predict such transitions, as they did not impose a cut-off on the number of Fourier harmonics required to resolve the photonic-mode dispersion~\cite{gorlach2018far,li2021experimental}. In contrast, our model is restricted to the first order of Fourier harmonics due to the resonant frequency selectivity imposed by Green's function resonant pole structure [see Eq.~\eqref{eq:4}]. Moreover, previous models predicted that the signs of $\Omega_{12}$ and $\Delta\Omega_{12}$ change simultaneously upon variation of $\eps_1,\eps_2,\eps_3$. The model developed in Eq.~\eqref{eq:20} shows that both trivial and topological phases can be achieved through variation of the dielectric modulation, since the curvature coefficient scales as $\Delta\Omega_{12}\propto\eps_1^2$; see Sec.~S1C of the SM for details.
\end{comment}
Equation~\eqref{eq:20} identifies an additional geometry-controlled inversion beyond the conventional breathing-honeycomb transition. Near $R=a/3$, the analytical plane-wave expansions of Refs.~\cite{gorlach2018far,li2021experimental} predict simultaneous sign reversals of the mass and quadratic diagonal coefficient. Their product therefore retains its sign, so the usual BHZ sign criterion alone does not distinguish the trivial and nontrivial domains within these approximations. Our model retains the resonant shell of guided harmonics selected by the pole of the slab Green's function [Eq.~\eqref{eq:4}]. At the additional inversion, the $\Gamma$ point splitting can reverse sign without forcing the quadratic diagonal coefficient to vanish. Radiative coupling further shifts the inversion boundary (Sec.~S3 of the SM).

In conclusion, we have developed a multimode effective Hamiltonian for resonant polaritonic metasurfaces that connects semiclassical excitonic polarization, guided-mode resonances, non-Hermitian radiative coupling, and topological band theory within a unified framework. The model identifies the minimal set of excitonic degrees of freedom required to describe polariton spectra, establishes symmetry- and polarization-imposed selection rules for light-matter coupling, and provides analytical control over band inversion through the Fourier components of the dielectric modulation. This approach predicts a geometry-controlled topological transition beyond the conventional breathing-honeycomb mechanism that allows polaritonic edge states at the topological interfaces. The developed framework offers a practical tool for designing polaritonic metasurfaces with engineered dispersion, radiative losses, and topology, with potential applications in chiral nanophotonics, nonlinear light-matter interactions, and topological polariton transport.

% \disclosures 
\subsection*{Supporting Information}
Derivation of the effective Hamiltonian including finite-wavevector corrections, TE and TM blocks along high-symmetry directions, full Hamiltonian with confined excitons, analytical derivation of topological transition points, model parameters for bulk WS$_2$ and monolayer WS$_2$ metasurfaces, Berry flux calculations and domain-wall mode propagation, and relation between light-matter coupling and oscillator strength.

\subsection*{Disclosures}
The authors declare that they have no competing financial interests or any other conflicts of interest.

\subsection*{Code, Data, and Materials Availability} 
The data used in this study are available from the corresponding author upon request.

\subsection*{Author Contributions}
D.S. and K.K. conceived the idea. P.P. and D.S. performed the numerical simulations and analyzed the data. D.S. and K.K. supervised the project and provided theoretical guidance, contributed to the interpretation of the physical mechanisms, and shaped the overall research direction. P.P prepared the initial draft, and all authors contributed to discussions, revisions, and the final preparation of the manuscript.

\section*{Acknowledgements}
K.K. and P.P. acknowledge financial support from the Australian Research Council via Discovery Early Career Researcher Award (grant no. DE250100419). D.S. acknowledges financial support from the Australian Research Council via Future Fellowship (grant no. FT230100058).

\bibliographystyle{elsarticle-num}
\bibliography{main}

\end{document}

\title{\SM\\
Band Engineering of Exciton Polaritons in Resonant Polaritonic Metasurfaces}

\author{Polina Pantiukhina}
\affiliation{Department of Electronic Materials Engineering, Research School of Physics, Australian National University, Canberra, ACT 2601, Australia
 }

\author{Daria Smirnova}
\email{daria.smirnova@anu.edu.au}
\affiliation{Department of Electronic Materials Engineering, Research School of Physics, Australian National University, Canberra, ACT 2601, Australia
 }
\affiliation{Department of Fundamental and Theoretical Physics, Research School of Physics, Australian National University, Canberra, ACT 2601, Australia
 }
 
\author{Kirill Koshelev}%
\email{kirill.koshelev@anu.edu.au}
\affiliation{Department of Electronic Materials Engineering, Research School of Physics, Australian National University, Canberra, ACT 2601, Australia
 }

\maketitle
\begin{quote}\textbf{Abstract:} Section S1A contains the derivation of finite-$\mathbf{k}$ corrections to the photonic harmonics, including the guided-mode projection for the resonant nonzero harmonics and the dyadic Green function treatment of the zero harmonic. Section S1B formulates the excitonic response in the envelope function approach, discusses the bulk and two-dimensional limits of the out-of-plane exciton profile, and defines the polarization basis used for the excitonic amplitudes. Section S1C derives the elimination of the zeroth photonic and excitonic harmonics and obtains the closed equations for the retained resonant amplitudes through second order in $\mathbf{k}$. Section S2A presents the reduced photon-exciton Hamiltonian, its angular-momentum representation, and the TE and TM blocks along the high-symmetry directions. Section S2B gives the full $\Gamma$ point Hamiltonian including confined excitons. Section S3A derives the geometrical origin of the photonic transition points from the Fourier coefficients of the six-hole unit cell. Section S3B illustrates these transition points using calculated photonic spectra. Section S4 describes the numerical implementation, including the rigorous coupled-wave analysis (RCWA) and COMSOL calculations and the fitting procedure. Section S4A lists the effective-Hamiltonian parameters for the bulk WS$_2$ metasurface of Figs.~2 and 3 of the main text and tests the elimination of the zeroth excitonic harmonic and the isotropic approximation for the WS$_2$ permittivity. Section S4B applies the model to a WS$_2$ monolayer on a patterned TiO$_2$ membrane. Section S4C introduces the effective dipole-quadrupole Hamiltonian for the large-hole domains of Fig.~5 of the main text, evaluates their finite-domain Berry fluxes, and calculates the propagation of the domain wall modes. Section S4D relates the light-matter coupling to the oscillator strength of the excitonic permittivity and compares the common and multipole dependent couplings.
\end{quote}

\tableofcontents

\section{Model description}
The supplementary derivation below provides the finite-$\bfk$ corrections to the effective Hamiltonian introduced in the main text and gives the corresponding symmetry-reduced blocks along the high-symmetry directions.
\subsection{Finite-$\bfk$ correction to the photonic harmonics}

We use $k=\omega/c$ for the spectral variable and $\bfk=(k_x,k_y)$ for the Bloch wave vector. The retained photonic subspace consists of the six first reciprocal vectors $\bfg$, with $|\bfg|=g$. The zero harmonic is eliminated using its outgoing dyadic Green's function. The approximation developed below retains the dispersion and polarization of a single guided branch, and both polarization channels of the zero harmonic.

\subsubsection{Guided-mode pole}

Let $k_{\rm TE_0}(\beta)$ denote the dispersion of the fundamental TE guided mode of the effective slab. We take $k_g$ to be a free real parameter specifying the reference frequency of the retained photonic subspace and define
\begin{equation}
k_{\rm wg}(\beta)
=k_g+k_{\rm TE_0}(\beta)-k_{\rm TE_0}(g),
\qquad k_{\rm wg}(g)=k_g.
\label{eq:kwg_ref}
\end{equation}
This parametrization allows a common frequency offset $k_g-k_{\rm TE_0}(g)$ while preserving the slope and curvature of the effective slab dispersion. The offset represents a phenomenological real-frequency renormalization beyond the retained guided-mode approximation. The zero-harmonic self-energy derived below is added separately and supplies its own frequency shifts and radiative losses.

The effective slab limit is reached as the periodic dielectric modulation vanishes at fixed $\veps(z)$. Provided the TE$_0$ branch remains spectrally isolated and no discarded photonic resonance approaches the reference frequency, mixing with the excluded photonic modes vanishes in this limit. Consistency with the effective slab then requires $k_g\to k_{\rm TE_0}(g)$. At finite modulation, $k_g$ is treated as an independent parameter, while the reference permittivity, guided-mode profile, and dispersion derivatives remain fixed by the chosen slab. The same renormalization affects $N_h$ and the radiative shift ${\rm Im}\,\gamma_{\rm ph}$ introduced below, which are treated as effective $\Gamma$-point parameters in the fits (Sec.~S4).
\\

For each retained harmonic we define
\begin{equation}
\beta_{{\bfg}}(\bfk)=|{\bfg}+\bfk|,\qquad
\hat{\mathbf e}^{\bfk}_{{\rm p},{\bfg}}
=\frac{{\bfg}+\bfk}{\beta_{{\bfg}}(\bfk)},\qquad
\hat{\mathbf e}^{\bfk}_{{\rm s},{\bfg}}
=\hat{\mathbf e}_z\times\hat{\mathbf e}^{\bfk}_{{\rm p},{\bfg}}.
\label{eq:qg_finite_k}
\end{equation}
The symbols $\rm s$ and $\rm p$ here specify local in-plane polarization directions. The retained photonic branch is TE$_0$ and has only the $\rm s$ component. We abbreviate $\hat{\mathbf e}^{\mathbf0}_{{\rm s},{\bfg}}=\hat{\mathbf e}_{{\rm s},{\bfg}}$ and $\hat{\mathbf e}^{\mathbf0}_{{\rm p},{\bfg}}=\hat{\mathbf e}_{{\rm p},{\bfg}}$.
\\

The scalar guided-mode profile $\theta_\beta(z)$ of the effective slab satisfies
$\left[\partial_z^2+k_{\rm TE_0}^2(\beta)\veps(z)-\beta^2\right]
\theta_\beta(z)=0.$
We choose real profiles for the lossless effective slab and fix their phase continuously. Throughout the effective model we use the single profile
\begin{equation*}
\theta(z)\equiv\theta_g(z),\qquad
N_h=\intz\theta^2(z).
\end{equation*}
At $\Gamma$, this common profile follows from the in-plane rotational symmetry of the slab and the equality $|\bfg|=g$. Away from $\Gamma$, using $\theta(z)$ for all $\beta_\bfg$ is an additional approximation. It is imposed consistently in the dielectric overlaps, the coupling to the zero harmonic, and the direct photon--exciton overlap.\\

With these conventions, the resonant Green's tensor and the field ansatz are
\begin{equation}
\opp{G}^{\bfk}_{{\bfg}}(z,z';k)
\simeq
\frac{\boldsymbol{\theta}^{\bfk}_{{\bfg}}(z)\otimes
\boldsymbol{\theta}^{\bfk}_{{\bfg}}(z')}
{k[k-k_{\rm wg}(\beta_{{\bfg}})]},
\qquad
\boldsymbol{\theta}^{\bfk}_{{\bfg}}(z)
=\theta(z)\hat{\mathbf e}^{\bfk}_{{\rm s},{\bfg}},
\label{eq:Gg_finite_k_corrected}
\end{equation}
\begin{equation*}
\bfe_{{\bfg}}(z)=\eta\hbar c\,a_{{\rm s},{\bfg}}\theta(z)
\hat{\mathbf e}^{\bfk}_{{\rm s},{\bfg}},
\qquad \eta=\sqrt{\frac{4\pi k_g}{\hbar c}}.
\end{equation*}
The rotation of the polarization enters every dielectric coupling term through
\begin{equation}
C_{{\bfg\bfg'}}(\bfk)
=\hat{\mathbf e}^{\bfk}_{{\rm s},{\bfg}}\cdot
\hat{\mathbf e}^{\bfk}_{{\rm s},{\bfg'}}
=\frac{({\bfg}+\bfk)\cdot({\bfg'}+\bfk)}
{\beta_{{\bfg}}\beta_{{\bfg'}}}.
\label{eq:Cgg_finite_k}
\end{equation}
Projecting the wave equation onto this ansatz gives
\begin{equation}
\begin{split}
[k-k_{\rm wg}(\beta_{{\bfg}})]a_{{\rm s},{\bfg}}={}&
-k_gN_h{\sum_{|\bfg'|=g}}
\Delta\eps_{{\bfg}-{\bfg'}}C_{{\bfg\bfg'}}(\bfk)a_{{\rm s},{\bfg'}}\\
&-\frac{k_g\eps_1}{\eta\hbar c}
\intz\theta(z)\hat{\mathbf e}^{\bfk}_{{\rm s},{\bfg}}\cdot\bfe_{\mathbf0}(z)
-\eta\intz\theta(z)\hat{\mathbf e}^{\bfk}_{{\rm s},{\bfg}}\cdot
\mathbf P_{{\bfg}}(z).
\end{split}
\label{eq:ag_eq_k_projected}
\end{equation}
Here $\Delta\eps_{\mathbf 0}\equiv0$, and $\Delta\eps_{{\bfg}-{\bfg'}}=\eps_m$, where $m=1,2,3$ labels $|{\bfg}-{\bfg'}|=g,\sqrt3g,2g$.

\subsubsection{Expansion of the guided-mode kinematics}
Introduce the wave-vector components
$k_{\parallel,{\bfg}}=\hat{\mathbf e}_{{\rm p},{\bfg}}\cdot\bfk$ and
$k_{\perp,{\bfg}}=\hat{\mathbf e}_{{\rm s},{\bfg}}\cdot\bfk$. Then
$\beta_{{\bfg}}=g+k_{\parallel,{\bfg}}
+{k_{\perp,{\bfg}}^2}/({2g})
+\mathcal O(\bfk^3).$
Define the derivatives of the effective slab TE$_0$ branch at $\beta=g$ by
\begin{equation}
v_{\rm g}=\left.\frac{\mathrm d k_{\rm TE_0}}{\mathrm d\beta}\right|_g,\qquad
\mu_{\parallel}=\frac12
\left.\frac{\mathrm d^2 k_{\rm TE_0}}{\mathrm d\beta^2}\right|_g,\qquad
\mu_{\perp}=\frac{v_{\rm g}}{2g}.
\label{eq:mu_parallel_perp}
\end{equation}
The pole shift is therefore
\begin{equation}
k_{\rm wg}(\beta_{{\bfg}})-k_g
=v_{\rm g} k_{\parallel,{\bfg}}
+\mu_{\parallel}k_{\parallel,{\bfg}}^2
+\mu_{\perp}k_{\perp,{\bfg}}^2+\mathcal O(\bfk^3).
\label{eq:guided_dispersion_expansion}
\end{equation}
The two quadratic coefficients have distinct origins: $\mu_{\parallel}$ is the curvature of the guided branch, whereas $\mu_{\perp}$ results from the change of the magnitude of ${\bfg}+\bfk$ under a transverse displacement. 

\subsubsection{Dyadic response of the zero photonic harmonic}
The outgoing Green's function of the zero photonic harmonic at a fixed spectral variable $k$ satisfies
\begin{equation}
\left[k^2\veps(z)-\nabla_{z,\bfk}\times\nabla_{z,\bfk}\times\right]
\opp{G}^{\bfk}_{\mathbf0}(z,z';k)
=\hat{\mathbf1}\delta(z-z'),
\qquad
\nabla_{z,\bfk}=(\iu k_x,\iu k_y,\partial_z).
\label{FG0}
\end{equation}
At $\bfk=\mathbf0$, its in-plane response is independent of the polarization direction
\begin{equation*}
\begin{aligned}
\opp{G}_{\mathbf0}(z,z';k)
=G_{\bf 0}(z,z';k)\hat{\mathbf1}_{\parallel}
+\frac{\delta(z-z')}{k^2\veps(z)}
\hat{\mathbf e}_z\otimes\hat{\mathbf e}_z,\quad
[\partial_z^2+k^2\veps(z)]G_{\bf 0}(z,z';k)
&=\delta(z-z').
\end{aligned}
\end{equation*}
Here $\hat{\mathbf1}_{\parallel}$ is the identity in the $(x,y)$ plane.
For $|\bfk|>0$, we extend the polarization convention of Eq.~\eqref{eq:qg_finite_k} to the zero harmonic 
$\hat{\mathbf e}^{\bfk}_{{\rm p},0}={\bfk}/{|\bfk|},$
$\hat{\mathbf e}^{\bfk}_{{\rm s},0}
=\hat{\mathbf e}_z\times\hat{\mathbf e}^{\bfk}_{{\rm p},0}.$
The TE radiation channel is polarized along
$\hat{\mathbf e}^{\bfk}_{{\rm s},0}$, perpendicular to $\bfk$.
The in-plane component of the TM radiation channel is along
$\hat{\mathbf e}^{\bfk}_{{\rm p},0}$, parallel to $\bfk$, and is coupled to an out-of-plane electric field.
The in-plane Green's tensor~\cite{sipe1987new} can therefore be written exactly as
\begin{equation*}
\begin{split}
\opp{G}^{\bfk}_{\mathbf0,\parallel}(z,z';k)
={}&G_{\rm s}(z,z';k)
\hat{\mathbf e}^{\bfk}_{{\rm s},0}\otimes
\hat{\mathbf e}^{\bfk}_{{\rm s},0}+G_{\rm p}(z,z';k)
\hat{\mathbf e}^{\bfk}_{{\rm p},0}\otimes
\hat{\mathbf e}^{\bfk}_{{\rm p},0}.
\end{split}
\end{equation*}
The scalar kernels satisfy
\begin{equation}
\begin{aligned}
[\partial_z^2+k^2\veps(z)-\bfk^2]&G_{\rm s}
=\delta(z-z'),\\
\partial_z\left[
\frac{k^2\veps(z)}{k^2\veps(z)-\bfk^2}
\partial_zG_{\rm p}\right]
+k^2&\veps(z)G_{\rm p}
=\delta(z-z').
\end{aligned}
\label{eq:zero_sp_operators}
\end{equation}
The TM kernel follows from the $z$ component of Eq.~\eqref{FG0}.
At fixed $k$, both kernels have the expansion
\begin{equation*}
G_\alpha(z,z';k)
=G_{\bf 0}(z,z';k)
+\bfk^2\mathcal G_\alpha^{(2)}(z,z';k)
+\mathcal O(\bfk^4),
\qquad \alpha={\rm s,p}.
\end{equation*}
The TE coefficient is
\begin{equation}
\mathcal G_{\rm s}^{(2)}(z,z';k)
=\int_{-\infty}^{\infty}\mathrm dz''\,
G_{\bf 0}(z,z'';k)G_{\bf 0}(z'',z';k),
\label{eq:G0_scalar_Dyson}
\end{equation}
whereas the TM coefficient is
\begin{equation}
\mathcal G_{\rm p}^{(2)}(z,z';k)
=\int_{-\infty}^{\infty}\mathrm dz''\,
\frac{\partial_{z''} G_{\bf 0}(z,z'';k)\,
\partial_{z''}G_{\bf 0}(z'',z';k)}
{k^2\veps(z'')}.
\label{eq:GTM2}
\end{equation}
The integrals over the semi-infinite claddings in Eqs.~\eqref{eq:G0_scalar_Dyson} and \eqref{eq:GTM2} are understood with $k\to k+\iu0$.
Using
$\hat{\mathbf e}^{\bfk}_{{\rm s},0}\otimes
\hat{\mathbf e}^{\bfk}_{{\rm s},0}
+\hat{\mathbf e}^{\bfk}_{{\rm p},0}\otimes
\hat{\mathbf e}^{\bfk}_{{\rm p},0}
=\hat{\mathbf1}_{\parallel}$,
the complete second-order in-plane tensor becomes
\begin{equation}
\begin{split}
\opp{G}^{\bfk}_{\mathbf0,\parallel}
={}G_{\bf 0}\hat{\mathbf1}_{\parallel}
+\bfk^2\mathcal G_{\rm s}^{(2)}
\hat{\mathbf1}_{\parallel}+\left(\mathcal G_{\rm p}^{(2)}
-\mathcal G_{\rm s}^{(2)}\right)
\bfk\otimes\bfk+\mathcal O(\bfk^4).
\end{split}
\label{eq:G0_TE_TM}
\end{equation}
The eliminated electric field is
\begin{equation}
\begin{split}
\bfe_{\mathbf0}(z)={}
-\eta\hbar c\,k_g^2\eps_1
{\sum_{|\bfg|=g}}a_{{\rm s},{\bfg}}
\intzp\opp{G}^{\bfk}_{\mathbf0}(z,z';k_g)
\theta(z')\hat{\mathbf e}^{\bfk}_{{\rm s},{\bfg}}-4\pi k_g^2\intzp
\opp{G}^{\bfk}_{\mathbf0}(z,z';k_g)
\mathbf P_{\mathbf0}(z').
\end{split}
\label{eq:lippmann_schwinger_full_green_finite_k}
\end{equation}
The sources and retained guided-mode fields are in-plane, so only
$\opp{G}^{\bfk}_{\mathbf0,\parallel}$ enters their overlap integrals.

\subsection{Excitonic response}
\subsubsection{Envelope wavefunction formulation of the excitonic response}
We use the same notation as in the main text: $F(z)$ is the normalized out-of-plane exciton envelope, and $\Phi_{\alpha}(x,y)$ is the in-plane center-of-mass envelope wavefunction, where $\alpha$ labels the exciton polarization, $\mathbf d_{\alpha}=d\hat{\mathbf e}_{\alpha}$ is the excitonic dipole moment. The exciton-induced polarization is written as $\mathbf P(\mathbf r)=F(z)\sum_{\alpha}\mathbf d_{\alpha}^{*}\Phi_{\alpha}(x,y)$. In the absence of in-plane exciton dynamics this definition is equivalent to the nonlocal susceptibility in Eq.~(7) of the main text, with 
\begin{equation*}
\Phi_{\alpha}(x,y)=\dfrac{\int_{-h}^{h}\mathrm dz\,F^*(z)\mathbf d_{\alpha}\cdot\mathbf E(x,y,z)}{\hbar(\omega_X-\omega-\iu\Gamma)}.
\end{equation*}
We now include the in-plane center-of-mass motion and the periodic potential $U(x,y)$ induced by the metasurface pattern. The driven equation then gives
\begin{equation}
\left[\hbar\omega-\hbar\omega_X+\iu\hbar\Gamma
+\frac{\hbar^2\nabla^2}{2M}-U(x,y)\right]\Phi_{\alpha}(x,y)=-\int_{-h}^{h}\mathrm dz\,F^*(z)\mathbf d_{\alpha}\cdot\mathbf E(x,y,z),
\label{eq:SM_exc_env}
\end{equation}
where $M$ is the exciton translational mass and $\nabla=(\partial_x,\partial_y)$. Since excitons exist only inside the material, the potential vanishes in the material and equals the barrier height $U_{\rm max}$ in the holes $U(x,y)=U_{\rm max}[1-\xi(x,y)]$, where the step function $\xi(x,y)$ equals 1 in the material and 0 in the holes. The barrier height, of the order of several eV, is much larger than all other energies of the problem and cannot be treated as a small perturbation. We therefore take the hard wall limit $U_{\rm max}\to\infty$, in which the potential reduces to the boundary condition $\Phi_\alpha=0$ in the holes, and Eq.~\eqref{eq:SM_exc_env} becomes
\begin{equation*}
\left[\hbar\omega-\hbar\omega_X+\iu\hbar\Gamma
+\frac{\hbar^2\nabla^2}{2M}\right]\Phi_{\alpha}(x,y)=-\xi(x,y)\int_{-h}^{h}\mathrm dz\,F^*(z)\mathbf d_{\alpha}\cdot\mathbf E(x,y,z),
\end{equation*}
with $\Phi_\alpha=0$ in the holes, the factor $\xi(x,y)$ on the right-hand side means that the excitons are driven only by the field inside the material. The same restriction is used in the full-wave calculations of Sec.~S4, where the excitonic permittivity is assigned to the material only. For a continuous excitonic layer on a patterned dielectric metasurface (Sec.~S4B) $\xi\equiv1$.

Next, we expand 
\begin{equation*}
\Phi_{\alpha}(x,y)=\dfrac{1}{\sqrt{S_u}}\sum_{\bfg}\Phi_{\alpha,\mathbf g}e^{\iu (\bfk+\bfg)\cdot{\bf r}}, \quad  \xi(x,y)=\sum_{\bfg}\xi_{\bfg}e^{\iu \bfg\cdot{\bf r}}.
\end{equation*}
For the excitonic metasurfaces the background permittivity of the slab is $\eps(x,y)=1+(\eps-1)\xi(x,y)$, where $\eps=\eps_\infty$ is the background permittivity of the material. The step function therefore has the same Fourier coefficients as the permittivity: $\xi_{\mathbf 0}=(\veps-1)/(\eps-1)$ is the area fraction of the material, and $\xi_{\bfg}=\Delta\eps_{\bfg}/(\eps-1)$ for $\bfg\neq\mathbf 0$. For $|\bfg|=g,\sqrt3g,2g$ we write $\xi_{\bfg}=\xi_1,\xi_2,\xi_3$, so that $\xi_m=\eps_m/(\eps-1)$. The confinement therefore introduces no additional parameters. The center-of-mass kinetic energy $\hbar^2|\bfk+\bfg|^2/(2M)$, together with the confinement energy induced by the holes, is relatively small for the structures considered here and is neglected at the scale of the photonic dispersion.

Dividing the projected equation by $\hbar c$ gives
\begin{equation}
(k-E_X)\Phi_{\alpha,\bfg}=-\frac{1}{\hbar c}\sum_{\bfg'}\xi_{\bfg-\bfg'}\int_{-h}^{h}\mathrm dz\,F^*(z)\mathbf d_{\alpha}\cdot\mathbf E_{\bfg'}(z),
\label{eq:SM_exc_fourier0}
\end{equation}
where $E_X=(\omega_X-\iu \Gamma)/c$. The term with $\bfg'=\bfg$ restricts the excitons to the material on average over the unit cell. The terms with $\bfg'\neq\bfg$ are of relative order $\xi_{\bfg-\bfg'}/\xi_{\mathbf 0}=\Delta\eps_{\bfg-\bfg'}/(\veps-1)$, i.e., of the order of the dielectric modulation. For the photons, the same $\Delta\eps_{\bfg-\bfg'}$ couples six degenerate guided modes in Eq.~\eqref{eq:ag_eq_k_projected} and therefore determines their splitting already in the first order. For the excitons at $|\bfg|=g$, they only correct the light-matter coupling by a small relative amount and make it weakly dependent on the multipole index (Sec.~S2B).  As in the main text, we neglect them and introduce the semiclassical excitonic amplitudes $b_{\alpha,\bfg}=\Phi_{\alpha,\bfg}/\sqrt{\xi_{\mathbf 0}}$, which gives
\begin{equation}
(k-E_X)b_{\alpha,\mathbf g}
=-\frac{{\sqrt{\xi_{\mathbf 0}}}}{\hbar c}
\int_{-h}^{h}\mathrm dz\,F^*(z)\mathbf d_{\alpha}\cdot\mathbf E_{\mathbf g}(z).
\label{eq:SM_exc_fourier}
\end{equation}
Equation~\eqref{eq:SM_exc_fourier} is therefore the equation for an unpatterned excitonic slab with the dipole moment $\sqrt{\xi_{\mathbf 0}}\,d$, i.e., with the area-averaged oscillator strength.

\subsubsection{Bulk and 2D limits of the out-of-plane exciton profile}
In the two-particle approximation, the total exciton wavefunction can be separated into $\Phi(x,y)$ and $\varphi(x_e-x_h,y_e-y_h,z_e,z_h)$ describing the center-of-mass in-plane electron-hole motion along $x,y$, relative electron and hole in-plane motion via  [$x_e-x_h$] and [$y_e-y_h$], and out-of-plane electron and hole motion via $z_e$ and $z_h$~\cite{ivchenko2005optical}. 

For the excitonic medium thickness much larger than the 3D Bohr radius $(2h\gg a_{\rm B})$, the out-of-plane motion of electron and hole can also be separated into the center-of-mass [$z=(z_e+z_h)/2$] and relative [$z_e-z_h$] degrees of freedom. Thus, the wavefunction further factorizes into $\varphi(x_e-x_h,y_e-y_h,z_e,z_h)=\varphi(x_e-x_h,y_e-y_h,z_e-z_h)F(z)$, where 
\begin{equation*}
\begin{aligned}
&\varphi(x_e-x_h,y_e-y_h,z_e-z_h)= \sqrt{\frac{1}{\pi a^3_{\rm B}}}\eu^{-r/a_{\rm B}},\\
&F(z)=\frac{1}{\sqrt{h}}\cos{\left(\frac{\pi z}{2h}\right)}.
\end{aligned}
\end{equation*}
Here, $r=\sqrt{(x_e-x_h)^2+(y_e-y_h)^2+(z_e-z_h)^2}$, and the wavefunctions obey the normalization $\int_{-h}^h {\rm d}z\, |F(z)|^2=1$ and $4\pi\int_0^\infty {\rm d}r\, r^2|\varphi(r)|^2=1$.  

For the excitonic medium thickness much smaller than the 3D Bohr radius $(2h\ll a_{\rm B})$ (e.g., a TMD monolayer integrated with a photonic metasurface), the out-of-plane motion of electron and hole is independent. Thus, the wavefunction further factorizes into $\varphi(x_e-x_h,y_e-y_h,z_e,z_h)=\varphi(x_e-x_h,y_e-y_h)\zeta(z_e)\zeta(z_h)$, where 
\begin{equation*}
\begin{aligned}
&\varphi(\rho)= \sqrt{\frac{8}{\pi a^2_{\rm B}}}\eu^{-2\rho/a_{\rm B}},\\
&\zeta(z)=\frac{1}{\sqrt{h}}\cos{\left(\frac{\pi z}{2h}\right)}H_\theta(|z|<h).
\end{aligned}
\end{equation*}
Here, $\rho=\sqrt{(x_e-x_h)^2+(y_e-y_h)^2}$, and the wavefunctions obey the normalization $2\pi\int_0^\infty {\rm d}\rho\, \rho|\varphi(\rho)|^2=1$ and $\int_{-h}^h {\rm d}z\, |\zeta(z)|^2=1$. In the 2D limit ($h\to0$), $\zeta(z_e)\zeta(z_h)\to\delta(z)$, where $z=z_e=z_h$. Thus, one can introduce $F(z)=\delta(z)$ and write $\varphi(x_e-x_h,y_e-y_h,z_e,z_h)=\varphi(x_e-x_h,y_e-y_h)F(z)$.

We also note that the effective dipole moment is defined by the parameter $\varphi(0)$ that depends on the exciton system dimensionality as
\begin{equation*}
\varphi(0)=\begin{cases}
\frac{1}{\sqrt{\pi a_{\rm B}^3}},\quad &2h \gg   a_{\rm B},\\
\frac{2\sqrt{2}}{\sqrt{\pi a_{\rm B}^2}},\quad &2h \ll   a_{\rm B}
\end{cases}.    
\end{equation*}

\subsubsection{Excitonic polarization basis}
Equation~\eqref{eq:SM_exc_fourier} is invariant under changes of basis in the in-plane polarization space. For a nonzero retained harmonic, use the real rotating directions
$\hat{\mathbf e}^{\bfk}_{{\rm s},{\bfg}}$ and
$\hat{\mathbf e}^{\bfk}_{{\rm p},{\bfg}}$ defined in Eq.~\eqref{eq:qg_finite_k}.
Write
\begin{equation*}
\mathbf b_{{\bfg}}=b_{x,{\bfg}}\hat{\mathbf e}_x+b_{y,{\bfg}}\hat{\mathbf e}_y
=b_{{\rm s},{\bfg}}\hat{\mathbf e}^{\bfk}_{{\rm s},{\bfg}}
+b_{{\rm p},{\bfg}}\hat{\mathbf e}^{\bfk}_{{\rm p},{\bfg}},\qquad
b_{\alpha,{\bfg}}=\hat{\mathbf e}^{\bfk}_{\alpha,{\bfg}}\cdot\mathbf b_{{\bfg}}.
\end{equation*}
With the kinetic term omitted, the local equation is
\begin{equation*}
\begin{split}
(k-E_X)b_{\alpha,{\bfg}}
=-\frac{\sqrt{\xi_{\bf0}}d}{\hbar c}\intz F^*(z)
\hat{\mathbf e}^{\bfk}_{\alpha,{\bfg}}\cdot\bfe_{{\bfg}}(z).
\end{split}
\end{equation*}
Because the retained optical field is along
$\hat{\mathbf e}^{\bfk}_{{\rm s},{\bfg}}$, its direct coupling is
$V a_{{\rm s},{\bfg}}\delta_{\alpha,{\rm s}}$ for a real $F(z)$ at every $\bfk$.
Since the excitons are not coupled to each other, the excitonic block $E_X\hat{\mathbf 1}$ is invariant under this rotation, and the local ${\rm p}$ excitons are not driven and remain dark.

For the zero harmonic we keep the global basis, $\mathbf b_0=b_{x,0}\hat{\mathbf e}_x+b_{y,0}\hat{\mathbf e}_y$, which is regular as $\bfk\to\mathbf0$
\begin{equation*}
(k-E_X)b_{x,0}
=-\frac{\sqrt{\xi_{\bf 0}}d}{\hbar c}\intz F^*(z)\hat{\mathbf e}_x\cdot\bfe_{\mathbf0}(z),
\end{equation*}
and similarly for the $y$ component.
At finite $\bfk$, this vector can also be decomposed into the TE and TM radiation directions.\\

\subsection{Elimination of the zeroth harmonics}
We eliminate the zero photonic and excitonic harmonics using the guided-mode approximation of Sec.~S1A and the excitonic response of Sec.~S1B. 
\\

We choose real $F(z)$ and $\theta(z)$ as above and define the overlap integrals
\begin{equation}
\begin{aligned}
V&=-\sqrt{\xi_{\bf0}}d\eta\intz\theta(z)F(z),\quad
J=\sqrt{\xi_{\bf0}}\eta k_g^2\eps_1
\iint_{-h}^{h}\mathrm dz\,\mathrm dz'\,
F(z)G_{\bf 0}(z,z';k_g)\theta(z'),\\
\gamma_{\rm ph}=\iu k_g^3\eps_1^2
\iint_{-h}^{h}&\mathrm dz\,\mathrm dz'\,
\theta(z)G_{\bf 0}(z,z';k_g)\theta(z'),\quad
\gamma_X=\iu\xi_{\bf0}\eta^2|d|^2k_g
\iint_{-h}^{h}\mathrm dz\,\mathrm dz'\,
F(z)G_{\bf 0}(z,z';k_g)F(z').
\end{aligned}
\label{eq:zero_overlap_Gamma}
\end{equation}
 The outgoing Green's function makes $J$, $\gamma_{\rm ph}$, and $\gamma_X$ generally complex. 
\\

For finite-$\bfk$, let $\alpha={\rm s,p}$ label the zero-harmonic TE and TM channels, respectively. The same correspondence applies to their second-order coefficients $\mathcal G_\alpha^{(2)}$
\begin{equation}
\Delta J_\alpha
=\sqrt{\xi_{\bf0}}\eta k_g^2\eps_1
\iint_{-h}^{h}\mathrm dz\,\mathrm dz'\,
F(z)\mathcal G_\alpha^{(2)}(z,z';k_g)\theta(z'),
\label{eq:J2}
\end{equation}
\begin{equation}
\Delta\gamma_{{\rm ph},\alpha}
=k_g^3\eps_1^2
\iint_{-h}^{h}\mathrm dz\,\mathrm dz'\,
\theta(z)\mathcal G_\alpha^{(2)}(z,z';k_g)\theta(z'),
\label{eq:gamma2}
\end{equation}
The finite-$\bfk$ correction to the renormalized detuning of the eliminated zeroth excitonic sector $D=k_g - (E_X -\iu \gamma_X)$ is defined as
\begin{equation}
\Delta D_\alpha
=-\xi_{\bf0}\eta^2|d|^2k_g
\iint_{-h}^{h}\mathrm dz\,\mathrm dz'\,
F(z)\mathcal G_\alpha^{(2)}(z,z';k_g)F(z').
\label{eq:D2}
\end{equation}
The accuracy of evaluating the zeroth-exciton denominator at $k_g$ is tested numerically in Sec. S4A.

For $|\bfk|>0$, it is convenient to resolve the zero-exciton amplitude along the same two directions as the zero photonic harmonic
$b_{\alpha,0}
=\hat{\mathbf e}^{\bfk}_{\alpha,0}\cdot\mathbf b_0,$
$\mathbf b_0=\sum_{\alpha={\rm s,p}}
b_{\alpha,0}\hat{\mathbf e}^{\bfk}_{\alpha,0}.$
This is a rotation of its global polarization basis. At $\Gamma$ the two channel coefficients coincide, and one may use any fixed orthonormal in-plane basis.
\\

Substituting Eq.~\eqref{eq:lippmann_schwinger_full_green_finite_k} into the guided-mode projection yields
\begin{equation*}
\begin{split}
-\frac{k_g\eps_1}{\eta\hbar c}
\intz\theta(z)\hat{\mathbf e}^{\bfk}_{{\rm s},{\bfg}}\cdot\bfe_{\mathbf0}(z)
={}{\sum_{|\bfg'|=g}}\sum_{\alpha}
\left(-\iu\gamma_{\rm ph}
+\Delta\gamma_{{\rm ph},\alpha}\bfk^2\right)&
\left(\hat{\mathbf e}^{\bfk}_{{\rm s},{\bfg}}\cdot
\hat{\mathbf e}^{\bfk}_{\alpha,0}\right)
\left(\hat{\mathbf e}^{\bfk}_{\alpha,0}\cdot
\hat{\mathbf e}^{\bfk}_{{\rm s},{\bfg'}}\right)a_{{\rm s},{\bfg'}}\\
&+d^*\sum_{\alpha}
\left(J+\Delta J_\alpha\bfk^2\right)
\left(\hat{\mathbf e}^{\bfk}_{{\rm s},{\bfg}}\cdot
\hat{\mathbf e}^{\bfk}_{\alpha,0}\right)b_{\alpha,0}.
\end{split}
\end{equation*}
The same substitution into the zero-exciton equation gives
\begin{equation*}
\begin{split}
\left[D+\Delta D_\alpha\bfk^2\right]b_{\alpha,0}
=d\left(J+\Delta J_\alpha\bfk^2\right)
{\sum_{|\bfg'|=g}}
\left(\hat{\mathbf e}^{\bfk}_{\alpha,0}\cdot
\hat{\mathbf e}^{\bfk}_{{\rm s},{\bfg'}}\right)a_{{\rm s},{\bfg'}}.
\end{split}
\end{equation*}
Eliminating the two amplitudes $b_{\alpha,0}$ gives
\begin{equation}
\begin{split}
[k-k_{\rm wg}(\beta_{{\bfg}})]a_{{\rm s},{\bfg}}
={}&-k_gN_h{\sum_{|\bfg'|=g}}
\Delta\eps_{{\bfg}-{\bfg'}}C_{{\bfg\bfg'}}(\bfk)a_{{\rm s},{\bfg'}}
+V^*b_{{\rm s},{\bfg}}\\
&+{\sum_{|\bfg'|=g}}\sum_{\alpha}
\left[
-\iu\gamma_{\rm ph}
+\Delta\gamma_{{\rm ph},\alpha}\bfk^2
+\frac{|d|^2(J+\Delta J_\alpha\bfk^2)^2}
{D+\Delta D_\alpha\bfk^2}
\right]
\left(\hat{\mathbf e}^{\bfk}_{{\rm s},{\bfg}}\cdot
\hat{\mathbf e}^{\bfk}_{\alpha,0}\right)
\left(\hat{\mathbf e}^{\bfk}_{\alpha,0}\cdot
\hat{\mathbf e}^{\bfk}_{{\rm s},{\bfg'}}\right)a_{{\rm s},{\bfg'}}.
\end{split}
\label{eq:ag_without0_k_SM}
\end{equation}
The corresponding equations for the two retained excitonic polarizations are
\begin{equation}
\begin{split}
(k-E_X)b_{{\rm s},{\bfg}}
={}Va_{{\rm s},{\bfg}},\quad
(k-E_X)b_{{\rm p},{\bfg}}
=0.
\end{split}
\label{eq:bpg_without0_k_SM}
\end{equation}
The p-polarized excitons are not driven and not enter the photonic equations, then form a dark manifold at $E_{\rm X}$.
The rational factors in the closed equations are understood through second order in $|\bfk|$.
For the photon self-energy this gives
\begin{equation*}
\begin{split}
\frac{|d|^2(J+\Delta J_\alpha\bfk^2)^2}
{D+\Delta D_\alpha\bfk^2}
={}\frac{|d|^2J^2}{D}+\Lambda_{X,\alpha}\bfk^2
+\mathcal O(\bfk^4),
\end{split}
\end{equation*}
where
$\Lambda_{X,\alpha}
=|d|^2\left(
{2J\Delta J_\alpha}/{D}
-{J^2\Delta D_\alpha}/{D^2}\right).$

\section{Effective Hamiltonian}
The closed equations~\eqref{eq:ag_without0_k_SM}--\eqref{eq:bpg_without0_k_SM} define an effective problem in the space of six photonic amplitudes $a_{{\rm s},{\bfg}}$ and twelve excitonic amplitudes $b_{{\rm s},{\bfg}}$ and $b_{{\rm p},{\bfg}}$. We now rewrite this system in Hamiltonian form. The reciprocal vectors are related by the $C_{6v}$ symmetry, and at the $\Gamma$ point the corresponding matrices are circulant, because the matrix elements depend only on the relative angle $\Delta\varphi_{{\bfg\bfg'}}=\varphi_{{\bfg}}-\varphi_{{\bfg'}}$. This makes it natural to use the discrete Fourier transform (DFT) to pass from the reciprocal-vector basis to the angular-momentum basis labelled by the multipole index $q$. The ${\rm p}$-polarized excitons are not driven (Eq.~\eqref{eq:bpg_without0_k_SM}) and the Hamiltonian reduces to a photon-exciton problem in the subspace of $a_{{\rm s},{\bfg}}$ and $b_{{\rm s},{\bfg}}$. We then restore the Fourier coefficients $\xi_m$ of the step function neglected in Eq.~\eqref{eq:SM_exc_fourier} and give the corresponding $\Gamma$ point Hamiltonian.

\subsection{Reduced photon-exciton Hamiltonian}
We begin with the minimal model in which the excitonic Fourier components are switched off, leaving a closed photon-exciton problem formed by the photonic harmonics and the corresponding ${\rm s}$-polarized excitonic amplitudes. 
\begin{comment}With $u_{\bfg}=0$, Eqs.~\eqref{eq:ag_without0_k_SM} and \eqref{eq:bsg_without0_k_SM} give, through second order in $|\bfk|$, the photonic matrix
\begin{equation}
\begin{split}
\left[\hat H_{\rm ph}^{(\bfg)}(\bfk)\right]_{jj'}
={}&k_{\rm wg}(\beta_j)\,\delta_{jj'}
+\left[-\iu\gamma_{\rm ph}+\frac{|d|^2J^2}{D}-k_gN_h\,\Delta\eps_{\bfg_j-\bfg_{j'}}\right]C_{jj'}(\bfk)\\
&+\left(\Delta\gamma_{{\rm ph},{\rm s}}+\Lambda_{X,{\rm s}}\right)k_{\parallel,j}k_{\parallel,j'}
+\left(\Delta\gamma_{{\rm ph},{\rm p}}+\Lambda_{X,{\rm p}}\right)k_{\perp,j}k_{\perp,j'}.
\end{split}
\label{eq:Hph_reciprocal_corrected}
\end{equation}
The last two terms follow from $|\bfk|\hat{\mathbf e}^{\bfk}_{{\rm s},0}=\hat{\mathbf e}_z\times\bfk$ and $|\bfk|\hat{\mathbf e}^{\bfk}_{{\rm p},0}=\bfk$, so that $|\bfk|\,\hat{\mathbf e}_{{\rm s},j}\cdot\hat{\mathbf e}^{\bfk}_{{\rm s},0}=k_{\parallel,j}$ and $|\bfk|\,\hat{\mathbf e}_{{\rm s},j}\cdot\hat{\mathbf e}^{\bfk}_{{\rm p},0}=k_{\perp,j}$; the rotation of $\hat{\mathbf e}^{\bfk}_{{\rm s},j}$ contributes to these terms only at order $\bfk^3$. The excitonic block and the direct photon-exciton coupling are $E_X\hat{\mathbf 1}_6$ and $V\hat{\mathbf 1}_6$, respectively.
\end{comment}
To make the $C_{6v}$ structure explicit, we transform from the reciprocal-vector basis to the angular-momentum basis using the DFT. We order the six reciprocal vectors counterclockwise,
as in the reciprocal-space schematic of the main text. Their polar angles $\varphi_{\bfg}$
therefore run from $\pi/2$ in steps of $\pi/3$. This gives the same DFT phase convention in vector-labelled form
\begin{equation}
{F_{q\bfg}}=\frac{1}{\sqrt6}
{\exp\!\left[\iu q\left(\varphi_{\bfg}-\frac{\pi}{2}\right)\right]},
\qquad q=0,1,2,3,-2,-1,
\qquad {|\bfg|=g}.
\label{eq:DFT}
\end{equation}
At the $\Gamma$ point, $C_{{\bfg\bfg'}}(\mathbf 0)=\cos\Delta\varphi_{{\bfg\bfg'}}$, the photonic block is circulant and is therefore diagonalized by this transformation. Its eigenvalues in the angular-momentum basis are
\begin{equation}
\Omega_q
=k_g
+3\delta_{q,\pm1}
\left[-\iu\gamma_{\rm ph}+\frac{|d|^2J^2}{D}\right]
+k_gN_h\left[
-\eps_1\cos\left(\frac{\pi q}{3}\right)
+\eps_2\cos\left(\frac{2\pi q}{3}\right)
+\eps_3(-1)^q
\right].
\label{eq:11}
\end{equation}
In the reduced photon-exciton subspace the transformation is
\begin{equation*}
\hat H=\hat{\mathcal F}\hat H_0\hat{\mathcal F}^{\dagger},
\qquad
\hat{\mathcal F}=\operatorname{diag}(\hat F,\hat F),
\end{equation*}
where $\hat H_0$ is the reduced Hamiltonian in the reciprocal-vector basis. Since the excitonic block and the direct photon-exciton coupling are proportional to identity matrices, they remain diagonal after this transformation.

Away from the $\Gamma$ point, we expand the overlap $C_{{\bfg\bfg'}}(\bfk)=\cos(\varphi_{{\bfg}}^{\bfk}-\varphi_{{\bfg'}}^{\bfk})$ using the polar angle of ${\bfg}+\bfk$, $\varphi_{{\bfg}}^{\bfk}=\varphi_{{\bfg}}+k_{\perp,{\bfg}}/g-k_{\parallel,{\bfg}}k_{\perp,{\bfg}}/g^2+\mathcal O(\bfk^3)$. Keeping terms up to second order in the in-plane wave vector, the reduced Hamiltonian in the angular-momentum basis takes the form
\begin{equation}
\hat H=
\left(
\begin{array}{cccccccccccc}
\tilde\Omega_0(\bfk) & \nu_{01}k_- & -\mu_{02}k_-^2 & 0 & -\mu_{02}k_+^2 & -\nu_{01}k_+ & V^* & 0 & 0 & 0 & 0 & 0\\
-\nu_{01}k_+ & \tilde\Omega_1(\bfk) & \nu_{12}k_- & -\mu_{13}k_-^2 & 0 & -\mu_{11}k_+^2 & 0 & V^* & 0 & 0 & 0 & 0\\
-\mu_{02}k_+^2 & -\nu_{12}k_+ & \tilde\Omega_2(\bfk) & \nu_{23}k_- & -\mu_{22}k_-^2 & 0 & 0 & 0 & V^* & 0 & 0 & 0\\
0 & -\mu_{13}k_+^2 & -\nu_{23}k_+ & \tilde\Omega_3(\bfk) & \nu_{23}k_- & -\mu_{13}k_-^2 & 0 & 0 & 0 & V^* & 0 & 0\\
-\mu_{02}k_-^2 & 0 & -\mu_{22}k_+^2 & -\nu_{23}k_+ & \tilde\Omega_2(\bfk) & \nu_{12}k_- & 0 & 0 & 0 & 0 & V^* & 0\\
\nu_{01}k_- & -\mu_{11}k_-^2 & 0 & -\mu_{13}k_+^2 & -\nu_{12}k_+ & \tilde\Omega_1(\bfk) & 0 & 0 & 0 & 0 & 0 & V^*\\
V & 0 & 0 & 0 & 0 & 0 & E_X & 0 & 0 & 0 & 0 & 0\\
0 & V & 0 & 0 & 0 & 0 & 0 & E_X & 0 & 0 & 0 & 0\\
0 & 0 & V & 0 & 0 & 0 & 0 & 0 & E_X & 0 & 0 & 0\\
0 & 0 & 0 & V & 0 & 0 & 0 & 0 & 0 & E_X & 0 & 0\\
0 & 0 & 0 & 0 & V & 0 & 0 & 0 & 0 & 0 & E_X & 0\\
0 & 0 & 0 & 0 & 0 & V & 0 & 0 & 0 & 0 & 0 & E_X
\end{array}
\right),
\label{eq:H_kp}
\end{equation}
where $k_\pm=(k_x\pm\iu k_y)/\sqrt2$. The diagonal terms are written as $\tilde\Omega_q(\bfk)=\Omega_q+\Delta\Omega_q\bfk^2$, $\tilde\Omega_{-q}(\bfk)=\tilde\Omega_q(\bfk)$, with
\begin{equation}
\begin{aligned}
\Delta\Omega_0&=\frac{\mu_\parallel+\mu_\perp}{2}
+\frac{k_gN_h(\eps_1-3\eps_2-4\eps_3)}{4g^2}
+\frac{3}{2g^2}\left[-\iu\gamma_{\rm ph}+\frac{|d|^2J^2}{D}\right],\\
\Delta\Omega_1&=\frac{\mu_\parallel+\mu_\perp}{2}
+\frac{k_gN_h(\eps_1+3\eps_2+8\eps_3)}{8g^2}
-\frac{3}{2g^2}\left[-\iu\gamma_{\rm ph}+\frac{|d|^2J^2}{D}\right]
+\frac32\left(\Delta\gamma_{{\rm ph},{\rm s}}+\Lambda_{X,{\rm s}}+\Delta\gamma_{{\rm ph},{\rm p}}+\Lambda_{X,{\rm p}}\right),\\
\Delta\Omega_2&=\frac{\mu_\parallel+\mu_\perp}{2}
-\frac{k_gN_h(\eps_1-3\eps_2+8\eps_3)}{8g^2}
+\frac{3}{4g^2}\left[-\iu\gamma_{\rm ph}+\frac{|d|^2J^2}{D}\right],\\
\Delta\Omega_3&=\frac{\mu_\parallel+\mu_\perp}{2}
-\frac{k_gN_h(\eps_1+3\eps_2-4\eps_3)}{4g^2}.
\end{aligned}
\label{eq:DeltaOmega_q}
\end{equation}
The linear couplings between neighbouring angular harmonics, $q-q'\equiv\pm1\pmod 6$, are
\begin{equation}
\begin{aligned}
\nu_{01}&=\nu-\frac{3\iu k_gN_h}{2\sqrt2\,g}(\eps_1+\eps_2)
+\frac{3\iu}{\sqrt2\,g}\left[-\iu\gamma_{\rm ph}+\frac{|d|^2J^2}{D}\right],\\
\nu_{12}&=\nu+\frac{3\iu k_gN_h}{\sqrt2\,g}\eps_2
-\frac{3\iu}{\sqrt2\,g}\left[-\iu\gamma_{\rm ph}+\frac{|d|^2J^2}{D}\right],\\
\nu_{23}&=\nu+\frac{3\iu k_gN_h}{2\sqrt2\,g}(\eps_1-\eps_2),
\end{aligned}
\label{eq:nu_qq}
\end{equation}
where $\nu={\iu v_{\rm g}}/{\sqrt2}=\iu\sqrt2\,g\mu_\perp,$
and the quadratic couplings between harmonics with $q-q'\equiv\pm2\pmod 6$ are
\begin{equation}
\begin{aligned}
\mu_{02}&=\frac{\mu_\parallel-\mu_\perp}{2}
+\frac{k_gN_h(7\eps_1-3\eps_2+8\eps_3)}{8g^2}
-\frac{3}{2g^2}\left[-\iu\gamma_{\rm ph}+\frac{|d|^2J^2}{D}\right],\\
\mu_{13}&=\frac{\mu_\parallel-\mu_\perp}{2}
-\frac{k_gN_h(7\eps_1+3\eps_2+8\eps_3)}{8g^2}
+\frac{9}{4g^2}\left[-\iu\gamma_{\rm ph}+\frac{|d|^2J^2}{D}\right],\\
\mu_{11}&=\frac{\mu_\parallel-\mu_\perp}{2}
+\frac{k_gN_h(7\eps_1+3\eps_2-4\eps_3)}{4g^2}
-\frac{3}{2g^2}\left[-\iu\gamma_{\rm ph}+\frac{|d|^2J^2}{D}\right]
+3\left(\Delta\gamma_{{\rm ph},{\rm s}}+\Lambda_{X,{\rm s}}-\Delta\gamma_{{\rm ph},{\rm p}}-\Lambda_{X,{\rm p}}\right),\\
\mu_{22}&=\frac{\mu_\parallel-\mu_\perp}{2}
-\frac{k_gN_h(7\eps_1-3\eps_2-4\eps_3)}{4g^2}.
\end{aligned}
\label{eq:mu_qq}
\end{equation}
 The guided-mode kinematics enters through $v_{\rm g}$, $\mu_\parallel$, and $\mu_\perp$, Eq.~\eqref{eq:mu_parallel_perp}. The terms proportional to $k_gN_h\eps_m$ and $-\iu\gamma_{\rm ph}+|d|^2J^2/D$ result from the rotation of the guided-mode polarization in $C_{\bfg\bfg'}(\bfk)$, and the terms with $\Delta\gamma_{{\rm ph},\alpha}+\Lambda_{X,\alpha}$ are the second-order TE and TM corrections of the eliminated zero harmonic. In the limit of a vanishing modulation $\eps_m\to0$ all linear couplings reduce to $\nu$, all quadratic couplings to $(\mu_\parallel-\mu_\perp)/2$, and all diagonal coefficients to $(\mu_\parallel+\mu_\perp)/2$, since $J,\Delta J_\alpha\propto\eps_1$ and $\gamma_{\rm ph},\Delta\gamma_{{\rm ph},\alpha}\propto\eps_1^2$.

We next reduce the angular-momentum Hamiltonian along the high-symmetry directions. For a direction with polar angle $\phi$ we write $k_\pm=|\bfk|\,e^{\pm\iu\phi}/\sqrt2$. Away from the $\Gamma$ point, the relevant group along a high-symmetry line contains a vertical mirror plane. It is therefore convenient to replace the angular-momentum basis by a basis with definite parity with respect to this mirror plane.

We introduce the unitary matrix
\begin{equation}
\hat U(\phi)=
\left(
\begin{array}{cccccc}
 1 & 0 & 0 & 0 & 0 & 0 \\
 0 & \dfrac{e^{\iu \phi }}{\sqrt{2}} & 0 & 0 & \dfrac{e^{\iu \phi }}{\sqrt{2}} & 0 \\
 0 & 0 & \dfrac{e^{2\iu \phi }}{\sqrt{2}} & 0 & 0 & \dfrac{e^{2\iu \phi }}{\sqrt{2}} \\
 0 & 0 & 0 & e^{3\iu \phi } & 0 & 0 \\
 0 & 0 & \dfrac{e^{-2\iu \phi }}{\sqrt{2}} & 0 & 0 & -\dfrac{e^{-2\iu \phi }}{\sqrt{2}} \\
 0 & -\dfrac{e^{-\iu \phi }}{\sqrt{2}} & 0 & 0 & \dfrac{e^{-\iu \phi }}{\sqrt{2}} & 0
\end{array}
\right).
\label{eq:U_phi}
\end{equation}
The columns of $\hat U(\phi)$ define even and odd combinations of the $|q\rangle$ and $|-q\rangle$ angular harmonics. The phase $\phi$ fixes the orientation of the mirror plane. For the high-symmetry directions considered below, $\phi=\pi n/6$ with $n=0,1$, this basis block-diagonalizes the Hamiltonian according to mirror parity. For the polaritonic Hamiltonian the corresponding transformation is
\begin{equation*}
\hat H_\phi=\hat{\mathcal U}^{\dagger}(\phi)\hat H\hat{\mathcal U}(\phi),
\qquad
\hat{\mathcal U}(\phi)=\operatorname{diag}[\hat U(\phi),\hat U(\phi)].
\end{equation*}
For the $\Gamma$-$K$ direction we use $\phi_K=0$. The Hamiltonian decomposes as
\begin{equation*}
\hat H_K=\hat H_K^{({\rm TE})}\oplus\hat H_K^{({\rm TM})},
\end{equation*}
with two six-dimensional blocks
\begin{equation}
\hat H_K^{({\rm TE})}=
\left(
\begin{array}{cccccc}
\tilde\Omega_0(\bfk) & \nu_{01}|\bfk| & -\dfrac{\mu_{02}}{\sqrt2}\bfk^2 & V^* & 0 & 0\\
-\nu_{01}|\bfk| & \tilde\Omega_1(\bfk)+\dfrac{\mu_{11}}{2}\bfk^2 & \dfrac{\nu_{12}}{\sqrt2}|\bfk| & 0 & V^* & 0\\
-\dfrac{\mu_{02}}{\sqrt2}\bfk^2 & -\dfrac{\nu_{12}}{\sqrt2}|\bfk| & \tilde\Omega_2(\bfk)-\dfrac{\mu_{22}}{2}\bfk^2 & 0 & 0 & V^*\\
V & 0 & 0 & E_X & 0 & 0\\
0 & V & 0 & 0 & E_X & 0\\
0 & 0 & V & 0 & 0 & E_X
\end{array}
\right)
\label{H_K_TE}
\end{equation}
and
\begin{equation}
\hat H_K^{({\rm TM})}=
\left(
\begin{array}{cccccc}
\tilde\Omega_3(\bfk) & -\dfrac{\mu_{13}}{\sqrt2}\bfk^2 & -\nu_{23}|\bfk| & V^* & 0 & 0\\
-\dfrac{\mu_{13}}{\sqrt2}\bfk^2 & \tilde\Omega_1(\bfk)-\dfrac{\mu_{11}}{2}\bfk^2 & \dfrac{\nu_{12}}{\sqrt2}|\bfk| & 0 & V^* & 0\\
\nu_{23}|\bfk| & -\dfrac{\nu_{12}}{\sqrt2}|\bfk| & \tilde\Omega_2(\bfk)+\dfrac{\mu_{22}}{2}\bfk^2 & 0 & 0 & V^*\\
V & 0 & 0 & E_X & 0 & 0\\
0 & V & 0 & 0 & E_X & 0\\
0 & 0 & V & 0 & 0 & E_X
\end{array}
\right).
\label{H_K_TM}
\end{equation}
According to Eqs.~\eqref{eq:DeltaOmega_q} and \eqref{eq:mu_qq}, the second-order zero-harmonic contribution to the $\bfk^2$ coefficient of the dipolar state is $3(\Delta\gamma_{{\rm ph},{\rm s}}+\Lambda_{X,{\rm s}})$ in the TE block and $3(\Delta\gamma_{{\rm ph},{\rm p}}+\Lambda_{X,{\rm p}})$ in the TM block: the dipolar state of each block couples only to the radiation channel of the same polarization.

For the $\Gamma$-$M$ direction we use $\phi_M=\pi/6$. The transformed Hamiltonian takes the form
\begin{equation*}
\hat H_M=\hat H_M^{({\rm TE})}\oplus\hat H_M^{({\rm TM})},
\end{equation*}
where the TE sector is eight-dimensional and the TM sector is four-dimensional
\begin{equation}
\hat H_M^{({\rm TE})}=
\left(
\begin{array}{cccccccc}
\tilde\Omega_0(\bfk) & \nu_{01}|\bfk| & -\dfrac{\mu_{02}}{\sqrt2}\bfk^2 & 0 & V^* & 0 & 0 & 0\\
-\nu_{01}|\bfk| & \tilde\Omega_1(\bfk)+\dfrac{\mu_{11}}{2}\bfk^2 & \dfrac{\nu_{12}}{\sqrt2}|\bfk| & -\dfrac{\mu_{13}}{\sqrt2}\bfk^2 & 0 & V^* & 0 & 0\\
-\dfrac{\mu_{02}}{\sqrt2}\bfk^2 & -\dfrac{\nu_{12}}{\sqrt2}|\bfk| & \tilde\Omega_2(\bfk)+\dfrac{\mu_{22}}{2}\bfk^2 & \nu_{23}|\bfk| & 0 & 0 & V^* & 0\\
0 & -\dfrac{\mu_{13}}{\sqrt2}\bfk^2 & -\nu_{23}|\bfk| & \tilde\Omega_3(\bfk) & 0 & 0 & 0 & V^*\\
V & 0 & 0 & 0 & E_X & 0 & 0 & 0\\
0 & V & 0 & 0 & 0 & E_X & 0 & 0\\
0 & 0 & V & 0 & 0 & 0 & E_X & 0\\
0 & 0 & 0 & V & 0 & 0 & 0 & E_X
\end{array}
\right)
\label{H_M_TE}
\end{equation}
and
\begin{equation}
\hat H_M^{({\rm TM})}=
\left(
\begin{array}{cccc}
\tilde\Omega_1(\bfk)-\dfrac{\mu_{11}}{2}\bfk^2 & \dfrac{\nu_{12}}{\sqrt2}|\bfk| & V^* & 0\\
-\dfrac{\nu_{12}}{\sqrt2}|\bfk| & \tilde\Omega_2(\bfk)-\dfrac{\mu_{22}}{2}\bfk^2 & 0 & V^*\\
V & 0 & E_X & 0\\
0 & V & 0 & E_X
\end{array}
\right).
\label{H_M_TM}
\end{equation}

\subsection{Full Hamiltonian with confined excitons}
We now restore the coefficients $\xi_m$ ($m=1,2,3$) neglected in Eq.~\eqref{eq:SM_exc_fourier} and give the corresponding $\Gamma$ point Hamiltonian. Since the envelope $\Phi_\alpha$ vanishes in the holes, its Fourier amplitudes $\Phi_{\alpha,\bfg}$ are the projections of the confined envelope on the plane waves restricted to the material. These waves are not orthogonal: their overlap matrix is $\xi_{\bfg-\bfg'}(\hat{\mathbf e}_{\alpha,\bfg}\cdot\hat{\mathbf e}_{\alpha',\bfg'})$. In the reduced model this matrix is replaced by its diagonal $\xi_{\mathbf 0}$, which is compensated by the factor $1/\sqrt{\xi_{\mathbf 0}}$ in $b_{\alpha,\bfg}=\Phi_{\alpha,\bfg}/\sqrt{\xi_{\mathbf 0}}$. For finite $\xi_m$ we define the excitonic amplitudes as the amplitudes of orthonormal excitonic states. The zero-harmonic excitons keep their amplitudes $\mathbf b_0$ of Sec.~S1B, with the components $b_{x,0}=\Phi_{x,0}/\sqrt{\xi_{\mathbf 0}}$ and $b_{y,0}=\Phi_{y,0}/\sqrt{\xi_{\mathbf 0}}$. The ${\rm s}$-polarized amplitudes at $|\bfg|=g$ are
\begin{equation}
b_{{\rm s},\bfg}=\frac16\sum_{|\bfg'|=g}\sum_{q}\frac{\cos[q(\varphi_{\bfg}-\varphi_{\bfg'})]}{\sqrt{\Xi_q}}\left[\Phi_{{\rm s},\bfg'}-\frac{\xi_1}{\sqrt{\xi_{\mathbf 0}}}\,\hat{\mathbf e}_{{\rm s},\bfg'}\cdot\mathbf b_0\right],
\label{eq:b_new}
\end{equation}
where $q=0,\pm1,\pm2,3$ and
\begin{equation}
\Xi_q=\xi_{\mathbf 0}+\xi_1\cos\left(\frac{\pi q}{3}\right)-\xi_2\cos\left(\frac{2\pi q}{3}\right)-\xi_3(-1)^q-3\delta_{q,\pm1}\frac{\xi_1^2}{\xi_{\mathbf 0}}.
\label{eq:Xi_q}
\end{equation}
These are the amplitudes of the confined ${\rm s}$-polarized waves $\xi(x,y)\,\hat{\mathbf e}_{{\rm s},\bfg}\,e^{\iu\bfg\cdot\mathbf r}$, from which the uniform component, described by $\mathbf b_0$, is removed and which are then orthonormalized symmetrically. $\Xi_q$ are the eigenvalues of their overlap matrix.  The ${\rm p}$-polarized amplitudes are obtained in the same way from the confined ${\rm p}$-polarized waves, after their uniform component and their projection on the ${\rm s}$-polarized excitons are removed.
 For $\xi_m\to0$, $\Xi_q\to\xi_{\mathbf 0}$ and $b_{\alpha,\bfg}$ reduce to $\Phi_{\alpha,\bfg}/\sqrt{\xi_{\mathbf 0}}$ of Sec.~S1B.

We obtain at the $\Gamma$ point
\begin{equation}
\begin{aligned}
(k-k_g)\,a_{{\rm s},\bfg}={}\sum_{|\bfg'|=g}\left[-\iu\gamma_{\rm ph}-k_gN_h\Delta\eps_{\bfg-\bfg'}\right]
&(\hat{\mathbf e}_{{\rm s},\bfg}\cdot\hat{\mathbf e}_{{\rm s},\bfg'})\,a_{{\rm s},\bfg'}\\
+\frac{V^*}{6}\sum_{|\bfg'|=g}&\sum_q\sqrt{\frac{\Xi_q}{\xi_{\mathbf 0}}}\cos[q(\varphi_{\bfg}-\varphi_{\bfg'})]\,b_{{\rm s},\bfg'}
+\Big(d^*J+\frac{\xi_1}{\xi_{\mathbf 0}}V^*\Big)\hat{\mathbf e}_{{\rm s},\bfg}\cdot\mathbf b_0,
\end{aligned}
\label{eq:xi_gamma1}
\end{equation}
\begin{equation}
\begin{aligned}
(k-E_X)\,b_{{\rm s},\bfg}={}\frac{V}{6}&\sum_{|\bfg'|=g}\sum_q\sqrt{\frac{\Xi_q}{\xi_{\mathbf 0}}}\cos[q(\varphi_{\bfg}-\varphi_{\bfg'})]\,a_{{\rm s},\bfg'},\quad
(k-E_X)\,b_{{\rm p},\bfg}={}0,\\
(k&-E_X+\iu\gamma_X)\,\mathbf b_0={}\Big(dJ+\frac{\xi_1}{\xi_{\mathbf 0}}V\Big)\sum_{|\bfg'|=g}\hat{\mathbf e}_{{\rm s},\bfg'}a_{{\rm s},\bfg'}.
\end{aligned}
\label{eq:xi_gamma}
\end{equation}
 With finite $\xi_m$ the coupling matrix is no longer diagonal in $\bfg$. The guided photons excite the zero-harmonic excitons not only radiatively, via $J$, but also directly, with the weight $\xi_1$, because the guided-mode field restricted to the material has a uniform component. The excitons are not coupled to each other, the ${\rm p}$-polarized excitons decouple and remain dark at $E_X$.
Eliminating $\mathbf b_0$ gives the photonic self-energy$|d|^2\left[J+\xi_1V/(\xi_{\mathbf 0}d)\right]^2$. 

The DFT~\eqref{eq:DFT}, applied to $a_{{\rm s},\bfg}$, $b_{{\rm s},\bfg}$, and $b_{{\rm p},\bfg}$, diagonalizes the coupling matrix in Eqs.~\eqref{eq:xi_gamma1},\eqref{eq:xi_gamma}, whose eigenvalues are $V\sqrt{\Xi_q/\xi_{\mathbf 0}}$. For each $q$, in the basis $(a_{{\rm s},q},b_{{\rm s},q},b_{{\rm p},q})$, we obtain
\begin{equation}
\hat H^{(q)}=
\left(
\begin{array}{ccc}
\Omega_q & V^*\sqrt{\Xi_q/\xi_{\mathbf 0}} & 0\\
V\sqrt{\Xi_q/\xi_{\mathbf 0}} & E_X & 0\\
0 & 0 & E_X
\end{array}
\right),
\label{eq:Hq_formB}
\end{equation}
where $\Omega_q$ is given by Eq.~\eqref{eq:11}, with $|d|^2J^2$ replaced by $|d|^2\left[J+\xi_1V/(\xi_{\mathbf 0}d)\right]^2$ for $q=\pm1$.

\section{Geometrical origin of the photonic transition points}
In this section, we provide the geometrical interpretation of the photonic transition condition used in the main text and illustrate how the two solutions of this condition appear in the calculated photonic spectra. 

\subsection{Fourier coefficients of the six-hole unit cell}
We first express the Fourier coefficients of the dielectric modulation using the same form-factor/structure-factor language that is commonly used in the Fourier analysis of a crystal basis~\cite{kittel2018introduction}. The same decomposition can be applied to the present photonic-crystal unit cell. The role of the atom is played by an individual triangular hole, while the role of the basis is played by the six-hole cluster shown in Fig.~2(a) (main text). Therefore, the Fourier coefficient of the dielectric modulation can be written as a product of a single-hole form factor and a structure factor determined by the positions of the holes in the unit cell.

For a nonzero reciprocal-lattice vector $\bfg$, we write
\begin{equation}
\Delta\varepsilon_{\bfg}=A_{\bfg}(r)F_{\bfg}(R).
\label{eq:eps}
\end{equation}
Here $A_{\bfg}(r)$ is the form factor of one triangular hole. It depends on the hole size and shape. The second factor, $F_{\bfg}(R)$, is the normalized structure factor of the six-hole basis. It depends only on the hole-center positions and, therefore, only on the radial displacement $R$. For the geometry considered here, the six hole centers are located at
$\mathbf R_j =
\pm R\hat{\mathbf a}_1,\,
\pm R\hat{\mathbf a}_2,\,
\pm R(\hat{\mathbf a}_2-\hat{\mathbf a}_1),$
where $\hat{\mathbf a}_{1,2}$ are unit vectors along the lattice vectors. The corresponding normalized structure factor is
\begin{equation}
F_{\bfg}(R)=\frac{1}{6}
\sum_{j=1}^{6}
e^{-\iu \bfg \cdot \mathbf R_j}.
\label{eq:Fg}
\end{equation}
This expression is the direct analogue of the usual structure factor of a crystal basis, with atomic positions replaced by the positions of triangular holes.
We now evaluate Eq.~\eqref{eq:Fg} for the three Fourier harmonics entering the effective Hamiltonian, namely for reciprocal vectors with absolute values $g$, $\sqrt{3}g$, and $2g$. We denote the corresponding Fourier coefficients by
$\varepsilon_1=\Delta\varepsilon_{g},\; \varepsilon_2=\Delta\varepsilon_{\sqrt{3}g},\; \varepsilon_3=\Delta\varepsilon_{2g}.$
The structure factors are
\begin{equation*}
F_1(R)=\frac{1+2\cos x}{3},
\qquad
F_2(R)=\frac{\cos 2x+2\cos x}{3},
\qquad
F_3(R)=\frac{1+2\cos 2x}{3},
\end{equation*}
where $x={2\pi R}/{a}$.
The coefficients $A_1(r)$, $A_2(r)$, and $A_3(r)$ are the triangular-hole form factors evaluated at reciprocal vectors with absolute values $g$, $\sqrt{3}g$, and $2g$, respectively. Their explicit analytical form is not required below, it is sufficient that they depend only on the individual hole shape and size.

The band-inversion condition derived in the main text is
\begin{equation}
\varepsilon_1+2\varepsilon_3\simeq0.
\label{eq:SM_transition_condition}
\end{equation}
Using Eq.~\eqref{eq:eps}, this condition becomes
\begin{equation*}
A_1(r)F_1(R)+2A_3(r)F_3(R)=0.
\end{equation*}
Substituting the explicit structure factors, we obtain
\begin{equation*}
1+2\cos x
+
2\frac{A_3(r)}{A_1(r)}
\left[
1+2\cos 2x
\right]
=0.
\end{equation*}
This equation has two solutions,
\begin{equation*}
\cos x=-\frac{1}{2},
\qquad
\cos x=\frac{1}{2}-\frac{A_1(r)}{4A_3(r)}.
\end{equation*}
The first solution gives $R=a/3.$
At this point, $F_1(a/3)=F_3(a/3)=0,$
and therefore both $\varepsilon_1$ and $\varepsilon_3$ vanish independently of the triangular-hole form factor. This transition is fixed entirely by the structure factor of the six-hole basis and corresponds to the conventional breathing-honeycomb transition.
The second solution is
\begin{equation*}
R(r)=\frac{a}{2\pi}\arccos\left[\frac{1}{2}-\frac{A_1(r)}{4A_3(r)}\right].
\end{equation*}
This solution depends on the ratio of the form factors and is therefore sensitive to the size and shape of the triangular holes. In the small-hole limit, $r\ll a$, the form factors at $g$ and $2g$ become close to each other, $A_3(r)/A_1(r)\simeq1$, and the second solution approaches
\begin{equation*}
R\simeq\frac{a}{2\pi}\arccos\left(\frac{1}{4}\right)\simeq 0.21a.
\end{equation*}

Equation~\eqref{eq:SM_transition_condition} keeps only the dielectric couplings in the splitting between the dipolar and quadrupolar states at the $\Gamma$ point. For the photonic problem, Eq.~\eqref{eq:11} gives the full splitting
\begin{equation*}
\Omega_1-\Omega_2=-3\iu\gamma_{\rm ph}-k_gN_h(\varepsilon_1+2\varepsilon_3).
\end{equation*}
The dipolar and quadrupolar states are therefore inverted at
\begin{equation}
\Omega_{12}=\frac{\Omega_1-\Omega_2}{2},\quad
{\rm Re}\,\Omega_{12}=\frac{3}{2}\,{\rm Im}\,\gamma_{\rm ph}-\frac{k_gN_h}{2}(\varepsilon_1+2\varepsilon_3)=0.
\label{eq:SM_transition_full}
\end{equation}
The additional term is the frequency shift of the dipolar states caused by their coupling to the radiative zero harmonic. With $\gamma_{\rm ph}$ from Eq.~\eqref{eq:zero_overlap_Gamma}, Eq.~\eqref{eq:SM_transition_full} takes the form
\begin{equation*}
\varepsilon_1+2\varepsilon_3=\frac{3k_g^2\varepsilon_1^2}{N_h}\,{\rm Re}\iint_{-h}^{h}\mathrm dz\,\mathrm dz'\,\theta(z)G_{\mathbf 0}(z,z';k_g)\theta(z').
\end{equation*}
The left-hand side is of first order in the dielectric modulation, whereas the right-hand side is of second order. For small holes, both $\varepsilon_1$ and $\varepsilon_3$ are proportional to the hole area, the right-hand side is smaller than each term on the left-hand side by a factor of order $|\varepsilon_1|$, and Eqs.~\eqref{eq:SM_transition_condition} and~\eqref{eq:SM_transition_full} give the same boundary in the limit $r\to0$. For larger holes, the radiative shift is no longer negligible and moves the second transition point. Equation~\eqref{eq:SM_transition_condition} is therefore the small-hole limit of the full condition~\eqref{eq:SM_transition_full}. At the first solution, $R=a/3$, both $\varepsilon_1$ and $\varepsilon_3$ vanish, $\gamma_{\rm ph}=0$, and the two conditions give the same boundary for any hole size.

\subsection{Photonic spectra at the transition geometries}
\begin{figure}
\centering
\includegraphics[width=0.4\linewidth]{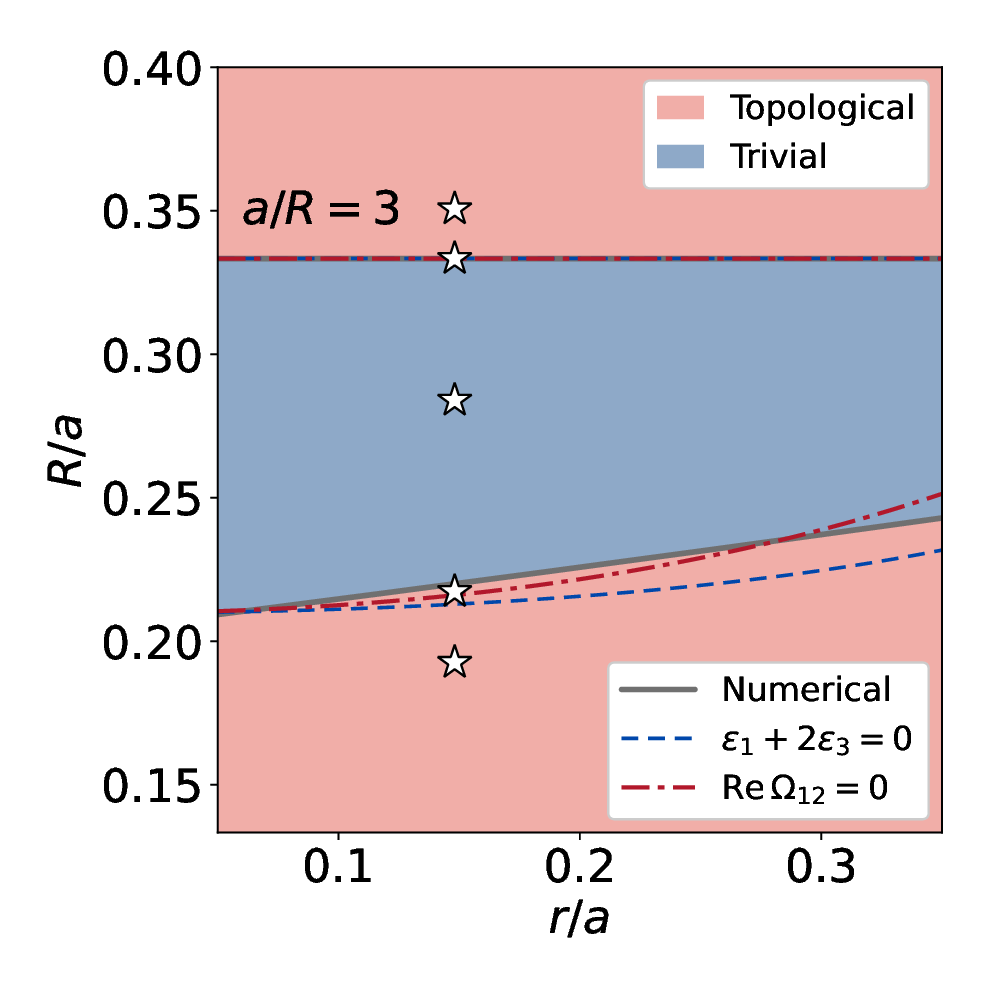}
\caption{The phase map in the $(r/a,R/a)$ parameter plane shows the topological and trivial regions determined by the numerical transition boundaries. The solid gray curves show the numerical boundaries. The blue dashed curves show the simplified analytical condition $\varepsilon_1+2\varepsilon_3=0$, Eq.~\eqref{eq:SM_transition_condition}, and the red dash-dotted curves shows the full condition ${\rm Re}\,\Omega_{12}=0$, Eq.~\eqref{eq:SM_transition_full}. The upper boundaries coincide at the conventional breathing-honeycomb transition $R=a/3$. The stars indicate five geometries at fixed
$r/a=60\; \text{nm}/405\; \text{nm}\simeq0.15$ whose full-wave band structures are shown in Fig.~\ref{fig:SM_trans}.}
\label{fig:SM_trans0}
\end{figure}

\begin{figure}
\centering
\includegraphics[width=1\linewidth]{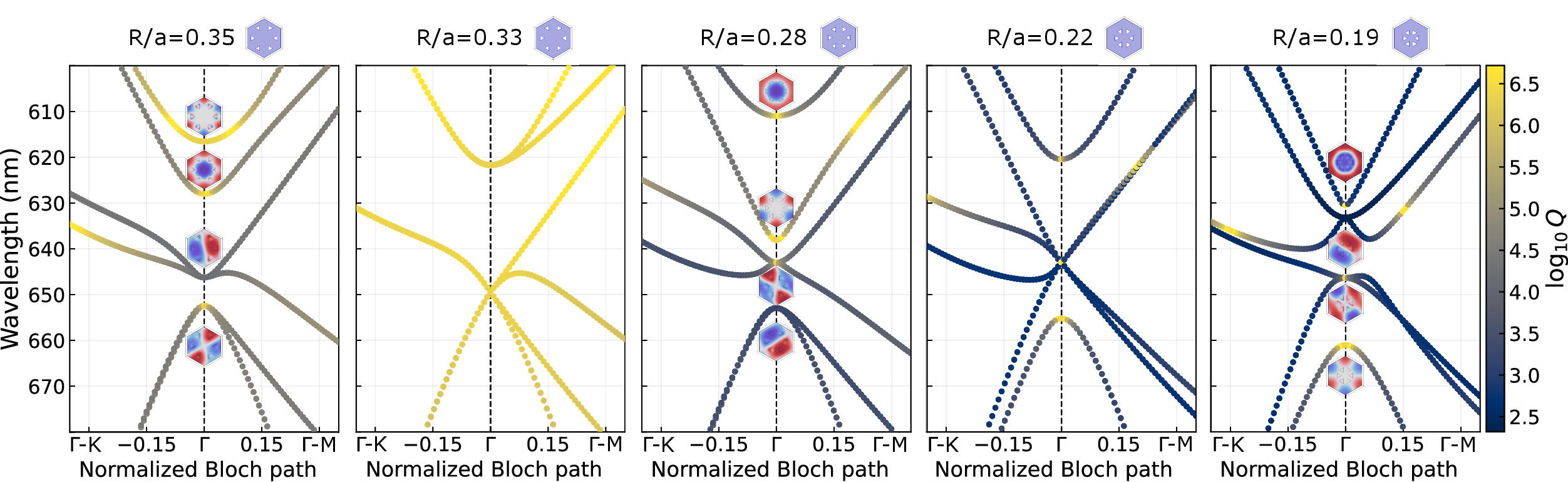}
\caption{Calculated band structures for five representative geometries describing topological, trivial, and transition regimes. The line color shows $ \log_{10}Q$, where $Q$ is the mode quality factor. The insets show the eigenmode $H_z$ field profiles at the $\Gamma$ point.
}
\label{fig:SM_trans}
\end{figure}

To verify the phase diagram in the small-hole regime, where the small-modulation approximation underlying the effective Hamiltonian is most accurate, we consider a fixed geometry with $r/a=60\; \text{nm}/405\; \text{nm}\simeq0.15$. Figure~\ref{fig:SM_trans0} compares the analytical transition boundaries with the numerical boundaries obtained from full-wave eigenfrequency calculations. The analytical boundaries contain no fitted parameters. The coefficients $\varepsilon_m$ are obtained from the Fourier transform of the hole pattern, while $k_g$, $N_h$, and $\gamma_{\rm ph}$ are evaluated in the effective slab limit of Sec.~S1A, $k_g=k_{\rm TE_0}(g)$, for the area-averaged slab of the same geometry. For small holes, the two analytical boundaries coincide with each other and with the numerical one. As the holes grow, the simplified condition~\eqref{eq:SM_transition_condition} increasingly underestimates $R$ at the second transition, whereas the full condition~\eqref{eq:SM_transition_full} follows the numerical boundary up to $r/a\simeq0.3$. For the largest holes both analytical boundaries deviate from the numerical one, consistent with the limited accuracy of the small-modulation approximation for large holes. The starred points correspond directly to the five band structures shown in Fig.~\ref{fig:SM_trans} and sample the two topological phases, the intermediate trivial phase, and the two transition points.


%Figure~\ref{fig:SM_trans} shows the photonic spectra for five representative values of $R/a$. The second panel corresponds to the conventional breathing-honeycomb point, $a/R=3$. The fourth panel, $R/a=0.22$, corresponds to the second solution of the transition condition controlled by the ratio of the triangular-hole form factors.

At $R=a/3$, the first and third Fourier coefficients vanish, $\varepsilon_1=\varepsilon_3=0$. As a result, the coupling of the guided harmonics to the zeroth radiative harmonic is absent. In the effective model this gives $\gamma_{\rm ph}=0$ and $\Delta\gamma_{{\rm ph},{\rm s}}=\Delta\gamma_{{\rm ph},{\rm p}}=0$. Within the model, the guided harmonics are therefore decoupled from the radiation continuum at all $\bfk$, which explains the large values of $Q$ over the whole dispersion in the second panel of Fig.~\ref{fig:SM_trans}. The same condition leaves only the Fourier harmonic $\varepsilon_2$ in the coupling matrix. The eigenfrequencies at the $\Gamma$ point reduce to
\begin{equation*}
\Omega_q=k_g+k_gN_h\varepsilon_2
\cos\left(\frac{2\pi q}{3}\right),
\end{equation*}
and the coefficients of Eqs.~\eqref{eq:DeltaOmega_q}--\eqref{eq:mu_qq} become
\begin{equation*}
\begin{aligned}
&\Delta\Omega_0=\Delta\Omega_3=\frac{\mu_\parallel+\mu_\perp}{2}-\frac{3k_gN_h\varepsilon_2}{4g^2},
&&\Delta\Omega_1=\Delta\Omega_2=\frac{\mu_\parallel+\mu_\perp}{2}+\frac{3k_gN_h\varepsilon_2}{8g^2},\\
&\nu_{01}=\nu_{23}=\nu-\frac{3\iu k_gN_h\varepsilon_2}{2\sqrt2\,g},
&&\nu_{12}=\nu+\frac{3\iu k_gN_h\varepsilon_2}{\sqrt2\,g},\\
&\mu_{02}=\mu_{13}=\frac{\mu_\parallel-\mu_\perp}{2}-\frac{3k_gN_h\varepsilon_2}{8g^2},
&&\mu_{11}=\mu_{22}=\frac{\mu_\parallel-\mu_\perp}{2}+\frac{3k_gN_h\varepsilon_2}{4g^2}.
\end{aligned}
\end{equation*}
Along the $\Gamma$-$K$ direction the TE and TM photonic blocks~\eqref{H_K_TE} and~\eqref{H_K_TM} then become equivalent. As a result, the six photonic branches reduce to three doubly degenerate dispersions, as seen in the second panel of Fig.~\ref{fig:SM_trans}. One of the two degenerate blocks can be written as
\begin{equation*}
\hat H_{\text{ph},K}
=
\left(
\begin{array}{ccc}
k_g(1+N_h\varepsilon_2)+\Delta\Omega_0\bfk^2
&
\nu_{01}|\bfk|
&
-\dfrac{\mu_{02}}{\sqrt{2}}\bfk^2
\\
-\nu_{01}|\bfk|
&
k_g\left(1-\dfrac{N_h\varepsilon_2}{2}\right)
+\left(\Delta\Omega_1+\dfrac{\mu_{11}}{2}\right)\bfk^2
&
\dfrac{\nu_{12}}{\sqrt{2}}|\bfk|
\\
-\dfrac{\mu_{02}}{\sqrt{2}}\bfk^2
&
-\dfrac{\nu_{12}}{\sqrt{2}}|\bfk|
&
k_g\left(1-\dfrac{N_h\varepsilon_2}{2}\right)
+\left(\Delta\Omega_1-\dfrac{\mu_{11}}{2}\right)\bfk^2
\end{array}
\right).
\end{equation*}

The second transition point has a different origin. Here $\varepsilon_1$ and $\varepsilon_3$ remain finite, and the radiative coupling is not suppressed, which is consistent with the smaller values of $Q$ in the fourth panel of Fig.~\ref{fig:SM_trans}.

\section{Numerical implementation}
The parameters of the effective Hamiltonian are obtained in the same way for all structures considered below. The Fourier coefficients $\varepsilon_1$, $\varepsilon_2$, and $\varepsilon_3$ are calculated from the in-plane permittivity profile, and the magnitude of the reciprocal vectors is $g=4\pi/(\sqrt{3}a)$. The effective slab of Sec.~S1A is a homogeneous layer of thickness $2h$ in air, with the permittivity $\veps$ equal to the area average of the permittivity of the patterned layer. Its TE$_0$ dispersion $k_{\rm TE_0}(\beta)$ is the lowest root of
\begin{equation}
\sqrt{k^2\veps-\beta^2}\,\tan\!\left(h\sqrt{k^2\veps-\beta^2}\right)=\sqrt{\beta^2-k^2},
\label{eq:SM_TE0_dispersion}
\end{equation}
which gives $v_{\rm g}$ and $\mu_\parallel$ in Eq.~\eqref{eq:mu_parallel_perp}. The common profile of Sec.~S1A is
\begin{equation*}
\theta(z)\propto
\begin{cases}
\cos(qz), & |z|\le h,\\
\cos(qh)\,\eu^{-p(|z|-h)}, & |z|>h,
\end{cases}
\qquad
q=\sqrt{k_{\rm TE_0}^2(g)\,\veps-g^2},\quad
p=\sqrt{g^2-k_{\rm TE_0}^2(g)},
\end{equation*}
normalized by $\int_{-\infty}^{\infty}\mathrm dz\,\veps(z)\theta^2(z)=1/2$. 

The scalar Green's function of the zero harmonic for a
homogeneous dielectric slab in air can be written in terms
of outgoing waves~\cite{muljarov2010brillouin} as
\begin{equation}
\begin{aligned}
G_0(z,z';k)
={}&
\frac{
\left[
e^{i\sqrt{\varepsilon_0}k(h+z_<)}
+\rho_\varepsilon e^{-i\sqrt{\varepsilon_0}k(h+z_<)}
\right]
\left[
e^{i\sqrt{\varepsilon_0}k(h-z_>)}
+\rho_\varepsilon e^{-i\sqrt{\varepsilon_0}k(h-z_>)}
\right]}{
2ik\sqrt{\varepsilon_0}
\left[
\rho_\varepsilon^2 e^{-2i\sqrt{\varepsilon_0}kh}
-e^{2i\sqrt{\varepsilon_0}kh}
\right]
},
\end{aligned}
\tag{S48}
\end{equation}
where $z_< = \min(z,z')$, $z_> = \max(z,z')$, and
$\rho_\varepsilon =
({\sqrt{\varepsilon_0}+1})/({\sqrt{\varepsilon_0}-1}).$ 
\\

With these definitions, $\varepsilon_m$, $\veps$, $v_{\rm g}$, $\mu_\parallel$, $\mu_\perp$, and $\Delta\gamma_{{\rm ph},\alpha}$ are calculated from the geometry and the effective slab, with the Green's function quantities evaluated at the fitted $k_g$. The radiative loss ${\rm Re}\,\gamma_{\rm ph}$ is fixed by the radiative linewidth of the dipolar mode at the $\Gamma$ point obtained from the full-wave calculation. The remaining $\Gamma$ point parameters $k_g$, $N_h$, and ${\rm Im}\,\gamma_{\rm ph}$ are fitted to the photonic resonance positions. For the polaritonic spectra, the photonic parameters are kept fixed, the complex exciton frequency $E_{\rm X}$ is fixed by $\lambda_{\rm X}$ and $\Gamma$ of the Lorentz permittivity, and only the light-matter coupling $V$ is fitted. The exciton-mediated zero-harmonic terms $|d|^2J^2/D$ and $\Lambda_{{\rm X},\alpha}$ are evaluated from Eqs.~\eqref{eq:zero_overlap_Gamma}--\eqref{eq:D2}.
\\

The analytical branches are the eigenvalues of the TE and TM blocks~\eqref{H_K_TE}--\eqref{H_M_TM}, calculated with a single parameter set for each structure. In the fits, the real parts of the eigenvalues are compared with the resonance positions extracted from the numerical spectra, and each extracted resonance is assigned to the closest analytical branch of the same symmetry block. Apart from the dipolar linewidth at the $\Gamma$ point, which fixes ${\rm Re}\,\gamma_{\rm ph}$, the imaginary parts of the eigenvalues are not fitted.

\subsection{Bulk WS$_2$ metasurface}

Using the Fourier expansion, we calculate the coefficients of $\varepsilon(x,y)$ and average them over the corresponding reciprocal-lattice vectors. Thus, $\varepsilon_1$, $\varepsilon_2$, and $\varepsilon_3$ are determined only by the in-plane hole geometry and the dielectric contrast between the slab and air.

The reflectance maps in Figs.~2 and 3 were calculated using RCWA implemented in FMMAX. For each wavelength and incidence angle, we computed the scattering matrix of the air/metasurface/air stack, and the $\Gamma$-$K$ and $\Gamma$-$M$ spectra were obtained by choosing the corresponding orientation of the unit cell. The analytical branches in the purely photonic case were obtained by diagonalizing the photonic parts of the TE and TM blocks~\eqref{H_K_TE}--\eqref{H_M_TM}. For the polaritonic spectra, we used the full light-matter blocks. The parameters $k_g$, $N_h$, and $\gamma_{\rm ph}$ were fitted to the photonic resonance positions and dipole linewidth extracted from the RCWA maps, and $V$ was then fitted to the polaritonic maps. The same parameters were used for the TE and TM validation spectra in Figs.~2 and 3. The root mean square (RMS) deviations of the analytical branches from the reflectance minima are $0.59$~nm for the photonic maps and $0.80$~nm for the polaritonic maps. The resulting parameter set is listed in Table~\ref{tab:fit_fig34}.

\begin{table}[h]
\centering
\caption{Effective-Hamiltonian parameters used for the analytical branches in Figs.~2 and 3 of the main text ($a=405$~nm, $2h=20$~nm, $r=60$~nm, $R=77$~nm). The upper part lists the quantities fixed by the geometry and the effective slab, and the lower part lists the fitted parameters.}
\label{tab:fit_fig34}
\begin{tabular}{lc}
\hline
Parameter & Value \\
\hline
$\varepsilon_1$ & $-0.7155$ \\
$\varepsilon_2$ & $-4.345\times10^{-3}$ \\
$\varepsilon_3$ & $0.1764$ \\
$\veps$ & $18.98$ \\
$v_{\rm g}$ & $0.3383$ \\
$\mu_\parallel$ & $-3.824~\mathrm{nm}$ \\
$\mu_\perp$ & $9.444~\mathrm{nm}$ \\
$\Delta\gamma_{{\rm ph},{\rm s}}$ & $(1.266+0.576\iu)\times10^{-2}~\mathrm{nm}$ \\
$\Delta\gamma_{{\rm ph},{\rm p}}$ & $(-1.129-0.798\iu)\times10^{-2}~\mathrm{nm}$ \\
\hline
$k_g$ & $9.6130\times10^{-3}~\mathrm{nm}^{-1}$ \\
$N_h$ & $0.01937$ \\
${\rm Re}\,\gamma_{\rm ph}$ & $3.084\times10^{-6}~\mathrm{nm}^{-1}$ \\
${\rm Im}\,\gamma_{\rm ph}$ & $2.277\times10^{-5}~\mathrm{nm}^{-1}$ \\
$V$ & $4.627\times10^{-4}~\mathrm{nm}^{-1}$ ($91.31$~meV) \\
\hline
\end{tabular}
\end{table}

The approximation $D(k)\to D(k_g)$, where $D(k)=k-E_{\rm X}+i\gamma_X$, was tested using the parameter set of Table~\ref{tab:fit_fig34}. We compared the reduced $12\times12$ Hamiltonian with the $14\times14$ system explicitly retaining both zeroth-harmonic exciton amplitudes. The overlap integrals and reference coefficients remained evaluated at $k_g$, and no parameters were refitted. When restoring the zeroth harmonic, its frozen self-energy was removed from the photonic block to avoid double counting. The wave vector dependent matrix elements were retained through second order, as in Sec. S1C. Corresponding polariton branches were matched within each symmetry sector using their complex eigenvalues and eigenvector overlaps. Over 590–690 nm and $|\mathbf{k}|\leq0.0042\, \mathrm{nm}^{-1}$ along both symmetry directions and for both polarizations, the maximum changes in resonance wavelength and energy HWHM are below 0.041 nm and 0.23 meV, respectively.\\

Bulk WS$_2$ is a uniaxial crystal. The A exciton has an in-plane transition dipole and appears only in the in-plane permittivity, whereas the out-of-plane permittivity is nonresonant and much smaller, $\varepsilon_\perp=6.5$--$6.7$ for $\lambda=585$--$690$~nm~\cite{munkhbat2022optical}. As in Ref.~\cite{bouteyre2025simultaneous}, from which the material parameters are taken, the RCWA calculations treat WS$_2$ as isotropic with the permittivity of Eq.~(17). This approximation is accurate here because the quasiguided modes originate from the s-polarized guided mode, whose electric field lies in the slab plane, so $\varepsilon_\perp$ enters only through the weak out-of-plane field near the hole sidewalls. We verified this by repeating the calculations for Figs.~2 and 3 with the uniaxial tensor ${\rm diag}(\varepsilon,\varepsilon,\varepsilon_\perp)$ and the tabulated $\varepsilon_\perp(\lambda)$~\cite{munkhbat2022optical}. For all photonic and polaritonic resonances at $|\mathbf k|=0$, $0.0015$, and $0.003$ nm$^{-1}$ along $\Gamma$-$K$ and $\Gamma$-$M$, for TE and TM incidence, the complex resonance frequencies extracted from the reflection coefficient shift by less than $0.006$~nm, and the linewidths change by less than $0.2\%$. These shifts are much smaller than the reported RMS deviations of the spectral fits.

\subsection{Monolayer WS$_2$ on a TiO$_2$ metasurface}

\begin{figure}[h!]
\centering
\includegraphics[width=0.6\linewidth]{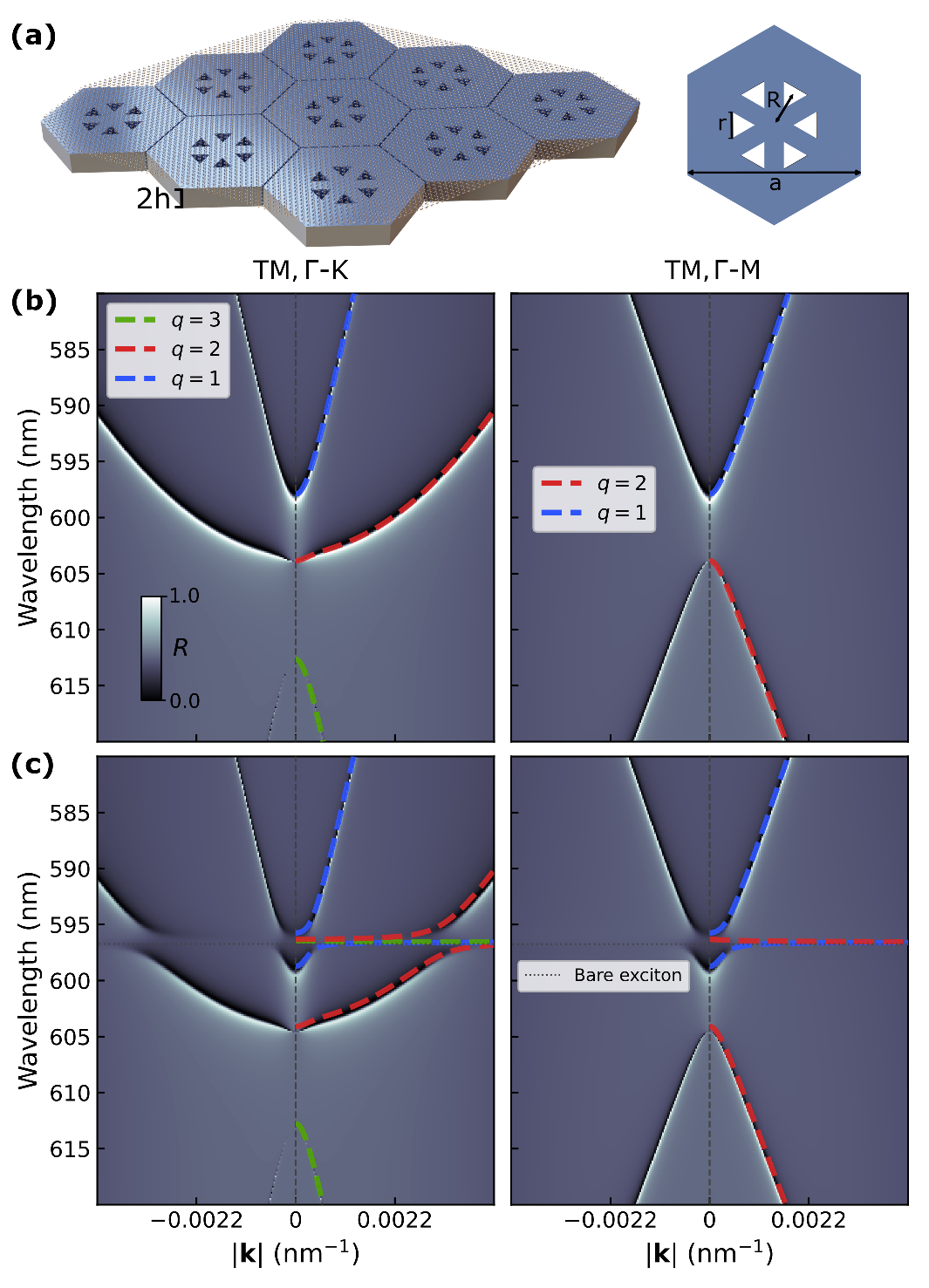}
\caption{Monolayer WS$_2$ calculations for the patterned TiO$_2$ membrane. (a) Schematic of a continuous WS$_2$ monolayer placed on top of the patterned TiO$_2$ membrane and of the corresponding hexagonal unit cell. The membrane thickness is $2h=85$~nm, the lattice period is $a=405$~nm, and holes with side length $r=60$~nm and radial displacement $R=77$~nm from the unit-cell center. The WS$_2$ monolayer is represented by a film of physical thickness $d_{\mathrm{ML}}=0.7$~nm.
(b) RCWA reflectance maps of the bare TiO$_2$ photonic structure for TM-polarized incidence along the $\Gamma$-$K$ and $\Gamma$-$M$ directions. Dashed lines show the analytical branches obtained by fitting the effective photonic Hamiltonian given in Sec.~S2A; the colors identify the corresponding angular-momentum indices $q$.
(c) Reflectance maps after adding the monolayer WS$_2$, whose in-plane permittivity is described by the single-pole Lorentz model of Eq.~(17) with $\varepsilon_{\infty}=2.7$, $\lambda_{\mathrm{X}}=596.6$~nm, $\Gamma=2.81$~ps$^{-1}$, and $f=0.065~\mu\mathrm{m}^{-2}$. The bare exciton wavelength is indicated by the dotted horizontal line. Dashed lines show the corresponding exciton-polariton branches obtained from the reduced light-matter Hamiltonian of Sec.~S2A, with the photonic parameters fixed to the fit in panel (b) and the light-matter coupling $V$ fitted to the monolayer spectra.}
\label{fig:SM_ML}
\end{figure}

Following the two-dimensional limit introduced in Sec.~S1B, we additionally consider a continuous monolayer WS$_2$ placed on the upper surface of a patterned TiO$_2$ membrane shown in  Fig.~\ref{fig:SM_ML}(a). In the RCWA calculations, the delta-like response is represented by an WS$_2$ film of physical thickness $d_{\mathrm{ML}}=0.7$~nm. Its in-plane permittivity is described by the same single-pole Lorentz model as in Eq.~(17), with $\varepsilon_{\infty}=2.7$, $\lambda_{\mathrm{X}}=596.6$~nm, $\Gamma=2.81$~ps$^{-1}$, and $f=0.065~\mu\mathrm{m}^{-2}$, reported for a WS$_2$ monolayer at $6.5$~K~\cite{cao2025}.
The photonic parameters were obtained by fitting the analytical TM branches to the bare-structure reflectance maps in Fig.~\ref{fig:SM_ML}(b). The Fourier coefficients $\varepsilon_1$, $\varepsilon_2$, and $\varepsilon_3$ were calculated from the in-plane dielectric profile of the TiO$_2$ membrane with the permittivity fixed at $\lambda=600$~nm, which is also the wavelength at which $\veps$ is quoted in Table~\ref{tab:SM_ML_fit}. The material dispersion of TiO$_2$ was retained in the effective slab, $\veps(k)$, which determines $v_{\rm g}$ and $\mu_\parallel$. The parameters  $k_g$, $N_h$, and $\gamma_{\rm ph}$ were fitted simultaneously along the $\Gamma$-$K$ and $\Gamma$-$M$ directions. For the polaritonic spectra in Fig.~\ref{fig:SM_ML}(c), the photonic parameters were kept fixed and only the light-matter coupling $V$ was fitted. The RMS deviations of the analytical branches from the reflectance minima are $0.075$~nm for the bare structure and $0.32$~nm for the monolayer spectra. The resulting parameters are listed in Table~\ref{tab:SM_ML_fit}.
\begin{table}[h]
\centering
\caption{Effective-Hamiltonian parameters used for the analytical branches in Fig.~\ref{fig:SM_ML}. The upper part lists the quantities fixed by the geometry and the effective slab, and the lower part lists the fitted parameters.}
\label{tab:SM_ML_fit}
\begin{tabular}{lc}
\hline
Parameter & Value \\
\hline
$\varepsilon_1$ & $-0.1810$ \\
$\varepsilon_2$ & $-1.099\times10^{-3}$ \\
$\varepsilon_3$ & $0.04463$ \\
$\veps$ & $5.550$ \\
$v_{\rm g}$ & $0.3973$ \\
$\mu_\parallel$ & $-4.566~\mathrm{nm}$ \\
$\mu_\perp$ & $11.09~\mathrm{nm}$ \\
$\Delta\gamma_{{\rm ph},{\rm s}}$ & $(8.881+2.044\iu)\times10^{-3}~\mathrm{nm}$ \\
$\Delta\gamma_{{\rm ph},{\rm p}}$ & $(-4.282-7.013\iu)\times10^{-3}~\mathrm{nm}$ \\
\hline
$k_g$ & $1.0443\times10^{-2}~\mathrm{nm}^{-1}$ \\
$N_h$ & $0.07897$ \\
${\rm Re}\,\gamma_{\rm ph}$ & $2.135\times10^{-6}~\mathrm{nm}^{-1}$ \\
${\rm Im}\,\gamma_{\rm ph}$ & $9.310\times10^{-6}~\mathrm{nm}^{-1}$ \\
$V$ & $2.357\times10^{-5}~\mathrm{nm}^{-1}$ ($4.651$~meV) \\
\hline
\end{tabular}
\end{table}

\newpage
\subsection{Topological structures}

The coefficients~\eqref{eq:11} and~\eqref{eq:DeltaOmega_q}--\eqref{eq:mu_qq} are obtained within the retained subspace of the six harmonics with $|\bfg|=g$ and to leading order in the dielectric modulation $\eps_m$. In this nearly-free-photon picture the holes weakly couple the six guided waves of the homogeneous reference slab, and their radiative coupling through the zero harmonic is kept to order $\eps_1^2$. The picture is accurate for the small holes of Figs.~2 and~3 ($r/a\sim0.15$). The domains of Fig.~5 have large holes ($r/a\sim0.31-0.33$ ) that remove a large part of the high-index material. There, the virtual coupling through the discarded harmonics and the higher orders in $\eps_m$ become comparable to the retained terms, the field concentrates in the dielectric between the holes, and the radiative coupling is no longer perturbative. For these domains we keep the symmetry-allowed form of the Hamiltonian and extract its coefficients from full-wave calculations.

The model~\eqref{eq:H_kp} acts on the photons $a_q$ and excitons $b_q$ with $q=0,1,2,3,-2,-1$. We split the states into the retained set $P=\{a_q,b_q\}_{q=\pm1,\pm2}$ and the singlets $Q=\{a_q,b_q\}_{q=0,3}$. For an eigenvalue $k$ the eigenproblem is equivalent to
\begin{equation}
\hat H_{\rm eff}(k)\psi_P=k\,\psi_P,\qquad
\hat H_{\rm eff}(k)=\hat H_{PP}+\hat H_{PQ}\left(k-\hat H_{QQ}\right)^{-1}\hat H_{QP}.
\label{eq:SM_schur}
\end{equation}
Each exciton couples only to the photon with the same $q$, and the photons $a_0$ and $a_3$ are not coupled to each other. The correction therefore acts only on the photonic block, through the photon element of the singlet resolvent,
$(k-E_X)/[(k-\Omega_q)(k-E_X)-|V|^2]=1/[k-\Omega_q-|V|^2/(k-E_X)]$, $q=0,3$.
Through second order in $\bfk$ only the linear couplings $\nu_{01}$ and $\nu_{23}$ contribute, and $\hat H_{QQ}$ can be taken at $\bfk=0$. In the basis $(a_1,a_2,a_{-2},a_{-1},b_1,b_2,b_{-2},b_{-1})$
\begin{equation}
\hat H_{\rm eff}=
\begin{pmatrix}
\Omega_1+\Delta\tilde\Omega_1\bfk^2&\nu_{12}k_-&0&-\tilde\mu_{11}k_+^2& V^*&0&0&0\\
-\nu_{12}k_+&\Omega_2+\Delta\tilde\Omega_2\bfk^2&-\tilde\mu_{22}k_-^2&0&0 &V^*&0&0\\
0&-\tilde\mu_{22}k_+^2&\Omega_2+\Delta\tilde\Omega_2\bfk^2&\nu_{12}k_-&0&0&V^*&0\\
-\tilde\mu_{11}k_-^2&0&-\nu_{12}k_+&\Omega_1+\Delta\tilde\Omega_1\bfk^2&0&0&0&V^*\\
V&0&0&0&E_X&0&0&0\\
0&V&0&0&0&E_X&0&0\\
0&0&V&0&0&0&E_X&0\\
0&0&0&V&0&0&0&E_X
\end{pmatrix}
\label{eq:SM_H8}
\end{equation}
with the singlet-renormalized coefficients
\begin{equation}
\begin{aligned}
\Delta\tilde\Omega_1&=\Delta\Omega_1-\frac{\nu_{01}^2}{2\left[k-\Omega_0-|V|^2/(k-E_X)\right]}, &\qquad
\tilde\mu_{11}&=\mu_{11}-\frac{\nu_{01}^2}{k-\Omega_0-|V|^2/(k-E_X)},\\
\Delta\tilde\Omega_2&=\Delta\Omega_2-\frac{\nu_{23}^2}{2\left[k-\Omega_3-|V|^2/(k-E_X)\right]}, &
\tilde\mu_{22}&=\mu_{22}-\frac{\nu_{23}^2}{k-\Omega_3-|V|^2/(k-E_X)}.
\end{aligned}
\label{eq:SM_H8_renorm}
\end{equation}
All photons couple to their excitons with the same $V$ and $E_X$, so eliminating $b_{\pm1}$ and $b_{\pm2}$ from Eq.~\eqref{eq:SM_H8} adds the same self-energy $|V|^2/(k-E_X)$ to every photonic state. Together with Eq.~\eqref{eq:SM_H8_renorm}, the problem then depends on $V$ only through the combination $k-|V|^2/(k-E_X)$: the polaritonic eigenproblem at the frequency $k$ is the photonic eigenproblem at the frequency $k-|V|^2/(k-E_X)$. Every photonic branch $\Omega^{\rm ph}(\bfk)$ therefore gives two polariton branches
\begin{equation}
\Omega^{\rm pol}_\pm(\bfk)=\frac{\Omega^{\rm ph}(\bfk)+E_X}{2}\pm\sqrt{\frac{\left[\Omega^{\rm ph}(\bfk)-E_X\right]^2}{4}+|V|^2}.
\label{eq:SM_pol_branches}
\end{equation}

For the large-hole domains the frequency-dependent coefficients~\eqref{eq:SM_H8_renorm} are replaced by constants. We fit the seven complex coefficients $\Omega_{1,2}$, $\Delta\tilde\Omega_{1,2}$, $\nu_{12}$, $\tilde\mu_{11}$, $\tilde\mu_{22}$ to the COMSOL eigenfrequencies of the isolated domains, calculated without the excitonic response along $\Gamma$-$K$ and $\Gamma$-$M$ for $|\bfk|\le0.12\,g$, which covers the angular range of the reflectance maps in Fig.~5. The modes are assigned to the TE and TM blocks by the parity of $H_z$ under the vertical mirror containing $\bfk$, and all four blocks are fitted simultaneously. By Eq.~\eqref{eq:SM_pol_branches}, the same constants give the polariton branches, with $V$ as the only additional parameter.

The fitted constants do not contain the exciton-mediated zero-harmonic terms $|d|^2J^2/D$ and $\Lambda_{X,\alpha}$. These terms are frequency-independent corrections proportional to $|V|^2$. Since the coefficients of the large-hole domains are not given by the perturbative expressions, we do not add these corrections to the fitted constants.

Along $\Gamma$-$K$ and $\Gamma$-$M$, Eq.~\eqref{eq:SM_H8} splits into the TE and TM blocks of Eqs.~\eqref{H_K_TE}--\eqref{H_M_TM} without the rows and columns of $q=0,3$ and with $\Delta\Omega_q$, $\mu_{qq}$ replaced by $\Delta\tilde\Omega_q$, $\tilde\mu_{qq}$, so that one parameter set describes both directions.

\begin{figure}[h]
\centering
\IfFileExists{figures/SM_photon5_012.pdf}{\includegraphics[width=0.8\linewidth]{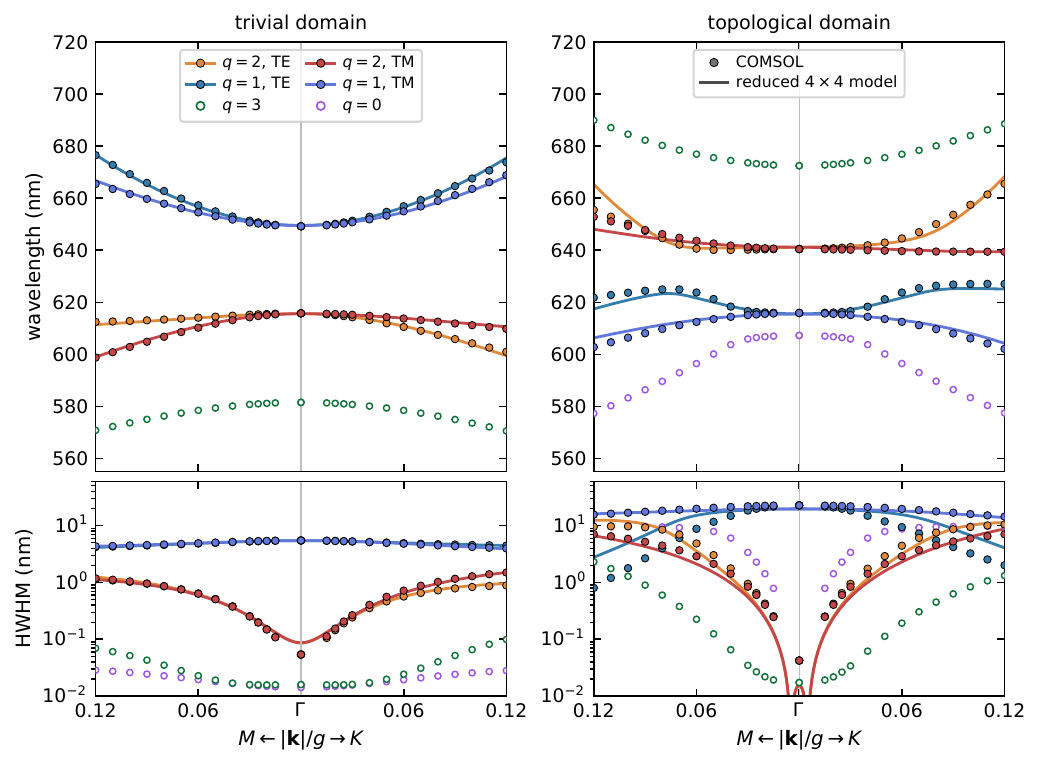}}{\fbox{\parbox{0.8\linewidth}{\centering\vspace{3cm}\texttt{figures/SM\_photon4.pdf}\vspace{3cm}}}}
\caption{Photonic spectra of the reduced Hamiltonian. The top panels show the resonance wavelengths and the bottom panels the half width at half maximum (HWHM) of the photonic eigenmodes of the isolated trivial (left) and topological (right) domains of Fig.~5. In each panel, the left half corresponds to the $\Gamma$-$M$ direction and the right half to $\Gamma$-$K$. Circles show the COMSOL eigenfrequencies. Filled circles mark the branches originating from the dipolar ($q=1$) and quadrupolar ($q=2$) doublets, which are assigned to the TE and TM blocks by mirror parity. Open circles mark the singlets $q=0,3$, which are eliminated in the reduced model. Solid lines show the eigenvalues of the photonic TE and TM blocks of Eq.~\eqref{eq:SM_H8}, labeled by their origin at the $\Gamma$ point. % Grey shading marks the region outside the fit window $|\bfk|\le0.12\,g$.
}
\label{fig:SM_eff4}
\end{figure}

The fitted parameters are listed in Table~\ref{tab:SM_eff4}, and the branches are compared with COMSOL in Fig.~\ref{fig:SM_eff4}. Within the fit window, the RMS deviation of the resonance wavelengths of the dipolar and quadrupolar branches from COMSOL is $0.44$~nm in the trivial domain and $1.85$~nm in the topological domain. The largest deviations occur for the TE branches of the topological domain, which hybridize strongly with the nearby singlet $q=0$.

The coupling $V$ is fitted to the TM $\Gamma$-$M$ reflectance maps of Fig.~5, calculated with the excitonic permittivity of Eq.~(17) of the main text. We fit $V$ to the minima of the upper and lower polariton branches of quadrupolar origin in both domains, using Eq.~\eqref{eq:SM_pol_branches} with the photonic TM $\Gamma$-$M$ branches. A common fit gives $V=4.38\times10^{-4}$~nm$^{-1}$ ($\hbar cV=86$~meV). 

\begin{table}[h]
\centering
\caption{Parameters of the reduced Hamiltonian~\eqref{eq:SM_H8} for the trivial ($r=150$~nm, $R=130$~nm) and topological ($r=140$~nm, $R=102$~nm) domains of Fig.~5 ($a=450$~nm, $2h=23$~nm).}
\label{tab:SM_eff4}
\begin{tabular}{lcc}
\hline
Parameter & Trivial domain & Topological domain \\
\hline
$\Omega_1$ &
$(9.6736-0.0810\iu)\times10^{-3}~\mathrm{nm}^{-1}$ &
$(10.2078-0.3269\iu)\times10^{-3}~\mathrm{nm}^{-1}$ \\

$\Omega_2$ &
$(10.2058-0.0014\iu)\times10^{-3}~\mathrm{nm}^{-1}$ &
$(9.7995+0.0002\iu)\times10^{-3}~\mathrm{nm}^{-1}$ \\

$\Delta\tilde\Omega_1$ &
$(-61.45-0.20\iu)~\mathrm{nm}$ &
$(-78.89+10.13\iu)~\mathrm{nm}$ \\

$\Delta\tilde\Omega_2$ &
$(23.44+1.07\iu)~\mathrm{nm}$ &
$(33.79-2.62\iu)~\mathrm{nm}$ \\

$\tilde\mu_{11}$ &
$(-34.83+0.70\iu)~\mathrm{nm}$ &
$(-161.52+39.95\iu)~\mathrm{nm}$ \\

$\tilde\mu_{22}$ &
$(-59.27+0.25\iu)~\mathrm{nm}$ &
$(44.52+0.64\iu)~\mathrm{nm}$ \\

$\nu_{12}$ &
$0.0198+0.2147\iu$ &
$-0.0081+0.2315\iu$ \\
\hline
$V$ & \multicolumn{2}{c}{$4.38\times10^{-4}~\mathrm{nm}^{-1}$ ($86$~meV)} \\
\hline
\end{tabular}
\end{table}

To characterize the topological properties of the hybrid system, we first consider the uncoupled photonic subsystem. Specifically, we extract the photonic block of the full effective Hamiltonian, corresponding to its first four rows and first four columns:
$
\hat H_{\rm ph}=\hat H_{\rm eff}[1:4,1:4].
$
and construct its Hermitianised form:
$
\hat H_{\rm ph}^{\rm H}
=\frac{1}{2}\left(
\hat H_{\rm ph}+\hat H_{\rm ph}^{\dagger}
\right).
$
We rearrange the photonic basis by interchanging the third and
fourth components, such that
$
(a_1,a_2,a_{-1},a_{-2})^T\equiv
(p_+,d_+,p_-,d_-)^T,
$
where $p_\pm$ and $d_\pm$ denote the circularly polarised
dipolar and quadrupolar modes, respectively. This permutation brings the Hamiltonian into the dipole--quadrupole representation used in the Wu--Hu description of shrink-expand hexagonal topological photonic crystals
\cite{WuHu2015,Gorlach2018}. In this basis, the pseudospin operator is
$
S_z=\operatorname{diag}(1,1,-1,-1).
$
In the absence of the quadratic inter-sector coupling terms,
$
\tilde{\mu}_{11}=\tilde{\mu}_{22}=0,
$
the photonic Hamiltonian separates into two independent pseudospin blocks,
$$
\hat H'_{\rm ph}
=
\left.\hat H_{\rm ph}\right|_{\tilde{\mu}_{11}=\tilde{\mu}_{22}=0}
=
H_+(\mathbf{k})\oplus H_-(\mathbf{k}).
$$
The two blocks are related by an effective pseudo-time-reversal symmetry constructed on the basis of the $C_{6v}$ point group symmetry~\cite{WuHu2015,Mei2016}. In the chosen basis, the corresponding antiunitary operator~\cite{Mei2016} is
$$
\Theta=(i\hat{\sigma}_y\otimes I_2)K
=
\begin{pmatrix}
0&I_2\\
-I_2&0
\end{pmatrix}K,
\qquad
\Theta^2=-1,
$$
where $\hat{\sigma}_y$ acts in pseudospin space, $I_2$ acts in the dipole--quadrupole subspace, and $K$ denotes complex conjugation. The symmetry condition is
\[
\Theta \hat H'_{\rm ph}(\mathbf{k})\Theta^{-1}
=\hat H'_{\rm ph}(-\mathbf{k}).
\]
This construction follows the effective pseudospin time-reversal description introduced for photonic and classical-wave systems
\cite{WuHu2015,Mei2016,Gorlach2018}.

In this limit, the Chern numbers of the two
pseudospin sectors can be calculated independently. For
a given photonic gap, we denote these by $C_+$ and $C_-$.
Time-reversal symmetry requires
$
C_-=-C_+.
$
The corresponding spin Chern number is
$
C_s=\left({C_+-C_-}\right)/{2}=C_+,
$
and the associated $\mathbb{Z}_2$ index is determined by
$
\nu_{\mathbb{Z}_2}=C_s\pmod 2.
$
Thus, in the pseudospin-conserving limit, the system
admits the usual spin-resolved topological description,
analogous to the continuum Bernevig-Hughes-Zhang (BHZ) model~\cite{bernevig2006quantum}. The Chern numbers are evaluated from the Berry curvature
of the corresponding bands by integrating the
Berry curvature over the momentum-space domain. In the simplest BHZ model, retaining only the dispersive diagonal mass term, the Berry curvature can be calculated analytically. Its momentum-space integration yields quantised Chern numbers, with their values determined by the model parameters and the ordering of the dipole-quadrupole bands at the $\Gamma$ point.
Numerically, we use a gauge-invariant discretised
Berry-phase reconstruction based on the
Fukui-Hatsugai-Suzuki method
\cite{Fukui2005}.

For finite $\tilde{\mu}_{11}$ and $\tilde{\mu}_{22}$, the quadratic terms couple states belonging to opposite pseudospin sectors. In this case, we use the projected-spin operator,
$
S_{\mathrm{proj}}(\mathbf{k})=
V_{\mathrm{o}}^\dagger(\mathbf{k})
S_z V_{\mathrm{o}}(\mathbf{k}),
$
where $V_{\mathrm{o}}(\mathbf{k})$ contains the eigenvectors belonging to the lower-frequency ``occupied'' bands. Its eigenvectors define two spin-derived subspaces. A nonzero projected-spin gap allows the occupied subspace to be continuously decomposed into two sectors. The Chern numbers of these two projected-spin sectors define the spin Chern number according to
$
C_s=\left({C_+-C_-}\right)/{2}.
$

For the fitted photonic dispersion, the quadratic momentum dependence in the topological domain produces a band inversion at approximately
$
k_{\mathrm{inv}}\simeq 0.118g,
$
which is within the momentum range used to obtain the local photonic fit near the $\Gamma$ point and provides the momentum-space scale relevant to the topological band rearrangement. The integration for quantifying Berry fluxes is performed over the momentum disk $\mathcal{D}$ with radius $0.5g$ centred at $\Gamma$, $k\leq 0.5g$. The disk is discretised using concentric radial rings and uniform angular sampling.

We then extend the analysis to the original non-Hermitian photonic Hamiltonian and apply a biorthogonal formulation. In this case, we calculate the left and right eigenvectors and impose the biorthogonal normalisation condition to evaluate the corresponding biorthogonal Berry connection and Berry curvature. This biorthogonal formulation is appropriate for complex-frequency band structures and follows the non-Hermitian topological framework~\cite{Esaki2011,Shen2018,Kawabata2019}.

A distinction in the non-Hermitian problem is that a band gap must be assessed using the complex spectrum. We therefore monitor the real-part direct gap between the occupied (lower) and unoccupied (upper) bands at every momentum point. We define the minimum direct real-part gap between the ordered bands 1--4 as
$
\Delta_{\mathrm{dir}}
=
\min_{\mathbf{k}\in\mathcal D}
\left[
\operatorname{Re}\omega_{3}(\mathbf{k})
-
\operatorname{Re}\omega_2(\mathbf{k})
\right],
$
whose positive value means that the local band gap remains open at every momentum in the integration domain. For completeness, a strict global line gap requires a common line in the complex-frequency plane that separates the two spectral groups (lower bands 1 and 2, and upper bands 3 and 4) at all momenta. For a vertical separating line, the corresponding condition is
$
\Delta_{\mathrm{global}}
=
\min_{\mathbf{k}\in\mathcal D}
\operatorname{Re}\omega_{3}(\mathbf{k})
-
\max_{\mathbf{k}\in\mathcal D}
\operatorname{Re}\omega_2(\mathbf{k})>0.
$
For both fitted photonic parameter sets, the local and global gaps remain open over the integration domain.

The analysis is performed and the integrals are evaluated numerically over the momentum disk $\mathcal{D}$, using successively refined momentum meshes. $C_s$ is obtained from the difference between the positive- and negative-spin-derived Berry fluxes. The biorthogonal Berry-flux calculations reveal a clear distinction between the two parameter regimes. In particular, the projected-spin-resolved Berry-flux combination exhibits strong numerical convergence within the integration domain $k<0.5g$ with increasing momentum-space resolution. For the topological parameter set, it approaches $C_s\simeq-1.046$, whereas the corresponding quantity for the trivial parameter set converges to approximately $+0.05$ at a polar radius-angle mesh of $(120,384)$. The projected-spin spectral separation remains finite throughout the integration domain in both cases. 
% These calculations demonstrate a distinction in the pseudospin-resolved Berry response of the two parameter sets. 
We emphasise that the integration is performed over a finite momentum-space disk, rather than a fully regularised Brillouin zone. Accordingly, the reported quantities are interpreted as converged finite-domain Berry-flux responses, rather than as exactly quantised Chern invariants.

Finally, we apply this analysis to the complete photon--exciton Hamiltonian $\hat H_{\rm eff}$, including the exciton linewidth and the finite photon--exciton coupling. In the eight-band photon--exciton model, we consider two distinct polariton gaps. The lower gap separates the first two bands from the remaining six bands. The upper gap separates the first six bands from the remaining two bands. The Berry fluxes are again calculated using a discretised biorthogonal version of the Fukui-Hatsugai-Suzuki method, applied separately to the two projected-spin subspaces.

We apply this procedure to the two fitted parameter sets, referred to as the trivial and topological domains. As a reference, we first set $\tilde{\mu}_{11}=\tilde{\mu}_{22}=0$ in the complete photon--exciton Hamiltonian. In this limit, the two pseudospin sectors decouple into independent $4\times4$ photon--exciton blocks. Each block has a lower polariton gap, with one occupied band, and an upper polariton gap, with three occupied bands. For the trivial domain, the resulting spin-resolved Berry fluxes give
$
C_s^{\mathrm{lower}}
=
C_s^{\mathrm{upper}}
=+0.036.
$
For the topological domain, the corresponding values are
$
C_s^{\mathrm{lower}}
=
C_s^{\mathrm{upper}}
=-0.967.
$
Thus, in the uncoupled-spin reference calculation, the two domains exhibit clearly different spin-resolved Berry-flux responses.

We next retain the finite quadratic inter-sector couplings and evaluate the complete non-Hermitian eight-band Hamiltonian. The resulting gap-resolved projected-spin Berry-flux differences are
$
C_s^{\mathrm{lower}}=+0.049$, 
$C_s^{\mathrm{upper}}=+0.232
$
in the trivial domain, whereas in the nontrivial domain they are
$
C_s^{\mathrm{lower}}=-1.053$,
$C_s^{\mathrm{upper}}=-1.034.
$
Thus, the spin Chern response remains close to zero in the trivial domain (with the lower-gap result being more numerically robust) and close to $-1$ in the nontrivial domain.

The present integration is performed over a finite momentum-space disk and does not employ a large-wavevector regularisation or a compactification of the continuum Hamiltonian to a complete Brillouin zone. Accordingly, even though the numerical plateaus already observed at $k<0.5g$ point to convergence of the finite-domain calculation, the calculated quantities are finite-domain biorthogonal Berry fluxes and projected-spin Berry-flux differences, rather than strictly quantised invariants.
In addition, the photonic parameters were constrained by best fitting over a momentum range of approximately $|\bfk|/g\lesssim0.12$. Nevertheless, within the specified effective-model calculation, the distinction between the two fitted configurations persists, surviving the inclusion of pseudo-spin mixing, exciton coupling, and loss.

\subsubsection{Propagation of the domain wall mode}
The supercell reflectance maps of Fig. 5 reveal the domain wall states only through their far-field response. To verify that these states propagate along the wall, we excite them directly in full-wave COMSOL calculations. We first calculate the eigenmodes of a ribbon with 10 topological and 10 trivial rows. As the source mode we take the upper domain wall branch at $|\bfk|/g = 0.02$ $(\lambda = 623$ nm). 

The propagation is calculated in a finite strip of 25 unit cells along the wall with the same $10 + 10$ rows. The source is an external current density in the patterned slab of a single unit cell $J \propto (E_x,\, E_y)$, where $E_x$ and $E_y$ are the in-plane fields of the ribbon eigenmode. For the polaritonic metasurface, the same strip is calculated with the excitonic permittivity of Eq. (17) of the main text. The photonic eigenmode profile is used as the source at the upper (598 nm, $+k$) and lower (653 nm, $-k$) polariton wavelengths, estimated from~Eq.\eqref{eq:SM_pol_branches}.

\begin{figure}[h]
\centering
\includegraphics[width=1\linewidth]{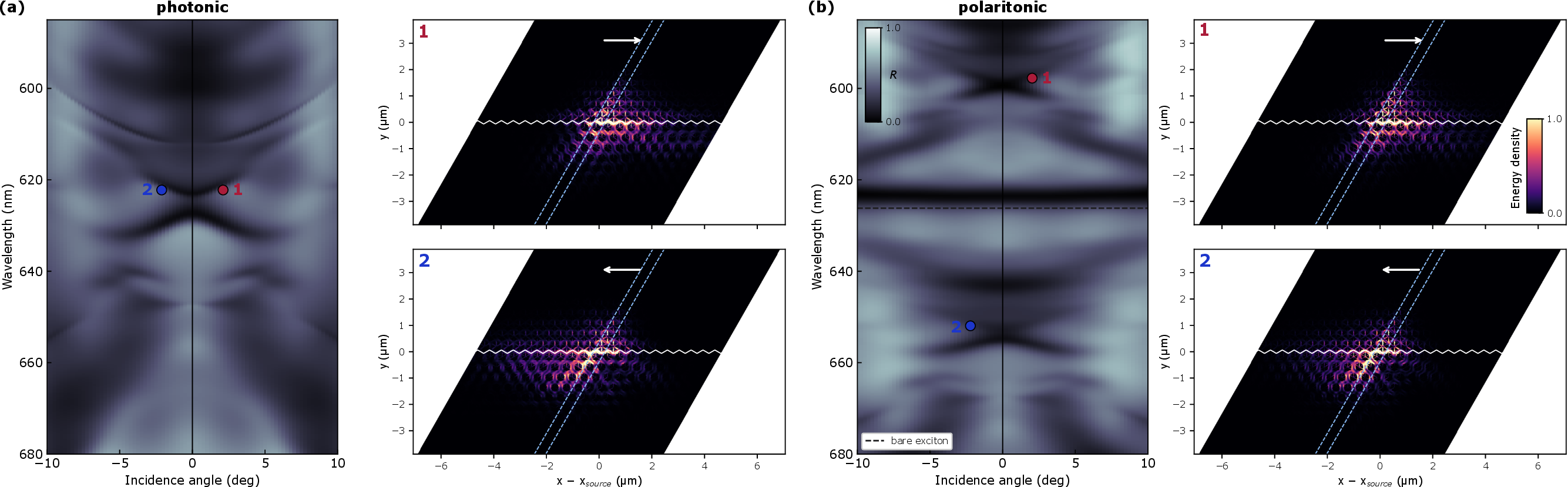}
\caption{Propagation of the domain wall mode of the photonic (a) and polaritonic (b) metasurface. The profile of the photonic domain wall eigenmode is used as a source in one unit cell at the wall (dashed), and the propagation maps show the resulting energy density in the slab mid-plane. Circles on the reflectance maps on the left mark the launched modes, 1 for $+k$ and 2 for $-k$. The energy decays along the wall over L = 1.34 $\mu$m $(+k)$ and 1.80 $\mu$m $(-k)$ at 623 nm, and over 1.12 $\mu$m for the upper (598 nm) and 1.20 $\mu$m for the lower polariton (653 nm).
}
\label{fig:SM_prop}
\end{figure}

Figure~\ref{fig:SM_prop} shows the resulting energy density in the slab mid-plane. We estimate the decay rate from the length $L$ over which the energy density at the wall drops by a factor of $e$. It is $L = 1.34$ and $1.80~\mu$m for $+k$ and $-k$ in the photonic case, and $1.12$ and $1.20~\mu$m for the upper and lower polariton. The propagation length is limited by the low quality factor of the modes in the large-hole domains, and for polaritons also by exciton absorption.

\subsection{Light-matter coupling and oscillator strength}

The light-matter coupling $V$ is the only parameter fitted to the polaritonic spectra. In this section we relate it to the oscillator strength $f$ of the Lorentz permittivity, Eq.~(17) of the main text, which is used in the full-wave calculations, and compare the result with the fitted couplings and with the couplings of the individual multipoles.

We substitute the guided-mode projection of the field, $\bfe_{\bfg}(z)=\eta\hbar c\,\theta(z)\hat{\mathbf e}_{{\rm s},\bfg}a_{{\rm s},\bfg}$, into Eq.~\eqref{eq:SM_exc_fourier} and use the excitonic polarization $\mathbf P_{\bfg}(z)=\sqrt{\xi_{\mathbf 0}}\,F(z)\sum_\alpha\mathbf d^*_\alpha b_{\alpha,\bfg}$. With $k-E_X=(\omega-\omega_X+\iu\Gamma)/c$, this gives
\begin{equation*}
\mathbf P_{\bfg}(z)=\frac{\xi_{\mathbf 0}|d|^2\eta c\,\hat{\mathbf e}_{{\rm s},\bfg}\,a_{{\rm s},\bfg}}{\omega_X-\omega-\iu\Gamma}\,F(z)\intzp F(z')\theta(z').
\end{equation*}
The polarization enters the photonic equation only through its projection $\int_{-h}^{h}\mathrm dz\,\theta(z)\,\hat{\mathbf e}_{{\rm s},\bfg}\cdot\mathbf P_{\bfg}(z)$ on the guided mode. Therefore, we can construct an equivalent local isotropic permittivity $\eps(\omega)$ that can be used in the full-wave calculations for the excitonic response as in Eq.~(17) of the main text, assigned to WS$_2$ only. With the coefficients $\xi_m$ ($m\geq1$) neglected, as in Eq.~\eqref{eq:SM_exc_fourier}, we can write within the effective model $\mathbf P_{\bfg}=\xi_{\mathbf 0}[\eps(\omega)-\eps_\infty]\bfe_{\bfg}/(4\pi)$, and its projection is $\xi_{\mathbf 0}[\eps(\omega)-\eps_\infty]N_h\eta\hbar c\,a_{{\rm s},\bfg}/(4\pi)$. This gives
\begin{equation*}
\eps(\omega)=\eps_\infty+\frac{4\pi|d|^2\left|\intz\theta(z)F(z)\right|^2}{\hbar N_h\left(\omega_{\rm X}-\omega-\iu\Gamma\right)}
=\eps_\infty+\frac{c\,|V|^2}{\xi_{\bf 0}k_gN_h\left(\omega_{\rm X}-\omega-\iu\Gamma\right)}.
\label{r11:eps_res}
\end{equation*}
The pole of this equation $\omega=\omega_{\rm X}-\iu\Gamma$ lies in the lower half plane and describes a decaying exciton for the $\eu^{-\iu\omega t}$ convention. 
Equation~\eqref{eq:SM_exc_fourier} is written in the resonant approximation. A real response requires $\eps(-\omega^*)=\eps^*(\omega)$, which adds the partner pole at $\omega=-\omega_{\rm X}-\iu\Gamma$:
\begin{equation*}
\eps(\omega)=\eps_\infty+\frac{c\,|V|^2}{\xi_{\bf 0}k_gN_h}\left[\frac{1}{\omega_{\rm X}-\omega-\iu\Gamma}+\frac{1}{\omega_{\rm X}+\omega+\iu\Gamma}\right]
=\eps_\infty+\frac{2c\,\omega_{\rm X}|V|^2}{\xi_{\bf 0}k_gN_h\left(\omega_{\rm X}^2-\omega^2-2\iu\Gamma\omega+\Gamma^2\right)}.
\end{equation*}
Both poles lie in the lower half plane. We neglect $\Gamma^2$ in the denominator and substitute $\omega=2\pi c/\lambda$ and $\omega_{\rm X}=2\pi c/\lambda_{\rm X}$. This gives 
\begin{equation}
\eps(\lambda)=\eps_\infty+\frac{f}{\lambda_{\rm X}^{-2}-\lambda^{-2}-\iu\Gamma\lambda^{-1}/(\pi c)},
\qquad
f=\frac{\omega_{\rm X}|V|^2}{2\pi^2c\,\xi_{\bf 0}k_gN_h}.
\label{eq:SM_fV}
\end{equation}

In the effective slab limit of Sec.~S1A, $k_g\to k_{\rm TE_0}(g)$, and $\theta(z)$ is the TE$_0$ profile of Sec.~S4, normalized by $\int\mathrm dz\,\veps(z)\theta^2(z)=1/2$. In this limit, Eq.~\eqref{eq:SM_fV} gives the coupling directly in terms of the parameters of the structure
\begin{equation}
|V_f|^2=\frac{2\pi^2c\,\xi_{\mathbf 0}f}{\omega_{\rm X}}\,k_{\rm TE_0}(g)\intz\theta^2(z).
\label{eq:SM_Vf}
\end{equation}

For the structure of Figs.~2 and~3 ($a=405$~nm, $2h=20$~nm, $\veps=18.98$, $\xi_{\mathbf 0}=0.934$), with $f=0.52~\mu$m$^{-2}$ and $\lambda_{\rm X}=626.2$~nm, we obtain $k_{\rm TE_0}(g)=9.626\times10^{-3}$~nm$^{-1}$ and $\int_{-h}^{h}\mathrm dz\,\theta^2(z)=0.02275$, the latter differs from the fitted $N_h=0.01937$ of Table~\ref{tab:fit_fig34}, which is an effective parameter of the in-plane dielectric couplings $k_gN_h\eps_m$ (Sec.~S1A). This gives $V_f=4.57\times10^{-4}$~nm$^{-1}$ ($\hbar cV_f=90.3$~meV), compared with the fitted $\hbar cV=91.3$~meV of Table~\ref{tab:fit_fig34}.

For the trivial and topological domains of Fig.~5 ($a=450$~nm, $2h=23$~nm, and $\xi_{\mathbf 0}=0.667$ and $0.710$, respectively), where the six-harmonic expressions do not apply (Sec.~S4C), Eq.~\eqref{eq:SM_Vf} still gives $\hbar cV_f=87.6$ and $87.3$~meV, close to the fitted $86.4$~meV (Sec.~S4C).

We also note that roughly we can approximate $k_{\rm TE_0}(g)\simeq\omega_{\rm X}/c$  and $\intz\theta^2(z)\simeq 1/(2\veps)$, so that
\begin{equation}
|V_f|^2\simeq\frac{\pi^2 \xi_{\mathbf 0}f}{\veps}.
\label{eq:SM_Vf}
\end{equation}

Table~\ref{tab:SM_Vcomp} collects, for Figs.~2 and~3, the common $V$ of Table~\ref{tab:fit_fig34}, fitted to the reflectance minima of the polaritonic RCWA maps with all photonic parameters fixed (Sec.~S4A), the coupling $V_f$, and the couplings of the individual multipoles (Sec.~S2B) fitted to the same minima. For the latter, the four $V_q$ are fitted simultaneously with the same assignment rule as for the common $V$. At the $\Gamma$ point the photon and the exciton of the multipole $q$ couple only to each other, so $V_q$ is set by the splitting of the two polariton branches originating from this multipole.
\begin{table}[h]
\centering
\setlength{\tabcolsep}{8pt}
\caption{Light-matter coupling $\hbar cV$ (meV) for the structure of Figs.~2 and~3 and the RMS deviation (nm) from the reflectance minima of the polaritonic RCWA maps, to which the common $V$ and the couplings $V_q$ are fitted. $V_f$ is given by Eq.~\eqref{eq:SM_Vf} without adjustable parameters.}
\label{tab:SM_Vcomp}
\begin{tabular}{lccccc}
\hline
Approach & $V_0$ & $V_{\pm1}$ & $V_{\pm2}$ & $V_3$ & RMS \\
\hline
Common fit $V$ (Table~\ref{tab:fit_fig34}) & \multicolumn{4}{c}{$91.3$} & $0.80$ \\
Oscillator strength $V_f$, Eq.~\eqref{eq:SM_Vf} & \multicolumn{4}{c}{$90.3$} & --- \\
Fit of $V_q$ & $90.7$ & $91.7$ & $89.3$ & $94.7$ & $0.78$ \\
\hline
\end{tabular}
\end{table}
\\

Thus, all couplings in Table~\ref{tab:SM_Vcomp} are close to the common $V$. The coupling $V_f$, obtained without adjustable parameters, practically coincides with it, and the individually fitted $V_q$ reduce the RMS deviation only from 0.80 to 0.78~nm despite three additional parameters. The common coupling $V$ used in the main text is therefore a good approximation.

\bibliography{SM}